\documentclass[11pt]{article}
\pdfoutput=1
\usepackage{amsmath,amssymb,epsfig,amsfonts}
\usepackage{graphicx}
\usepackage{subfig}
\usepackage[usenames, dvipsnames]{color}
\usepackage{hyperref}
\usepackage{cite}
\usepackage{multirow}
\usepackage{slashed}
\usepackage{verbatim} 
\usepackage{enumitem}
\usepackage{rotating}
\usepackage{xcolor}
\usepackage{multirow}
\usepackage{bm}
\usepackage[normalem]{ulem}
\usepackage{cleveref}
\usepackage{booktabs}

\usepackage{tikz}
\usepackage{diagbox}
\usetikzlibrary{positioning}
\usetikzlibrary{calc}
\usetikzlibrary{decorations.pathreplacing,calligraphy}

\usetikzlibrary{arrows}

\usepackage{xstring}
\usetikzlibrary{decorations.pathmorphing} 
\usetikzlibrary{decorations.markings} 
\usetikzlibrary{arrows} 
\usetikzlibrary{shapes} 
\usetikzlibrary{matrix} 
\usetikzlibrary{positioning} 
\usepackage[english]{babel} 
\usepackage[autostyle]{csquotes}

\tikzstyle{brane}=[draw]
\tikzset{D7/.style={circle, draw=black, inner sep=0pt, fill=white, minimum size=3mm}}
\tikzset{hasse/.style={circle, fill,inner sep=2pt}}
\tikzset{flavor/.style={regular polygon,regular polygon sides=4,inner sep=2.5pt, draw}}
\tikzset{gauge/.style={circle, draw,inner sep=2.5pt}}
\tikzset{gaugeb/.style={circle, draw,fill=black,inner sep=2.5pt}}
\tikzset{gaugecyan/.style={circle, draw,fill=cyan,inner sep=2.5pt}}
\tikzset{gaugegreen/.style={circle, draw,fill=green,inner sep=2.5pt}}
\tikzset{gaugeblue/.style={circle, draw,fill=blue,inner sep=2.5pt}}
\tikzset{gaugeorange/.style={circle, draw,fill=orange,inner sep=2.5pt}}
\tikzset{bd/.style={circle, draw=black, inner sep=0pt, fill=black, minimum size=2mm}}
\tikzset{wd/.style={circle, draw=black, inner sep=0pt, fill=white, minimum size=2mm}}
\tikzset{Dynkin/.style={circle, draw=black, inner sep=0pt, fill=white, minimum size=2mm}}
\tikzstyle{ligne}=[draw, thick] 
\tikzset{doublearrow/.style={ draw=black!75, color=black!75, thick, double distance=3pt, }}

\makeatletter

\newtheorem{definition}{Definition}

\newtheorem{proposition}{Fact}
\newtheorem{conjecture}{Conjecture}

\newcommand{\example}{\noindent{\bf Example. }}

\newtheorem{algorithm}{Algorithm}

\newcommand{\TD}{P} 

\newcommand{\be}{\begin{equation}}
\newcommand{\ee}{\end{equation}}
\newcommand{\ba}{\begin{aligned}}
\newcommand{\ea}{\end{aligned}}

\newcommand{\bea}{\begin{eqnarray}}
\newcommand{\eea}{\end{eqnarray}}

\newcommand{\R}{{\mathbb R}}
\newcommand{\Z}{{\mathbb Z}}

\def\half{{\frac{1}{2}}}

\def\p{\partial}

\def\unit{{1\kern-.65ex {\rm l}}}
\def\1{{1\kern-.65ex {\rm l}}}

\newcount\hour \newcount\minute
\hour=\time \divide \hour by 60
\minute=\time
\def\now{%
\ifnum \hour<13
  \ifnum \hour=0 \advance \hour by 12 \number\hour:\else \number\hour:\fi%
     \ifnum \minute<10 0\fi%
     \number\minute%
\ A.M.%
\else \advance \hour by -12 \number\hour:%
  \ifnum \minute<10 0\fi%
  \number\minute%
  \ P.M.%
\fi%
}

\makeatother

\begin{document}

\baselineskip=18pt  
\numberwithin{equation}{section}  
\allowdisplaybreaks  


%
%

\begin{flushright}{\tt Imperial/TP/20/AB/01}
\end{flushright}


\thispagestyle{empty}

\vspace*{1.2cm} 
\begin{center}

{
\huge {(Symplectic) Leaves and (5d Higgs) Branches   \\
\bigskip 
in the Poly(go)nesian Tropical Rain Forest
}}

 \vspace*{1.5cm}
Marieke van Beest$^1$, Antoine Bourget$^2$, Julius Eckhard$^1$,  Sakura Sch\"afer-Nameki$^1$\\

 \vspace*{1.0cm} 
{\it ${}^1$ Mathematical Institute, University of Oxford, \\
Andrew-Wiles Building,  Woodstock Road, Oxford, OX2 6GG, UK}\\
\smallskip
{\it ${}^2$ Theoretical Physics Group, The Blackett Laboratory, Imperial College London, \\
Prince Consort Road London, SW7 2AZ, UK}\\

\vspace*{0.8cm}
\end{center}
\vspace*{.5cm}

\noindent
We derive the structure of the Higgs branch of 5d superconformal field theories or gauge theories from their realization as a generalized toric polygon (or dot diagram). This approach is motivated by a dual, tropical curve decomposition of the $(p,q)$ 5-brane-web system. 
We define an edge coloring, which provides a decomposition of the generalized toric polygon into a refined Minkowski sum of sub-polygons, from which we compute the magnetic quiver. The Coulomb branch of the magnetic quiver is then conjecturally identified with the 5d Higgs branch. 
Furthermore, from partial resolutions, we identify the symplectic leaves of the Higgs branch and thereby the entire foliation structure. In the case of strictly toric polygons, this approach reduces to the description of deformations of the Calabi-Yau singularities in terms of Minkowski sums. 

\newpage

\tableofcontents


\section{Introduction}

The geometry of singular Calabi-Yau three-folds, so-called canonical singularities, is intimately related with the 
physics of 5d superconformal field theories (SCFTs). The moduli spaces of such singularities 
reflect the Higgs and Coulomb branches of the SCFT. The relation between these is furnished by M-theory: M-theory on a (non-compact) canonical Calabi-Yau three-fold singularity gives rise to a 5d SCFT, whereas resolving the singularity, i.e. K\"ahler deformations, correspond to the (extended) Coulomb branch, of vacuum expectation values (vevs)
of adjoint scalars in the vector multiplet. Deformations on the other hand map out the Higgs branch, i.e. the parameter space of vevs of hypermultiplets. 

From the geometry of the Calabi-Yau singularities, the resolutions are fairly well understood and mapped to the extended Coulomb branch of numerous theories (for the first studies see \cite{Morrison:1996xf}, and recent developments \cite{Intriligator:1997pq, Hayashi:2013lra, Hayashi:2014kca, DelZotto:2017pti, Jefferson:2017ahm, Closset:2018bjz, Jefferson:2018irk, Bhardwaj:2018vuu, Apruzzi:2019vpe, Apruzzi:2019opn, Apruzzi:2019enx, Bhardwaj:2019jtr, Apruzzi:2019kgb, Bhardwaj:2019fzv, Bhardwaj:2019xeg, Eckhard:2020jyr, Bhardwaj:2020kim}). On the contrary the deformations and their relation to the Higgs branch have been somewhat less systematically studied, and only recently a systematic study for isolated hyper-surface singularities \cite{yau2005classification, Xie:2017pfl} was initiated \cite{Closset:2020scj}.
A precise map can be achieved for strictly convex toric polygons realizing Calabi-Yau singularities \cite{Altmann,Altmann2}. 

On the other hand, for many 5d SCFTs, including those having a description in terms of 5-brane-webs \cite{Aharony:1997ju,Aharony:1997bh,Benini:2009gi,Bergman:2013aca}, there has been recent progress towards a comprehensive description of the Higgs branch using the so-called \emph{magnetic quivers} (MQ) and \emph{Hasse diagrams} \cite{Cremonesi:2015lsa,Ferlito:2017xdq,Cabrera:2018ann,Cabrera:2018jxt,Cabrera:2019izd,Bourget:2019aer,Bourget:2019rtl,Cabrera:2019dob,Grimminger:2020dmg, Eckhard:2020jyr,  Bourget:2020gzi, Bourget:2020asf, Akhond:2020vhc}. 
Magnetic quivers are graphs which, under certain conditions, provide a combinatorial description of a class of algebraic varieties. The key step of this construction is to interpret these graphs as 3d $\mathcal{N}=4$ quiver gauge theories, whose 3d Coulomb branch give a physical realization of these varieties. These 3d Coulomb branches can be quantitatively studied using the monopole formula of \cite{Cremonesi:2013lqa}. 
For the 5d SCFTs studied in the present paper, the magnetic quiver is a 3d $\mathcal{N}=4$ quiver gauge theory with unitary gauge groups, whose Coulomb branch is conjectured to be the same symplectic singularity as the Higgs branch of the 5d theory. 
From a geometric point of view, studying the realization of 5d SCFTs in M-theory on canonical singularities, magnetic quivers were derived in \cite{Closset:2020scj}.
The Hasse diagram, introduced in \cite{Bourget:2019aer}, is a depiction of the partially ordered set corresponding to the foliation of the Higgs branch by \emph{symplectic leaves}. Any pair of leaves which can be compared in the partial order defines a \emph{transverse slice} between the two leaves, which is again a symplectic singularity, and to which one can associate a magnetic quiver. The leaves correspond physically to phases of the SCFT, while the transverse slices are associated to new theories obtained from the original theory by (a generalization of) the Higgs mechanism. 

Ideally these insights should have a counterpart in the M-theory construction of said 5d SCFTs. 
Geometrically, there exists so far no comprehensive analysis. 
To counteract this, we propose in this paper a reverse approach: utilizing the map of 5-brane-webs to generalized toric polygons (GTP), or dot-diagrams \cite{Aharony:1997bh,Benini:2009gi,Bao:2011rc,Taki:2014pba} (obtained by dualizing the web), we identify the rules for determining the magnetic quiver and Hasse diagrams in terms of the polygons.

This proposal for computing the magnetic quiver from the GTP has several advantages: it first of all seems to be simpler to implement than the procedure in the brane-webs. Secondly, and more importantly, we hope this provides the first step to understanding the Higgs branch from a geometric point of view: when the GTP is a conventional toric diagram, there is a known map to an actual Calabi-Yau geometry. Furthermore, in the specialization to strictly convex toric diagrams our prescription agrees with the Minkowski sum decomposition approach of Altmann \cite{Altmann,Altmann2}. 

Let us briefly summarize the logic:  
The starting point is the derivation of the magnetic quiver and Hasse diagrams in the brane-webs, which relies on identifying sub-webs, which can move freely in the directions orthogonal to the 5-branes. Using insights from tropical geometry, an intersection between such sub-webs can be defined consistently \cite{Cabrera:2018jxt}. The resulting graph is identified with a magnetic quiver. 

In turn, in our approach, we identify the sub-webs in the dual GTP as an edge coloring, which requires that edges of one color define a closed sub-polygon. Furthermore, each sub-polygon is a refined Minkowski sum of multiple copies of an irreducible GTP, which obeys the so-called s-rule minimally. These sub-polygons play the role of supersymmetric sub-webs in the 5-brane-web. 
To a GTP with a consistent edge coloring we then associate a Tropical Quiver (TQ), where each color maps to a node in the tropical quiver with labels given by the multiplicity of the corresponding sub-polygon, whereas the edges receive contributions from the mixed volume \cite{MaclaganSturmfels}\footnote{The mixed volume and its role in the intersection of tropical curves features in the chapter ``Tropical Rain Forest" in \cite{MaclaganSturmfels}, which inspired our title.} (which maps to the stable intersection of two tropical curves in the brane-web) and  from sharing external edges in the GTP (corresponding to the 7-brane contribution in the brane-web). In addition, there are nodes in the tropical quiver that arise from vertices along an edge of the GTP, that are not associated to a colored sub-polygon. In general there exist several inequivalent colorings of a given GTP. Each coloring gives rise to a tropical quiver, which is identified with the magnetic quiver of one of the cones, that comprise the 5d Higgs branch of the theory realized by the GTP. The union of all the cones \cite{Ferlito:2016grh,Bourget:2019rtl}, intersecting in a pattern that can be read equivalently either from the construction of the tropical quivers via colorings, or from the construction of the magnetic quivers via the brane-webs, gives the full Higgs branch. 

Our approach is motivated from various points of view:
although the map to webs is one-to-one, much redundant information, such as the specific values of the Coulomb branch parameters, is not encoded in the generalized toric polygons. As such, several operations are far easier in the GTP. 
The process of flop transitions of curves connecting to external edges, which in physics terms corresponds to decoupling hypermultiplet matter, is realized much simpler in the generalized toric polygons, and will be used in \cite{BBESII}. Similarly, the process of ``pruning'' is the polygon analog of brane creations/annihilations, or Hanany-Witten moves \cite{Hanany:1996ie}, and is far simpler to implement. 

As we emphasized already, this approach provides a first step to generalizing the deformation theory of strictly convex Calabi-Yau singularities. 
Applied to strictly convex polygons, our approach reproduces precisely the mathematical results of Altmann \cite{Altmann,Altmann2}: the deformations are parametrized in terms of the Minkowski sum decomposition of the polygon. 
The Altmann algorithm then further maps these to the algebraic deformations of the canonical singularity. This last step requires to first find the map from GTPs to canonical singularities. This is a very interesting question to which we hope to return in the near future. 

In addition to deriving the magnetic quiver from the GTP, we also construct the Hasse diagram, and thereby the symplectic leaves of the Higgs branch. 
The Hasse diagram is obtained by introducing internal edges, which in 5d gauge theory corresponds to turning on Coulomb branch (not extended Coulomb branch) parameters. These identify sub-polygons, which define Minkowski sum decompositions involving the GTP of the theories that comprise the leaves in the Higgs branch. By successive application of this process of opening up a partial Coulomb branch, we can reconstruct the full Hasse diagram from the GTP.

Finally we should remark on some insights we gained from studying general webs/GTPs. When studying these constructions from a general point of view as we do here, a natural question is whether any web/GTP gives rise to a 5d SCFT. Clearly there are several basic consistency requirements that need to be satisfied: charge conservation (which is implemented in the GTP by these forming polygons) and supersymmetry, which is implemented in terms of the s-rule. 
We in fact provide a slight generalization of the standard s-rule in section \ref{sec:sub-webs}, which is manifestly $SL(2,\mathbb{Z})$-invariant. 
However we argue that these conditions alone still can be insufficient to guarantee that the theory realized on the 5-brane system is in fact a 5d SCFT. 
We propose that any web/GTP should in addition satify the {\it r-rule} (rank-rule), which states that 
\be
r \geq 0 \,,
\ee 
where $r$ is computed in terms of the web/GTP data. In fact the r-rule is strictly stronger than the s-rule, which is implied by the former when applied to triangle GTPs/trinion webs. It would be indeed very interesting to derive this r-rule from first principles.

The plan of this paper is as follows: we begin with a brief summary and overview of magnetic quivers and Hasse diagrams in section \ref{sec:Review}. 
We then present in section \ref{sec:ColoringRules} our proposed edge coloring and its relation to the magnetic quiver for 5d SCFTs, i.e. GTPs without internal edges. In section \ref{sec:Internal} we generalize to GTPs with internal edges and provide an algorithm to determine symplectic leaves and Hasse diagram, which is based on the introduction of internal edges. 
In section \ref{sec:Examples} we provide an extensive list of examples. Finally in section \ref{sec:BraneWebs} we give a derivation of our rules via the brane-webs and tropical geometry. Appendix \ref{app:Webs} provides a lightning review of brane-webs and the rules associated to them. 
The paper is accompanied by an ancillary {\tt Mathematica} file, which allows the reader to input GTPs and compute automatically the magnetic quivers. A documentation of this {\tt Mathematica} code is provided in appendix \ref{app:MMA}. Enjoy!

\section{Strategy: Higgs Branches and Magnetic Quivers from GTP}
\label{sec:Review}

In this paper we consider 5d SCFTs defined by so-called generalized toric polygons (GTP) $P$ (or dot diagrams) introduced in  \cite{Benini:2009gi}. We denote these theories by $\mathcal{T}_P$. 
The GTPs are lattice polygons, which generalize the concept of a toric fan for a Calabi-Yau three-fold. They map one-to-one to a 5-brane-web $\mathcal{W}_P$ (which in the toric case corresponds equivalently to a tropical geometry)
\be\label{PwP}
P \quad \longleftrightarrow \quad \mathcal{W}_P \,.
\ee
In the case when $P$ is a toric polygon, we can associate an actual Calabi-Yau three-fold geometry $\mathbf{X}_P$ to it, and the dual web is associated to a tropical geometry \cite{MaclaganSturmfels}. 

The moduli spaces of such 5d SCFTs are parametrized by the vevs of scalar fields, either in the vector multiplet -- the Coulomb branch (CB) -- or the hypermultiplet -- Higgs branch (HB). One of the challenges has been to compute the Higgs branch from a geometric approach to 5d SCFTs, though recent progress has been made in \cite{Closset:2020scj} for hypersurface singularities. Unlike the CB in 5d, the HB receives quantum corrections from instantons -- in M-theory on a canonical singularity, these are M2-brane instantons.
Computing these directly in 5d is a formidable task. 

In the 5-brane-webs, an alternative proposal was made that identifies the HB in 5d with the CB of a magnetic quiver (MQ), which is a 3d $\mathcal{N}=4$ quiver gauge theory (and in the current context, with $U(N)$ gauge nodes). 
The conjecture in \cite{Cabrera:2018jxt} is that the Coulomb branch of the magnetic quiver $\text{MQ}_P$ associated to a GTP $P$ can be identified with the Higgs branch of the 5d SCFT
\be
\text{CB} [\text{MQ}_P]= \text{HB} [\mathcal{T}_P] \,.
\ee
Both of these spaces are hyper-K\"ahler cones, and the isomorphism is as such. The advantage however is that the CB of 3d $\mathcal{N}=4$ theories is much better under control, using the monopole formula to compute their (refined) Hilbert series \cite{Cremonesi:2013lqa}, and subsequent work \cite{Bullimore:2015lsa, Nakajima:2015txa, Braverman:2016pwk,Nakajima:2017bdt}. 
The dressed monopole operators in 3d resum the 5d instanton corrections, and conjecturally this yields the correct hyper-K\"ahler metric for the 5d HB.  

A derivation using the M-theory geometry makes use of dualities in string theory, relating the 5d SCFT to a 4d SCFT obtained by compactifying Type IIB on the same geometry \cite{Closset:2020scj}. Reducing to 3d and applying mirror symmetry, realized as T-duality, identifies the magnetic quiver in certain instances.

\begin{table}
\centering
\begin{tabular}{|c | c|}\hline 
Singularity & Magnetic quiver\\
\hline\hline 
\begin{tabular}{l}
$\mathfrak{a}_n$
\end{tabular}
&
\begin{tabular}{l}
\begin{tikzpicture}[x=.8cm,y=.8cm]
\node (g1) at (0,0) [gauge,label=left:\large{1}] {};
\node (g2) at (1,0) [] {$\cdots$};
\node (g3) at (2,0) [gauge,label=right:\large{1}] {};
\node (g4) at (1,1) [gauge,label=left:\large{1}] {};
\draw (g1)--(g2)--(g3)--(g4)--(g1);
\end{tikzpicture}
\end{tabular}
\\
\hline
\begin{tabular}{l}
$\mathfrak{d}_n$
\end{tabular}
&
\begin{tabular}{l}
\begin{tikzpicture}[x=.8cm,y=.8cm]
\node (g1) at (0,0) [gauge,label=left:\large{1}] {};
\node (g2) at (1,0) [gauge,label=above right:\large{2}] {};
\node (g3) at (2,0) [] {$\cdots$};
\node (g4) at (3,0) [gauge,label=above left:\large{2}] {};
\node (g5) at (4,0) [gauge,label=right:\large{1}] {};
\node (g6) at (1,1) [gauge,label=left:\large{1}] {};
\node (g7) at (3,1) [gauge,label=right:\large{1}] {};

\draw (g1)--(g2)--(g3)--(g4)--(g5);
\draw (g2)--(g6);
\draw (g4)--(g7);
\end{tikzpicture}
\end{tabular}
\\
\hline
\begin{tabular}{l}
$\mathfrak{e}_6$
\end{tabular}
&
\begin{tabular}{l}
\begin{tikzpicture}[x=.8cm,y=.8cm]
\node (g1) at (0,0) [gauge,label=below:\large{1}] {};
\node (g2) at (1,0)[gauge,label=below:\large{2}] {};
\node (g3) at (2,0) [gauge,label=below:\large{3}] {};
\node (g4) at (3,0) [gauge,label=below:\large{2}] {};
\node (g5) at (4,0) [gauge,label=below:\large{1}] {};
\node (g6) at (2,1) [gauge,label=left:\large{2}] {};
\node (g7) at (3,1) [gauge,label=right:\large{1}] {};
\draw (g1)--(g2)--(g3)--(g4)--(g5);
\draw (g3)--(g6)--(g7);
\end{tikzpicture}
\end{tabular}
\\
\hline
\begin{tabular}{l}
$\mathfrak{e}_7$
\end{tabular}
&
\begin{tabular}{l}
\begin{tikzpicture}[x=.8cm,y=.8cm]
\node (g1) at (0,0) [gauge,label=below:\large{1}] {};
\node (g2) at (1,0)[gauge,label=below:\large{2}] {};
\node (g3) at (2,0) [gauge,label=below:\large{3}] {};
\node (g4) at (3,0) [gauge,label=below:\large{4}] {};
\node (g5) at (4,0) [gauge,label=below:\large{3}] {};
\node (g6) at (5,0) [gauge,label=below:\large{2}] {};
\node (g7) at (6,0) [gauge,label=below:\large{1}] {};
\node (g8) at (3,1) [gauge,label=left:\large{2}] {};
\draw (g1)--(g2)--(g3)--(g4)--(g5)--(g6)--(g7);
\draw (g4)--(g8);
\end{tikzpicture}
\end{tabular}
\\
\hline
\begin{tabular}{l}
$\mathfrak{e}_8$
\end{tabular}
&
\begin{tabular}{l}
\begin{tikzpicture}[x=.8cm,y=.8cm]
\node (g1) at (0,0) [gauge,label=below:\large{1}] {};
\node (g2) at (1,0)[gauge,label=below:\large{2}] {};
\node (g3) at (2,0) [gauge,label=below:\large{3}] {};
\node (g4) at (3,0) [gauge,label=below:\large{4}] {};
\node (g5) at (4,0) [gauge,label=below:\large{5}] {};
\node (g6) at (5,0) [gauge,label=below:\large{6}] {};
\node (g7) at (6,0) [gauge,label=below:\large{4}] {};
\node (g8) at (7,0) [gauge,label=below:\large{2}] {};
\node (g9) at (5,1) [gauge,label=left:\large{3}] {};
\draw (g1)--(g2)--(g3)--(g4)--(g5)--(g6)--(g7)--(g8);
\draw (g6)--(g9);
\end{tikzpicture}
\end{tabular}
\\
\hline
\begin{tabular}{l}
$A_{N-1}$
\end{tabular}
&
\begin{tabular}{l}
\begin{tikzpicture}[x=.8cm,y=.8cm]
\node (g1) at (0,0) [gauge,label=left:\large{1}] {};
\node (g2) at (1.5,0) [gauge,label=right:\large{1}] {};
\draw (g1)--(g2) node[midway,below] {$N$};
\end{tikzpicture}
\end{tabular}
\\\hline 
\end{tabular}
\caption{Summary of the elementary transverse slices that can appear in the Hasse diagram of a symplectic variety defined by a unitary and simply laced magnetic quiver. For the first five lines, these are closures of minimal nilpotent orbits of type $\mathfrak{g}$. The quivers for $\mathfrak{a}_n$ and $\mathfrak{d}_n$ have $n+1$ nodes in all. In the last line are the Kleinian singularities of type $A_{N-1}$, simply denoted by $A_{N-1}$.  For each transverse slice, the second column gives a magnetic quiver.  \label{tab:SymSing}}
\end{table}

The approach taken in this paper makes use of the 1-1 map in (\ref{PwP}), which allows us to identify in the GTP $P$ each of the steps in the construction of the magnetic quiver in the brane-web: the strategy, which will be expanded in section \ref{sec:BraneWebs}, is: 
\be\label{overviewpic}
\begin{tikzpicture}[->]
\node(1)  at (0,0) {$P$}; 
\node (2) at (2,0){$\longrightarrow \quad \mathcal{W}_P \quad \longrightarrow $}; 
\node (3)  at (6,0) {$\boxed{\ba &\text{\footnotesize Sub-web Decomposition} \cr +&\text{\footnotesize Tropical Curve Intersection}\ea}$};
\node (4) at (10,0) {$\longrightarrow \ \text{MQ}_P$};
\path (1) edge[bend right=45] node [below] {$ \boxed{
\ba &\text{\footnotesize Refined Minkowski Sum} \cr 
+ & \text{\footnotesize Edge-Coloring and Mixed Volume}
 \ea}$} (4);
\end{tikzpicture}
\ee
In the following we identify the direct map from $P$ to MQ$_P$. 
We use the rules derived from the brane-webs, and translate these into the language of the generalized toric polygons. We find that in the case of strictly convex toric polygons, we make contact with the work of Altmann on versal deformations \cite{Altmann,Altmann2}. 

In addition we extract the foliation structure of this hyper-K\"ahler cone in terms of symplectic leaves, which is achieved combinatorially in terms of the quiver subtractions \cite{Cabrera:2018ann}. The partially ordered set of such leaves is the Hasse diagram. 

The elementary transverse slices of the Hasse diagram of the 5d Higgs branches for the theories we consider here can be closures of minimal nilpotent orbits $\mathfrak{g}$ or Kleinian singularities -- see table \ref{tab:SymSing}, which includes their magnetic quivers \cite{kraft1980minimal,kraft1982geometry}. In some other instances, there can also be elementary slices of "rank-0" \cite{Bourget:2019aer, Bourget:2020asf}, or of more exotic type (see for instance \cite{malkin2005minimal,fu2017generic}). However a full classification of possible leaves is unknown. In our discussion of the Hasse diagram, which is distinct from the derivation of the magnetic quiver, we focus on the symplectic leaves that appear in table \ref{tab:SymSing}. 

Although we propose a way to derive the MQ from $P$ directly, a first principle derivation from geometry is of course missing, except for the strictly convex toric case. In fact for $P$ not a toric, but a generalized toric polygon, it is thus far unknown what the associated canonical singularity is. Building this dictionary should now be strongly motivated, given the efficiency of how we can determine the MQs from $P$, using very simple combinatorial rules -- which simplify not only the brane-web based analysis, but also connect to the known geometric constructions in toric geometry.


\section{Higgs Branches for 5d SCFTs from Edge Colorings}
\label{sec:ColoringRules}

We start our analysis by presenting the direct construction of the MQ$_P$ from the GTP $P$ in (\ref{overviewpic}). We review properties of toric polygons and their composition and introduce generalizations that apply to GTPs. Furthermore, we provide criteria for a GTP (or a sub-polygon thereof) to  satisfy  the so-called {\it s-rule}, which in the dual brane-web ensures that the configuration preserves supersymmetry. 

With this background, we then introduce the coloring of a GTP, which amounts to a generalization of a Minkowski sum decomposition for toric polygons. Finally, we give a short algorithm that associates a tropical quiver (TQ) to each coloring of a given GTP. We conjecture that the tropical quiver can then be identified with the magnetic quiver characterizing the 5d Higgs branch of the SCFT -- and in cases when there are multiple cones of the Higgs branch, i.e. multiple colorings, the union of these tropical quivers comprise the full Higgs branch of the 5d SCFT.

\subsection{GTPs and Minkowski Sums}
\label{sec:GTPMink}

The concept of a generalized toric polygon was first introduced in \cite{Benini:2009gi} as a dual graph to a 5-brane-web with multiple 5-branes ending on a single 7-brane. This generalization was motivated, as such diagrams furnish 5d SCFTs.

We define a generalized toric polygon $P$ as a lattice polygon, i.e. terms of a set of vertices $\bm{v}_i\in \mathbb{Z}^2$, and  edges, $E_\alpha$, $\alpha = 1, \cdots, n_E$, which connect a subset of the vertices \footnote{A toric polygon, is a special case of this, where each lattice point on an edge is also a vertex. This is not necessarily the case for a GTP. The absent lattice points are sometimes referred to as `white dots'. }. The GTP is the convex hull of the vertices. Each vertex lies on at least one edge.
The set of all vertices and edges will be denoted by 
\be
V_b=  \{ \bm{v}_i \in \mathbb{Z}^2\} \,,\qquad E = \cup_\alpha E_\alpha \,.
\ee
A subset of the edges are the external edges, which are the boundary of the GTP,
\be 
	\partial \TD=\cup_\alpha E_\alpha^{\partial}\equiv E^{\partial} \subset E \,,
\ee
and likewise the complement of these in $E$ are the internal edges, $E \backslash E^{\partial} \equiv E^{\text{in}}$. For the rest of this section we will assume $E^{\text{in}}=\emptyset$. We will return to the treatment of GTPs with internal edges in section \ref{sec:IntEdges}.

The set of vertices is in general a subset of the set of points 
\be
V_b \subset E \cap \mathbb{Z}^2 \,.
\ee
The set of points that are in the complement $(E^\partial\cap \mathbb{Z}^2)\backslash V_b \equiv V_w$, will sometimes be denoted by white dots/vertices, and 
\be \label{AllVertices}
V= V_b \cup V_w \,,
\ee
denotes the complete set of vertices -- black and white. 
The vertices along an edge $E_{\alpha}$, will be labeled by $\bm{v}_{\alpha, i} \in V_{b}$, $i= 0, \cdots, b_\alpha +1$. 
If a vertex is the boundary of an edge, 
\be
\partial E_\alpha = \bm{v}_{\alpha, 0}^{\text{ex}} \cup \bm{v}_{\alpha, b_\alpha+1}^{\text{ex}} \,,
\ee
we refer to it as an extreme vertex, which are always $\bm{v}_{\alpha, i}$ with $i=0$ and $i= b_\alpha+1$.  
Note that extreme vertices cannot lie in the interior of any other edge and (for now, in the absence of internal edges) must be part of the boundary of  two edges. 

Given that an edge $E_\alpha = \overline{\bm{v}_{\alpha, 0}^{\text{ex}} \bm{v}_{\alpha, b_\alpha+1}^{\text{ex}}}$ is a vector in $\mathbb{R}^2$ connecting lattice points, it can be identified with a pair of integers. We call the greatest common divisor of this pair of integers  
\be
\lambda_{\alpha} = \mathrm{gcd}(E_\alpha)\,,
\ee 
where we note that this is $SL(2,\mathbb{Z})$ invariant. We then define the reduced vector 
\be
\label{eq:reducedvec}
L_\alpha = \frac{E_\alpha}{\lambda_{\alpha}}\equiv (x_\alpha,y_\alpha) \, . 
\ee
Furthermore, we define an orientation of the external edges by
\be 
\label{eq:orient}
\bm{v}_{\alpha+1, 0}^{\text{ex}} = \bm{v}_{\alpha, b_\alpha+1}^{\text{ex}}\,,
\ee
where we understand the periodicity in the $\alpha$-indices.
Because of closedness of the boundary of $P$
\be
0 = \sum_{\alpha} \lambda_\alpha L_\alpha^{\partial}\,,
\ee
where the $L_\alpha^{\partial}$ are the line segments of the external edges and we take the sum with the orientation implied by \eqref{eq:orient}.

Finally, we define a partition of $\lambda_\alpha$ in terms of the vertices along an external edge $E_\alpha^\partial$ as 
\be \label{defmu}
\lambda_\alpha = \sum_{i=1 }^{b_\alpha +1} \mu_{\alpha,i} \,,\qquad 
\mu_{\alpha,i} L^{\partial}_\alpha  =\overline{\bm{v}_{\alpha, i} \bm{v}_{\alpha, i-1}}  \,,
\ee
with $b_\alpha$ the number of non-extreme vertices along $E^\partial_\alpha$. Note that for a toric polygon these partitions are always given by $\{1^{b_\alpha+1}\}$. We order the partition in descending magnitude
\be
\mu_{\alpha, x} \geq \mu_{\alpha, x+1} \,,
\ee
where the index $x$ is introduced as an ordered version of the index $i$. This ordered partition will also be referred to as $\{\mu_\alpha\}$. In the following we will make use of the standard partial order pertaining to a set of integer partitions of an integer $N \geq 1$, called the \emph{dominance ordering}. We therefore review it here for the convenience of the reader: 

{\definition[Dominance Ordering]
Let $\rho$ and $\rho'$ be two partitions of $N$, which means that they are sequences of integers satisfying 
\begin{equation}
N = \sum\limits_{1 \leq i} \rho_i =  \sum\limits_{1 \leq i} \rho'_i \qquad \textrm{and} \qquad \rho_1 \geq \rho_2 \geq \cdots \geq 0 \, , \quad \rho'_1 \geq \rho'_2 \geq \cdots  \geq 0  \, . 
\end{equation}
The dominance order is defined as follow: we say $\{\rho\} \geq \{\rho'\}$, if for all $j \geq 1$, we have 
\begin{equation}
\sum\limits_{1 \leq i \leq j} \rho_i \geq \sum\limits_{1 \leq i \leq j} \rho'_i \,,
\end{equation}
i.e. the smallest partition is given by $\{1^N\}$, whereas the largest partition is $\{N\}$.
}

{\example \\
In tandem with the general analysis in this section, we go through one example that illustrates all the salient points. The example is the SCFT associated to an IR description given by 
\be
SU(4)_4 +4\bm{F} \,,
\ee 
 which has vertices
\be
V_b=((0,0),(1,-1),(2,-1),(3,-1),(4,-1),(6,0),(6,3),(3,3))\,,
\ee
with no internal lines. From this we can read off the white dots
\be
V_w=((6,1),(6,2),(5,3),(4,3),(2,2),(1,1))\,,
\ee
which we draw as
\be \label{MainExStart}
\begin{tikzpicture}[x=.5cm,y=.5cm]

\draw[step=.5cm,gray,very thin] (0,-1) grid (6,3);

\draw[ligne,black] (0,0)--(1,-1)--(4,-1)--(6,0)--(6,1)--(6,2)--(6,3)--(5,3)--(4,3)--(3,3)--(2,2)--(1,1)--(0,0);
\node[] at (-4,1) {$P=$};
\node[] at (-1.3,0) {$(0,0)$};
\node[bd] at (0,0) {}; 
\node[bd] at (1,-1) {}; 
\node[bd] at (2,-1) {}; 
\node[bd] at (3,-1) {}; 
\node[bd] at (4,-1) {}; 
\node[bd] at (6,0) {}; 
\node[wd] at (6,1) {}; 
\node[wd] at (6,2) {}; 
\node[bd] at (6,3) {}; 
\node[wd] at (5,3) {}; 
\node[wd] at (4,3) {}; 
\node[bd] at (3,3) {}; 
\node[wd] at (2,2) {}; 
\node[wd] at (1,1) {};

\end{tikzpicture}
\ee
We label the edges in counter-clockwise order, usually beginning on the lower left
\be
\ba
&E_\alpha=((1,-1),(3,0),(2,1),(0,3),(-3,0),(-3,-3))\,, \qquad \lambda_\alpha=(1,3,1,3,3,3)\\
&L_\alpha=((1,-1),(1,0),(2,1),(0,1),(-1,0),(-1,-1))\,.
\ea
\ee
From the distribution of black and white vertices we can read off
\be
b_\alpha=(0,2,0,0,0,0)\,, \qquad \{\mu_\alpha\}=\left(\{1\},\{1^3\},\{1\},\{3\},\{3\},\{3\}\right)\,.
\ee
}

For a conventional toric polygon there exists a composition rule, the Minkowski sum, which is instrumental in analyzing the deformations of a toric polygon \cite{Altmann,Altmann2}. We give the definition of a Minkowski sum of two polygons below, as well as a generalization that we introduce for GTPs, which we call the {\it refined Minkowski sum}, from which we define a notion of colored GTPs, that underlie the construction of the magnetic quiver. 

{\definition[Minkowski Sum]
A Minkowski sum of two polygons $P_a$ and $P_b$ is defined by 
\be
P_a + P_b = \left\{\bm{v}_a+\bm{w}_b\right\}\,,
\ee 
where $\bm{v}_a\in V_{P_a}$ and $\bm{w}_b\in V_{P_b}$ are the vertices of $P_a$ and $P_b$. }

Note that this definition does not distinguish between black and white vertices. This means that the Minkowski sum is not uniquely defined for a set of GTPs. We will thus introduce a refinement of the Minkowski sum, which can accommodate the presence of white dots in a GTP.

{\definition[Refined Minkowski Sum for GTPs]
Let $P_a$ and $P_b$ be GTPs. We define their refined Minkowski sum (or partition sum) as
\be
P_a \oplus P_b\,,
\ee
such that the edges agree with the ones of $P_a + P_b$, i.e. the ordinary Minkowski sum, and the partitions are
\be
\mu_{\alpha,x}^{P_a \oplus P_b} = \mu_{\alpha,x}^{P_a} + \mu_{\alpha,x}^{P_b}\,.
\ee
}

Contrary to the Minkowski sum, the partition sum uniquely determines the partition of all the edges of the resulting GTP. In other words, it imposes a unique configuration of black and white vertices along the edges (up to an irrelevant reordering) of the resulting GTP.

{\example (Continued)
We can write $P$ in \eqref{MainExStart} as a refined Minkowski sum
\be
\begin{tikzpicture}[x=.5cm,y=.5cm]
\draw[step=.5cm,gray,very thin] (0,-1) grid (6,3);
\draw[ligne,black] (0,0)--(1,-1)--(4,-1)--(6,0)--(6,1)--(6,2)--(6,3)--(5,3)--(4,3)--(3,3)--(2,2)--(1,1)--(0,0);
\node[bd] at (0,0) {}; 
\node[bd] at (1,-1) {}; 
\node[bd] at (2,-1) {}; 
\node[bd] at (3,-1) {}; 
\node[bd] at (4,-1) {}; 
\node[bd] at (6,0) {}; 
\node[wd] at (6,1) {}; 
\node[wd] at (6,2) {}; 
\node[bd] at (6,3) {}; 
\node[wd] at (5,3) {}; 
\node[wd] at (4,3) {}; 
\node[bd] at (3,3) {}; 
\node[wd] at (2,2) {}; 
\node[wd] at (1,1) {}; 
\end{tikzpicture}
\qquad 
\begin{tikzpicture}[x=.5cm,y=.5cm]
\node[] at (4,0) {$=$};
\draw[ligne,black] (5,1)--(6,0)--(6,1)--(5,1);
\node[bd] at (5,1) {}; 
\node[bd] at (6,0) {}; 
\node[bd] at (6,1) {}; 
\node[] at (7,0) {$\oplus$};
\draw[ligne,black] (8,0)--(10,1)--(9,1)--(8,0);

\node[bd] at (8,0) {}; 
\node[bd] at (10,1) {}; 
\node[bd] at (9,1) {}; 

\node[] at (11,0) {$\oplus$};

\draw[step=.5cm,gray,very thin] (12,0) grid (15,2);

\draw[ligne,black] (12,0)--(15,0)--(15,1)--(15,2)--(14,2)--(13,1)--(12,0);

\node[bd] at (12,0) {}; 
\node[bd] at (13,0) {}; 
\node[bd] at (14,0) {}; 
\node[bd] at (15,0) {}; 
\node[wd] at (15,1) {}; 
\node[bd] at (15,2) {}; 
\node[bd] at (14,2) {}; 
\node[wd] at (13,1) {};

\end{tikzpicture} \,.
\ee
Note that the third summand cannot be further decompose due to the $\{1^3\}$ partition on the lower edge:
\be
\ba
 \{\mu_\alpha^P\}&=\left(\{1\},\{1^3\},\{1\},\{3\},\{3\},\{3\}\right)\\
 &=\left(\{1\},-,-,\{1\},\{1\},-\right)+\left(-,-,\{1\},-,\{1\},\{1\}\right)+\left(-,\{1^3\},-,\{2\},\{1\},\{2\}\right)\,.
\ea
\ee
}

Finally, we review a well-known concept in tropical geometry, the mixed volume of a set of summands inside a polygon. The mixed volume is defined purely in terms of the edges of the summands and so can readily be computed for any GTP. We will use the mixed volume in the following to construct the magnetic quiver.

{\definition[Mixed Volume]
\label{def:MV}

Let $P$ be a $d$-dimensional (refined) Minkowski sum
\be 
P=\sum_{c=1}^d  S^c\,.
\ee Consider the object
\be
P(\delta^c)= \sum_{c=1}^d \delta^c S^c\,,
\ee
where the $\delta^c\in \R$ are scaling parameters. Then, the volume of $P$ is a formal polynomial in the $\delta^c$ of degree $d$, denoted by $\text{vol}_P(\delta^c)$. The mixed volume of $d$ Minkowski summands inside $P$ is given by the coefficient
\be
\text{MV}(S^{1},\dots,S^{d}) = \left.\text{vol}_P(\delta^c)\right|_{\delta^{1}\dots\delta^{d}}\,.
\ee
}

{\proposition[Mixed Volume for $d=2$]
Let $P$ be a (refined) Minkowski sum as above for $d=2$. The mixed volume of two Minkowski summands is given by
\be \label{MVAreas}
\text{MV}(S^{1},S^{2}) = \text{Area}\left(S^{1}+S^{2}\right) -  \text{Area}\left(S^{1}\right) -  \text{Area}\left(S^{2}\right)\,,
\ee
where we use the Euclidean metric to compute the areas.
}

{\it Proof:} Without loss of generality we take $P=S^1 + S^2$ and consequently $P(\delta^1,\delta^2)=\delta^1 S^1 + \delta^2 S^2$. By definition we have, for some constants $V^1$ and $V^2$,
\be
\text{vol}_P(\delta^1,\delta^2)= \text{V}^1 \left(\delta^1\right)^2 + \text{MV}(S^1,S^2) \delta^1 \delta^2 +  \text{V}^2 \left(\delta^2\right)^2\,,
\ee
where the volume in $d=2$ is just the euclidean area.
Consider the cases $P(1,0)=S^1$ and $P(0,1)=S^2$. Then, we have
\be
\text{Area}(S^1)=\text{V}^1\,,\qquad \text{Area}(S^2)=\text{V}^2\,.
\ee
Furthermore, in the case $P(1,1)=P$ we then have
\be
\text{Area}(P) = \text{Area}(S^1) + \text{MV}(S^1,S^2) +  \text{Area}(S^2)\,,
\ee
from which \eqref{MVAreas} immediately follows.

\subsection{Pruning}
\label{sec:Pruning}

The central concept in our construction is the coloring of a GTP. A coloring will be defined in terms of the data of the GTP that we detailed in the previous section. However, before determining a coloring of a given GTP, it will typically be very useful to first consider, whether there exists a different GTP, representing the same physical theory, that might simplify the construction, or identify equivalent theories. {This so-called pruning will simplify substantially the following analysis. It is not essential, but very useful in practice. }

The following maps can be applied to a GTP without changing the underlying physics: 
\begin{enumerate}
\item Global translations. 
\item Global $SL(2,\mathbb{Z})$ transformations. 
\item Local $SL(2,\mathbb{Z})$ transformations on two consecutive edges.
\item Crossing of an edge from one side of a polygon to the opposing side. In the web this operation corresponds to a Hanany-Witten move. 
\end{enumerate}
The first two are rather elementary transformations on a lattice polygon. 
The third transformation relates convex polygons to non-convex ones, and because a GTP is by definition a convex polygon, we do not use this transformation in the present paper. 
The fourth operation on the above list has proved exceedingly useful for producing simple GTPs that can be straightforwardly colored. We will refer to repeated application of this transformation as \emph{pruning}. Precise definitions for how this affects a GTP are given in appendix \ref{app:MMA}.

The idea is that, starting from a given GTP, one can produce a new polygon, giving rise to the same physics,\footnote{The theories are equivalent up to free hyper-multiplets. } by selecting an edge and moving it around the polygon, so to speak, to the opposite side. In this process, one chooses whether to move the edge through the polygon in the clockwise or counter-clockwise direction. Whichever direction is chosen, the slopes of the edges along this side will change, whereas the other half of the polygon is unaltered. In terms of the brane-web, this operation corresponds to picking out a 7-brane, which sits on one side of the brane-web, and pulling it in through the whole brane-web until it reaches the other side. Doing so will give rise to 5-brane creations and annihilations on the 7-brane that is being moved, along with monodromy transformations of the branes that are crossed. This is reflected in the polygon as a multiplicity change of the edge that is moved, and changes in the slopes of the edges it crosses.

The advantage of pruning is precisely this change of the slopes of the edges of a GTP, or equivalently the Euclidean length of the $L_\alpha$, denoted $|L_\alpha|$. It turns out the smaller $\mathrm{max}_{\alpha} |L_\alpha|$, the easier and the more intuitive it is to check whether the s-rule is satisfied (as detailed in the next subsection). For instance, if $\mathrm{max}_{\alpha} |L_\alpha| \leq 2$ then the edges of the polygon are horizontal, vertical, or at a 45 degrees slope, and in this situation the s-rule is straightforward to check. 
Therefore, although this step is not strictly necessary for the algorithm presented here, we use pruning to reduce $\mathrm{max}_{\alpha} |L_\alpha|$ as much as possible. 
\\
\\

{\example (Continued) Recall our example from (\ref{MainExStart}),  repeated here for convenience
\be
\begin{tikzpicture}[x=.5cm,y=.5cm]

\draw[step=.5cm,gray,very thin] (0,-1) grid (6,3);

\draw[ligne,black] (0,0)--(1,-1)--(4,-1)--(6,0)--(6,1)--(6,2)--(6,3)--(5,3)--(4,3)--(3,3)--(2,2)--(1,1)--(0,0);
\node[] at (-2,1) {$P=$};
\node[bd] at (0,0) {}; 
\node[bd] at (1,-1) {}; 
\node[bd] at (2,-1) {}; 
\node[bd] at (3,-1) {}; 
\node[bd] at (4,-1) {}; 
\node[bd] at (6,0) {}; 
\node[wd] at (6,1) {}; 
\node[wd] at (6,2) {}; 
\node[bd] at (6,3) {}; 
\node[wd] at (5,3) {}; 
\node[wd] at (4,3) {}; 
\node[bd] at (3,3) {}; 
\node[wd] at (2,2) {}; 
\node[wd] at (1,1) {}; 
\end{tikzpicture}
\ee
We move the horizontal edge $E_5$ at the top to the bottom in the clockwise direction (i.e. through the right-hand-side of the polygon), changing the multiplicity of $E_5$ and the slopes of $E_3$ and $E_4$, but leaving $E_1$, $E_2$ and $E_6$ unaltered. We arrive at
\be \label{MainExPruned}
\begin{tikzpicture}[x=.5cm,y=.5cm]

\draw[step=.5cm,gray,very thin] (0,-1) grid (6,3);

\draw[ligne,black] (0,0)--(1,-1)--(5,-1)--(6,0)--(5,1)--(4,2)--(3,3)--(2,2)--(1,1)--(0,0);

\node[] at (-2,0) {$P'=$};
 
\node[bd] at (0,0) {}; 
\node[bd] at (1,-1) {}; 
\node[bd] at (2,-1) {}; 
\node[bd] at (3,-1) {}; 
\node[bd] at (4,-1) {}; 
\node[bd] at (5,-1) {}; 
\node[bd] at (6,0) {}; 
\node[wd] at (5,1) {}; 
\node[wd] at (4,2) {}; 
\node[bd] at (3,3) {}; 
\node[wd] at (2,2) {}; 
\node[wd] at (1,1) {};

\end{tikzpicture}
\ee
defined by the data
\be
\ba
L_\alpha&=((1,-1),(1,0),(1,1),(-1,1),(-1,-1))\,,\\
 \lambda_\alpha&=(1,4,1,3,3)\,,\qquad \{\mu_\alpha\}=(\{1\},\{1^4\},\{1\},\{3\},\{3\})\,.
\ea
\ee
The details of how this is done are given in appendix \ref{app:pruning}. 
This polygon is used to illustrate the next steps of the algorithm. }

\subsection{The s-Rule}
\label{sec:s-rule}

Not all GTPs give rise to a supersymmetric theory in 5d (unlike any convex toric polygon). It is possible for a GTP to have an insufficient distribution of vertices (in a sense, too many white dots). This is the GTP equivalent of a web configuration that is non-supersymmetric because it has too many 5-branes ending on a single 7-brane -- this is referred to as a web that does not satisfy the so-called {\it s-rule}. Determining whether a GTP satisfies the s-rule is a highly non-local problem. In \cite{Benini:2009gi} it was argued that a GTP $P$ obeys the s-rule, if there exists a consistent resolution, i.e. a tessellation, of $P$ into tiles. Here we generalize the definition of such a tile and introduce notions of minimality and irreducibility, related to the s-rule. We should add a word of caution regarding the s-rule. The rule argued for in \cite{Benini:2009gi} seems to not apply in general, in particular it is not invariant under $SL(2, \mathbb{Z})$ transformations. We propose a generalization of this in section \ref{sec:BraneWebs} on brane-webs, and it is this generalization of the s-rule that we implement here.

{\definition[Tiles] 
\label{def:Tiles}
A tile $T_R$ is a GTP such that
\begin{enumerate}
\item $E_\alpha=(E_1,E_2,E_3,E_4)$, where $L_4=-L_2$ and $\lambda_2\geq\lambda_4$ where we allow $\lambda_4=0$.
Thus, $T_R$ is either a trapezoid or a triangle.
\item
The partitions are
\be
\{\mu_\alpha\} = \{\lambda_\alpha\}\,,
\ee
i.e. $T_R$ has no non-extreme vertices.
\item
Define the auxiliary GTP $\widetilde{T}_R$ with
\be
\widetilde{L}_\alpha=(L_1,L_2,L_3)\,, \qquad \widetilde{\lambda}_\alpha = (\lambda_1,\lambda_2-\lambda_4,\lambda_3)\,, \qquad \{\widetilde{\mu}_\alpha\}=\{\widetilde{\lambda}_\alpha\}\,,
\ee
such that $T_R$ is the refined Minkowski sum of the triangle $\widetilde{T}_R$ and the line of length $\lambda_4$ along $E_4$. Then,
\be
\label{sruletiles}
\widetilde{\lambda}_\alpha \widetilde{\lambda}_\beta\left|\det (\widetilde{L}_\alpha,\widetilde{L}_\beta)\right| \geq (\widetilde{\lambda}_\gamma)^2\,, \qquad \forall \alpha \neq \beta\,, \ \alpha \neq \gamma\,, \ \beta \neq \gamma\,.
\ee
\end{enumerate}
}
With this general definition of a tile, we use the requirement of \cite{Bergman:1998ej, Benini:2009gi} to determine whether a GTP respects the s-rule. We furthermore define concepts of minimality and irreducibility that are related to the s-rule, and which will be essential input for our definition of a coloring.

{\definition[s-Rule, Minimality, Irreducibility, and IMPs]
\label{DefinitionSrule}
Let $P$ be a GTP with edges $E_\alpha$ of length $\lambda_\alpha$ and partitions $\{\mu_\alpha\}$. 
\begin{enumerate}
\item $P$ is said to obey the s-rule, if $P$ can, by the inclusion of internal edges,  be resolved\footnote{Here we mean that the polygon can be tessellated by tiles by including internal edges.} into resolution tiles.
\item Let $\widetilde{P}$ be a GTP with edges $E_\alpha$  of length $\lambda^\alpha$ and partitions
\be
\label{RemoveVertices}
\{\widetilde{\mu}_\alpha\} > \{\mu_\alpha\}\,,
\ee
i.e. $\widetilde{P}$ is obtained from $P$ by removing vertices along the edges.
$P$ is said to obey the s-rule minimally (or we say that $P$ itself is minimal) if $\widetilde{P}$ does not obey the s-rule for any choice of $\{\widetilde{\mu}_\alpha\}$.
\item $P$ is irreducible if there is no decomposition
\be
P= P_1 \oplus P_2\,,
\ee
such that $P_1$ and $P_2$ obey the s-rule.
\end{enumerate}
If P is irreducible and obeys the s-rule minimally, we say that P is an IMP, for Irreducible and Minimal Polygon. 
}

The idea is that, given any GTP as a starting point, one can always find a minimal GTP by removing vertices (i.e. converting black vertices into white ones) in the original polygon. The minimal polygon(s) has the largest possible number of white dots, whilst still satisfying the s-rule, i.e. there is no possible way to remove another vertex from the GTP without breaking supersymmetry.
On the other hand, a GTP is said to be irreducible if it cannot be further decomposed into a partition sum of two other polygons. It is possible for a GTP to be minimal but not irreducible, or irreducible but not minimal. In the following section, we will require both conditions to be met by the building blocks of the coloring. 
To avoid notational clutter, an irreducible GTP, that obeys the s-rule minimally is called an irreducible minimal polygon (IMP).

{\example (Continued) Let us return to the example introduced in \eqref{MainExStart}. We will continue with the pruned GTP $P'$ in (\ref{MainExPruned}). Consider a sub-polygon $T$ that sits inside $P'$, given by
\be \label{MainExTile}
\begin{tikzpicture}[x=.5cm,y=.5cm]

\draw[step=.5cm,gray,very thin] (0,-1) grid (4,2);

\draw[ligne,black] (0,0)--(1,-1)--(3,-1)--(4,0)--(3,1)--(2,2)--(1,1)--(0,0);

\node[] at (-2,0) {$T=$};
 
\node[bd] at (0,0) {}; 
\node[bd] at (1,-1) {}; 
\node[bd] at (2,-1) {}; 
\node[bd] at (3,-1) {}; 
\node[bd] at (4,0) {}; 
\node[wd] at (3,1) {}; 
\node[bd] at (2,2) {}; 
\node[wd] at (1,1) {};

\end{tikzpicture}
\ee
A way to tesselate $T$ is
\be \label{MainExTileRes}
\begin{tikzpicture}[x=.5cm,y=.5cm]

\draw[step=.5cm,gray,very thin] (0,-1) grid (4,2);

\draw[ligne,black] (0,0)--(1,-1)--(3,-1)--(4,0)--(3,1)--(2,2)--(1,1)--(0,0);
\draw[ligne,black] (0,0)--(1,0)--(2,1)--(2,2);
\draw[ligne,black] (2,1)--(3,0)--(4,0);

\node[] at (-2,0) {$T\to$};
 
\node[bd] at (0,0) {}; 
\node[bd] at (1,-1) {}; 
\node[bd] at (2,-1) {}; 
\node[bd] at (3,-1) {}; 
\node[bd] at (4,0) {}; 
\node[wd] at (3,1) {}; 
\node[bd] at (2,2) {}; 
\node[wd] at (1,1) {}; 

\node[bd] at (1,0) {};
\node[bd] at (2,1) {};
\node[bd] at (3,0) {};

\end{tikzpicture}
\ee
where we have omitted the triangulation of the part of the GTP that is completely surrounded by black vertices, as these can always be triangulated in a standard toric way. We can easily check that the two symmetric trapezoids are tiles. Thus, $T$ obeys the s-rule. We check whether $T$ is minimal by defining $\widetilde{T}$ in which we have removed the only allowed vertex, which sits on the lower edge. In terms of partitions we have
\be
\{\widetilde{\mu}_2\}=\{2\}>\{1^2\}=\{\mu_2\}\,,
\ee
satisfying \eqref{RemoveVertices}.
 We find that $\widetilde{T}$ still obeys the s-rule, since there exists a valid resolution of $\widetilde{T}$ given by
\be \label{MainExTileResTilde}
\begin{tikzpicture}[x=.5cm,y=.5cm]

\draw[step=.5cm,gray,very thin] (0,-1) grid (4,2);

\draw[ligne,black] (0,0)--(1,-1)--(3,-1)--(4,0)--(3,1)--(2,2)--(1,1)--(0,0);
\draw[ligne,black] (0,0)--(1,0)--(2,1)--(2,2);
\draw[ligne,black] (2,1)--(3,0)--(4,0);
\draw[ligne,black] (1,-1)--(2,1)--(3,-1);

\node[] at (-2,0) {$\widetilde{T}= $};
 
\node[bd] at (0,0) {}; 
\node[bd] at (1,-1) {}; 
\node[wd] at (2,-1) {}; 
\node[bd] at (3,-1) {}; 
\node[bd] at (4,0) {}; 
\node[wd] at (3,1) {}; 
\node[bd] at (2,2) {}; 
\node[wd] at (1,1) {}; 

\node[bd] at (1,0) {};
\node[bd] at (3,0) {};
\node[bd] at (2,1) {};

\end{tikzpicture}
\ee
Thus, while $T$ is not a minimal polygon, $\widetilde{T}$ is, since there are no more vertices that can be removed in $\widetilde{T}$. We can also check that $\widetilde{T}$ is irreducible. For example, we could try to write $\widetilde{T}$ as a partition sum
\be
\begin{tikzpicture}[x=.5cm,y=.5cm]


\draw[ligne,black] (5,-1)--(6,0);

\node[bd] at (5,-1) {}; 
\node[bd] at (6,0) {};

\node[] at (7,0) {$\oplus$};

\draw[step=.5cm,gray,very thin] (8,-1) grid (11,1);

\draw[ligne,black] (8,0)--(9,-1)--(10,-1)--(11,-1)--(10,0)--(9,1)--(8,0);

\node[bd] at (8,0) {}; 
\node[bd] at (9,-1) {}; 
\node[wd] at (10,-1) {}; 
\node[bd] at (11,-1) {}; 
\node[wd] at (10,0) {}; 
\node[bd] at (9,1) {};

\end{tikzpicture} \,,
\ee
but then the last summand does not obey the s-rule itself. Thus, $\widetilde{T}$ is an IMP. }

\subsection{The r-Rule}
\label{sec:r-rule}

We now turn to the discussion of the rank of a GTP/web. As we alluded to in the introduction, we propose that in addition to the charge conservation and s-rule, another condition needs to be satisfied for a GTP/web to give rise to a 5d SCFT, namely that its rank be non-negative.
We define the rank of a GTP $P$ as
\be
r(P)=r_0-\half\sum_{\alpha}\sum_{x=1}^{b_\alpha} \mu_{\alpha,x} \left(\mu_{\alpha,x}-1\right)\,,
\ee
where $r_0$ is the number of internal nodes of $P$. We can use Pick's theorem \cite{Pick} to write this as
\be
\label{Pick}
r_0=\text{Area}(P)+1-\half\sum_\alpha \lambda_\alpha\,,
\ee
where $\text{Area}(P)$ is the Euclidean area of $P$. Putting this together, and using \eqref{defmu}, this gives
\be \label{rankP}
r(P)=\text{Area}(P) + 1 -\half \sum_{\alpha}\sum_{x=1}^{b_\alpha} \mu_{\alpha,x}^2 \,.
\ee
We conjecture that this coincides with the gauge rank of the 5d gauge theory or SCFT that $P$ represents. This is substantiated by agreement in all examples considered in the present paper, as well as in \cite{BBESII}, and by the invariance of \eqref{rankP} under edge-moves and pruning. The rank of a 5d gauge theory or SCFT, realised as a brane-web, is given by the number of local deformations of the web \cite{Aharony:1997bh}. 
Therefore, to make sense of a GTP as associated to a 5d theory, its rank $r$ should be non-negative. Note that this condition is not automatically satisfied for webs/GTPs that satisfy the s-rule. We therefore propose the following additional condition:

{\definition[r-rule]
\label{DefinitionRrule}
Let $P$ be a GTP with edges $E_\alpha$ of length $\lambda_\alpha$ and partitions $\{\mu_\alpha\}$. $P$ is said to obey the r-rule, if 
\be \label{r-rule}
r(P)\geq 0\,.
\ee
}

It is straight forward to show that the r-rule implies the s-rule for triangle GTP or trinion webs, and is thus strictly stronger:

{\proposition[r-rule implies s-rule for triangles]
Let $P$ be a triangular IMP\footnote{We write IMP instead of GTP to ensure multiplicity one.} with edges $E_\alpha, \alpha=1,2,3$, of length $\lambda_\alpha$ and partitions $\{\mu_\alpha\}=\{\lambda_\alpha\}$, i.e. $P$ has no non-extreme vertices. The r-rule, given in definition \ref{DefinitionRrule}, for $P$, implies the s-rule, given in definition \ref{DefinitionSrule}, for $P$.
}
\\

\paragraph{Proof.} To show this fact, consider a triangular IMP $P$ as above. The area of $P$ is given by
\be
\text{Area}(P) = \half \lambda_\alpha \lambda_\beta |\det(L_\alpha,L_\beta)|\,,
\ee
for any choice of $\alpha\neq \beta$. Thus, the rank of $P$ is
\be
2 r=\lambda_1 \lambda_2 |\text{det}(L_1,L_2)| - \lambda_1^2-\lambda_2^2-\lambda_3^2+2\,,
\ee
and equivalently for all other permutations of (1,2,3). From the above it immediately follows that
\be
\left(\lambda_1 \lambda_2 |\text{det}(L_1,L_2)| - \lambda_3^2\right) = 2r + (\lambda_1^2-1)+ (\lambda_2^2-1)\,.
\ee
Since $\lambda_\alpha \geq 1$, the r-rule implies
\be
\lambda_1 \lambda_2 |\text{det}(L_1,L_2)| \geq \lambda_3^2 \,,
\ee
and similarly for the other permutations.

\example 
(Continued) Let us illustrate these concepts on the example introduced in \eqref{MainExStart}. The number of internal dots is $r_0 = 12$, which can be computed using Pick's theorem \eqref{Pick} as $r_0 = 18 + 1 - \frac{1}{2}(1+3+1+3+3+3)=12$. This would be the rank of the theory if all the vertices on the edges of \eqref{MainExStart} were black. However the presence of white dots reduces the rank to $r = r_0 - \frac{1}{2} (3 \cdot 3(3-1)) = 3$. Alternatively, the rank can be computed directly from \eqref{rankP} as $r = 18+1-\frac{1}{2} (5 \cdot 1^2 + 3 \cdot 3^2)=3$. The conclusion is that the GTP \eqref{MainExStart} satisfies the r-rule. One can also check on that example that the rank is preserved under pruning: the rank of the GTP \eqref{MainExPruned} is $r=3$.  

\example 
Let us consider an example of a GTP which satisfies the s-rule but violates the r-rule, i.e. it has negative rank. One such GTP is
\be
\begin{tikzpicture}[x=.5cm,y=.5cm]

\draw[step=.5cm,gray,very thin] (0,0) grid (4,4);

\draw[ligne] (0,0)--(4,0)--(0,4)--(0,0);

\node[bd] at (0,0) {}; 
\node[bd] at (4,0) {}; 
\node[wd] at (1,0) {}; 
\node[wd] at (2,0) {}; 
\node[bd] at (0,4) {}; 
\node[wd] at (3,0) {}; 
\node[wd] at (0,1) {}; 
\node[wd] at (0,2) {}; 
\node[wd] at (0,3) {}; 
\node[wd] at (1,3) {}; 
\node[wd] at (2,2) {}; 
\node[wd] at (3,1) {};

\node[] at (6,.5) {$=$};

\foreach \x in {0,1,2,3}
\draw[step=.5cm,gray,very thin] (8+3*\x,0) grid (9+3*\x,1);
\foreach \x in {0,1,2,3}
\draw[ligne] (8+3*\x,0)--(9+3*\x,0)--(8+3*\x,1)--(8+3*\x,0);
\foreach \x in {0,1,2,3}
\node[bd] at (8+3*\x,0) {};
\foreach \x in {0,1,2,3}
\node[bd] at (9+3*\x,0) {};
\foreach \x in {0,1,2,3}
\node[bd] at (8+3*\x,1) {};
\foreach \x in {0,1,2}
\node[] at (10+3*\x,.5) {$\oplus$};

\end{tikzpicture}
\ee
with 
\be 
L_\alpha=((1,0),(-1,1),(0,-1))\,, \qquad \lambda_\alpha=(4,4,4)\,, \qquad  \mu_{\alpha,x}=(\{4\},\{4\},\{4\})\,.
\ee
Clearly, this GTP satisfies the s-rule as, according to definition \ref{def:Tiles}, it is itself a resolution tile, given that it satisfies \ref{sruletiles} as
\be
4 \cdot 4 \cdot 1 \geq 4^2\,.
\ee
The rank $r$ of this GTP is 
\be 
r=\frac{4 \cdot 4}{2}+1-\frac{3}{2} \cdot 4^2=-15\,.
\ee

\example 
In the previous example, the GTP was a Minkowski multiple of a simpler GTP. We now show that there also exist irreducible GTPs which violate the r-rule while satisfying the s-rule.  Consider 
\be
\begin{tikzpicture}[x=.5cm,y=.5cm]

\draw[step=.5cm,gray,very thin] (0,0) grid (6,9);

\draw[ligne] (0,0)--(6,9)--(0,5)--(0,0);

\node[bd] at (0,0) {}; 
\node[bd] at (6,9) {}; 
\node[bd] at (0,5) {}; 
\node[wd] at (2,3) {}; 
\node[wd] at (4,6) {}; 
\node[wd] at (3,7) {}; 
\node[wd] at (0,4) {}; 
\node[wd] at (0,3) {}; 
\node[wd] at (0,2) {}; 
\node[wd] at (0,1) {}; 

\end{tikzpicture}
\ee
for which the data is
\be 
L_\alpha=((2,3),(-3,-2),(0,-1))\,, \qquad \lambda_\alpha=(3,2,5)\,, \qquad  \mu_{\alpha,x}=(\{3\},\{2\},\{5\})\,.
\ee
Again, this GTP is itself a resolution tile, which satisfies \ref{sruletiles} for each edge:
\be
3 \cdot 2 \cdot 5 \geq 5^2\,, \qquad 3 \cdot 5 \cdot 2 \geq 2^2\,, \qquad 2 \cdot 5 \cdot 3 \geq 3^2\,.
\ee
However the rank $r$ of this GTP is 
\be
r=\frac{5 \cdot 6}{2}+1-\half (3^2+2^2+5^2)=-3\,.
\ee

\subsection{Colorings of GTPs}
\label{sec:ColoredPolygons}

In this section we define the notion of \emph{coloring} of a GTP, which is the crucial step in determining the magnetic quiver. A coloring of a polygon is dual to a maximal subdivision of the 5-brane-web into consistent, supersymmetric sub-webs.
The definition relies on the data of the polygon, as well as the generalized decomposition rule, which was introduced in section \ref{sec:GTPMink}. Notice that the practical implementation of the following definition is usually significantly simplified by applying it to a pre-pruned GTP (see section \ref{sec:Pruning}). Furthermore, an essential feature of the building blocks of the coloring is that they represent minimal supersymmetric configurations. This criterion is implemented by requiring irreducibility and minimality with respect to the s-rule (see section \ref{sec:s-rule}).


{\algorithm[Colored Polygon] \label{def:Coloring}
Let $P$ be a GTP (with no internal edges), with edge lengths  
$\lambda_\alpha$. Let 
 $\{\lambda^c_\alpha\in\mathbb{N}\}$ be a partition of the edge lengths, such that 
\be
\sum_{c=1}^{n_c} \lambda^c_\alpha = \lambda_\alpha\,,
\ee
where $n_c$ is the number of colors, and along each edge $E_\alpha$ of $P$, $\lambda^c_\alpha$ segments $L_\alpha$ are colored by $c$.
A partition defines a colored GTP $(P, \{\lambda^c_\alpha\})$, if the following conditions are met:
\begin{enumerate}
\item For each $c=1, \cdots, n_c$ the associated line segments form a polygon, i.e.
\be
0=\sum_{\alpha} \lambda^c_\alpha L_\alpha\,.
\ee
We denote these color sub-polygons by $S^{c}$. 
\footnote{This also include the case of two parallel lines, as e.g. in (\ref{MainExColor2}).}
\item Each $S^c$ is a refined Minkowski sum of $m_c$ times the same IMP $T^c$, 
\be \label{ColorMultiplicity}
S^c = \underbrace{T^c \oplus \dots \oplus T^c}_{m^c}\,,
\ee
and we require that the IMP $T^c$ satisfies the r-rule
\be 
r(T^c) \geq 0\,.
\ee
Furthermore, the IMPs used for different colors must be distinct, 
\be
T^c \neq T^d \,, \qquad c\neq d\,.
\ee
\item
For each $\alpha$, 
\be 
\label{constraintSrule}
\{\mu_\alpha\} \leq \sum\limits_{c=1}^{n_c} \{\mu^c_\alpha\}\,,
\ee 
where we use the dominance partial order of the partitions defined by \eqref{defmu} for $P$ and $S^c$ respectively.
\end{enumerate}
We then write 
\be
P \leq \bigoplus_{c=1}^{n_c} S^c\,, \qquad P =  S^1 +\cdots +  S^{n_c}\,.
\ee
For strict inequality of the partitions in \eqref{constraintSrule} we also write $P < \bigoplus_{c=1}^{n_c} S^c$.
}

In other words, we color the edges of a GTP such that the edges pertaining to a single color form a closed sub-polygon $S^c$. A closed sub-polygon must be made up of a number $m^c$ of identical IMPs $T^c$. We require that no two distinct sub-polygons are comprised of the same IMP (in such a situation the two sub-polygons are identified and the resulting multiple of tiles $m^c$ is the sum). The final condition essentially ensures that all the sub-polygons fit simultaneously into the GTP. 
The above conditions were derived by using the map from brane-web to GTP to identify the dual concept of a maximal subdivision of the web into sub-webs. The details of this map are explained in section \ref{sec:BraneWebs}.

\example (Continued)
We discuss the coloring of the example \eqref{MainExStart}. Since GTPs connected by pruning are equivalent, we can use the pruned GTP $P'$ in \eqref{MainExPruned} for simplicity. We need to determine the colored sub-polygons that are multiples of IMPs. We already know of one IMP, $\widetilde{T}$ in \eqref{MainExTileResTilde}, defining a blue sub-polygon $S^b$ with $m^b=1$. The data of the blue sub-polygon inside $P'$ is
\be \label{MainExSb}
\lambda^b_\alpha=(1,2,1,2,2)\,,\qquad \{\mu^b_\alpha\}=(\{1\},\{2\},\{1\},\{2\},\{2\})\,.
\ee
We can check that another IMP defines a green sub-polygon $S_g$ with $m_g=1$ and
\be \label{MainExSg}
\lambda^g_\alpha=(0,2,0,1,1)\,,\qquad \{\mu^g_\alpha\}=(-,\{1^2\},-,\{1\},\{1\})\,.
\ee
Combining the two colors we obtain
\be
\{\mu^b_\alpha\}+\{\mu^g_\alpha\}=(\{1\},\{3,1\},\{1\},\{3\},\{3\}) > (\{1\},\{1^4\},\{1\},\{3\},\{3\})=\{\mu_\alpha\}\,,
\ee
so indeed
\be
P' < S^b \oplus S^g\,.
\ee
We can draw this as
\be \label{MainExColor1}
\begin{tikzpicture}[x=.5cm,y=.5cm]

\draw[step=.5cm,gray,very thin] (0,-1) grid (6,3);

\draw[ligne,blue] (2,2)--(1,1)--(0,0)--(1,-1)--(2,-1);
\draw[ligne,blue] (4,-1)--(5,-1)--(6,0)--(5,1)--(4,2);
\draw[ligne,green] (2,-1)--(4,-1);
\draw[ligne,green] (4,2)--(3,3)--(2,2);

\node[] at (-2,0) {$S^b \oplus S^g=$};
 
\node[bd] at (0,0) {}; 
\node[bd] at (1,-1) {}; 
\node[wd] at (2,-1) {}; 
\node[wd] at (3,-1) {}; 
\node[bd] at (4,-1) {}; 
\node[bd] at (5,-1) {}; 
\node[bd] at (6,0) {}; 
\node[wd] at (5,1) {}; 
\node[wd] at (4,2) {}; 
\node[bd] at (3,3) {}; 
\node[wd] at (2,2) {}; 
\node[wd] at (1,1) {}; 

\node[] at (7,0) {$>$};

\draw[step=.5cm,gray,very thin] (8,-1) grid (14,3);

\draw[ligne,blue] (10,2)--(9,1)--(8,0)--(9,-1)--(10,-1);
\draw[ligne,blue] (12,-1)--(13,-1)--(14,0)--(13,1)--(12,2);
\draw[ligne,green] (10,-1)--(12,-1);
\draw[ligne,green] (12,2)--(11,3)--(10,2);

\node[bd] at (8,0) {}; 
\node[bd] at (9,-1) {}; 
\node[bd] at (10,-1) {}; 
\node[bd] at (11,-1) {}; 
\node[bd] at (12,-1) {}; 
\node[bd] at (13,-1) {}; 
\node[bd] at (14,0) {}; 
\node[wd] at (13,1) {}; 
\node[wd] at (12,2) {}; 
\node[bd] at (11,3) {}; 
\node[wd] at (10,2) {}; 
\node[wd] at (9,1) {}; 

\node[] at (15.5,0) {$=P'$};

\end{tikzpicture}
\ee
Note that the two GTPs have the same coloring but different partitions. From now on, we only draw the partitions of the full GTP (in this case $P'$).
We can actually check that there is a second coloring of $P'$, given by
\be \label{MainExColor2}
\begin{tikzpicture}[x=.5cm,y=.5cm]

\draw[step=.5cm,gray,very thin] (0,-1) grid (6,3);

\draw[ligne,blue] (0,0)--(1,-1);
\draw[ligne,blue] (6,0)--(5,1);
\draw[ligne,green] (1,-1)--(5,-1);
\draw[ligne,green] (5,1)--(4,2)--(3,3)--(2,2)--(1,1);
\draw[ligne,cyan] (5,-1)--(6,0);
\draw[ligne,cyan] (1,1)--(0,0);

\node[] at (-3,0) {$S^b \oplus S^g \oplus S^c>$};
 
\node[bd] at (0,0) {}; 
\node[bd] at (1,-1) {}; 
\node[bd] at (2,-1) {}; 
\node[bd] at (3,-1) {}; 
\node[bd] at (4,-1) {}; 
\node[bd] at (5,-1) {}; 
\node[bd] at (6,0) {}; 
\node[wd] at (5,1) {}; 
\node[wd] at (4,2) {}; 
\node[bd] at (3,3) {}; 
\node[wd] at (2,2) {}; 
\node[wd] at (1,1) {};

\end{tikzpicture}
\ee
In this coloring, 
\be
S^g=T^g \oplus T^g\,,
\ee
so $m^g=2$, in other words, the green sub-polygon has multiplicity 2.\\

In the convex toric case, our definition of a coloring exactly reduces to identifying the decomposition of a given polygon into Minkowski summands, which was shown in \cite{Altmann,Altmann2} to parametrize the deformations of the geometry.

{\proposition
For a convex toric polygon $P$, its coloring $(P, \{\lambda^c_\alpha\})$ defines a Minkowski sum decomposition 
\be \label{MinkSum}
P = S^1 + \cdots + S^{n_c} \,,
\ee
where the $S^c$ are the sub-polygons, that are defined by the coloring $\{\lambda^c_\alpha\}$. 
Conversely, any Minkowski sum decomposition of $P$ into $S^c$ defines a coloring  $\{\lambda^c_\alpha\}$, if 
the $S^c$ are inequivalent, i.e. there is no $n\in \mathbb{Q}$ such that $S^c = n S^d$ for any $c, d$, and the decomposition is maximal, i.e. none of the sub-polygons can be further decomposed to a set of distinct sub-polygons (where two polygons that are related by scaling are not understood to be distinct). 
}

\paragraph{Color-Subdivided GTPs.}
Before we show how to associate a quiver to a colored GTP, we will explain how to graphically determine the mixed volume of definition  \ref{def:MV}, between two colored sub-polygons. 
To this end, we define a \textit{color sub-division} of a colored GTP  $(P, \{\lambda^c_\alpha\})$, in terms of uni-colored polygons and bi-colored parallelograms. The sub-division agrees with the coloring of the edges of $P$ on the boundary, and extends this to the interior. The mixed volume of two color sub-polygons is dual to the tropical intersection of two sub-webs in the brane-web. In particular, two sub-webs can be interpreted as a pair of tropical curves whose intersection is determined by pulling apart the curves and adding up the intersections of individual 5-branes. The pulling apart of the tropical curves is exactly dual to the color sub-division of the colored GTP.

{\definition[Color-subdivided GTP.]   
A color-subdivision of a GTP is a tiling of  $(P, \{\lambda^c_\alpha\})$ by two types of polygons:
\begin{enumerate}
\item Parallelograms $G^{c_1 c_2}$, where parallel edges have the same color
\item Polygons $G^c$ where all edges have the same color
\end{enumerate}
such that as a set 
\be
P= \bigcup_{c_1, c_2} G^{c_1 c_2} \cup \bigcup_c G^{c} \,,
\ee
with the sub-polygons intersecting each other at most in edges. Furthermore, the color sub-division agrees with the edge-coloring, i.e. 
\be
\partial \left.\left(\bigcup_{c_1, c_2} G^{c_1 c_2} \cup \bigcup_c G^{c}\right)\right|_{E_\alpha} = \{\lambda^c_\alpha\}\,.
\ee
}

Due to the last requirement, a colored polygon $(P, \{\lambda^c_\alpha\})$ determines a color-subdivision of $P$, albeit not uniquely. Yet the sum of the area of all parallelograms $G^{c_1, c_2}$ (that are bi-colored in $c_1$ and $c_2$) is an invariant. 
For each pair $c_1,c_2$ we write this invariant
\be
A^{c_1,c_2}=\sum \text{Area}(G^{c_1,c_2})\,,
\ee
where the sum is over all parallelograms with colors $c_1$ and $c_2$, and we use the standard flat metric in $\mathbb{R}^2$ to compute the area. This area is exactly the mixed volume of the sub-polygons $S^{c_1}$ and $S^{c_2}$.

{\proposition[Color-subdivided Polygons and the Mixed Volume]
Let $(P,\{\lambda^c_\alpha\})$ be a colored GTP corresponding to a Minkowski sum $P=\sum_c S^c$. Then, for each choice of color-subdivision
\be
A^{c_1,c_2}=\text{MV}(S^{c_1},S^{c_2})\,,
\ee
for all pairs $c_1, c_2$.

Proof: \cite{MaclaganSturmfels}, section 4.6.
}

\example
Consider the two colorings of $P'$ in \eqref{MainExColor1} and \eqref{MainExColor2}. The color sub-division is given by introducing virtual lines, dividing the respective polygons as 
\be \label{MainExSubColor2}
\begin{tikzpicture}[x=.5cm,y=.5cm]

\node[] at (-8,4) {$\text{Coloring 1:}$};
\node[] at (2,4) {$\text{Coloring 2:}$};

\draw[step=.5cm,gray,very thin] (-10,-1) grid (-4,3);

\draw[ligne,blue] (-8,2)--(-9,1)--(-10,0)--(-9,-1)--(-8,-1);
\draw[ligne,blue] (-6,-1)--(-5,-1)--(-4,0)--(-5,1)--(-6,2);
\draw[ligne,green] (-8,-1)--(-6,-1);
\draw[ligne,green] (-6,2)--(-7,3)--(-8,2);

\draw[thick, densely dotted,blue] (-8,-1)--(-8,2);
\draw[thick, densely dotted,blue] (-6,-1)--(-6,2);

\draw[thick, densely dotted,green] (-8,2)--(-6,2);

\node[bd] at (-10,0) {}; 
\node[bd] at (-9,-1) {}; 
\node[bd] at (-8,-1) {}; 
\node[bd] at (-7,-1) {}; 
\node[bd] at (-6,-1) {}; 
\node[bd] at (-5,-1) {}; 
\node[bd] at (-4,0) {}; 
\node[wd] at (-5,1) {}; 
\node[wd] at (-6,2) {}; 
\node[bd] at (-7,3) {}; 
\node[wd] at (-8,2) {}; 
\node[wd] at (-9,1) {};

\draw[step=.5cm,gray,very thin] (0,-1) grid (6,3);

\draw[ligne,blue] (0,0)--(1,-1);
\draw[ligne,blue] (6,0)--(5,1);
\draw[ligne,green] (1,-1)--(5,-1);
\draw[ligne,green] (5,1)--(4,2)--(3,3)--(2,2)--(1,1);
\draw[ligne,cyan] (5,-1)--(6,0);
\draw[ligne,cyan] (1,1)--(0,0);

\draw[thick, densely dotted,blue] (3,-1)--(2,0);
\draw[thick, densely dotted,blue] (4,0)--(3,1);

\draw[thick, densely dotted,cyan] (3,-1)--(4,0);
\draw[thick, densely dotted,cyan] (2,0)--(3,1);

\draw[thick, densely dotted,green] (0,0)--(2,0);
\draw[thick, densely dotted,green] (4,0)--(6,0);
\draw[thick, densely dotted,green] (1,1)--(5,1);

\node[bd] at (0,0) {}; 
\node[bd] at (1,-1) {}; 
\node[bd] at (2,-1) {}; 
\node[bd] at (3,-1) {}; 
\node[bd] at (4,-1) {}; 
\node[bd] at (5,-1) {}; 
\node[bd] at (6,0) {}; 
\node[wd] at (5,1) {}; 
\node[wd] at (4,2) {}; 
\node[bd] at (3,3) {}; 
\node[wd] at (2,2) {}; 
\node[wd] at (1,1) {};

\end{tikzpicture}
\ee
such that the respective areas are
\be\label{MainExAreas}
\ba \text{Coloring 1:} &\qquad A^{bg}= 6\cr 
\text{Coloring 2:} &\qquad A^{bg}=A^{cg}=4\,,\qquad A^{bc}=2\,.
\ea
\ee

\subsection{Magnetic Quivers from Colored GTPs}
\label{sec:MQ}

Associated to a colored GTP we now define a quiver, which we refer to as the tropical quiver. If the GTP in question allows for more than a single consistent coloring, then each coloring gives rise to its own quiver. The tropical quiver will be identified with the magnetic quiver, i.e. the 3d $\mathcal{N}=4$ theory, whose Coulomb branch is isomorphic to the Higgs branch of the 5d theory $\mathcal{T}_P$.

{\definition[Tropical Quiver]

The tropical quiver $\text{TQ}(P,\{\lambda^c_\alpha\})$ of a colored GTP is given by a set of nodes with labels $m_I$ and symmetric intersections $k_{IJ}$. Given a colored GTP $(P,\{\lambda^c_\alpha\})$ we define it as follows.
\begin{enumerate}
\item Color nodes: Each color maps to a node in the tropical quiver. 
The labels of the nodes are the $m^c$ defined in \eqref{ColorMultiplicity}. In general, this is given by
\be
\label{eq:multiplicity}
m^c = \gcd_{\alpha}\left(\lambda^c_\alpha\right)\,.
\ee
The intersections between the nodes of color $c_1$ and $c_2$ are determined from two parts. The first is the mixed volume $\text{MV}(S^{c_1}, S^{c_2})$ of definition \ref{def:MV}.
The second contribution, which is negative, comes from configurations where an edge $E_\alpha$ is colored in both colors, i.e. $\lambda_{\alpha}^{c_1},\lambda_{\alpha}^{c_2}>0$.
Thus,
\be \label{kColor}
k^{c_1, c_2} = \frac{1}{m^{c_1} m^{c_2}} \left(\text{MV}(S^{c_1}, S^{c_2}) - \sum_\alpha \sum_{x=1}^{b_\alpha} \mu^{c_1}_{\alpha, x} \mu^{c_2}_{\alpha, x} \right)\,.
\ee
In particular, the self-intersection of a color node is given by
\be \label{eq:self-int}
k^{cc}=\frac{1}{(m^c)^2} \left( 2 \cdot \text{Area}(S^c) - \sum_\alpha \sum_{x=1}^{b_\alpha} (\mu^{c}_{\alpha, x})^2 \right)\,,
\ee
where $\text{Area}(S^c)$ is the Euclidean area of the $c$-colored sub-polygon. The self-intersection determines the number of edges beginning and ending on the $c$-colored node (i.e. loops) as 
\be 
\ell^c=1+\frac{k^{cc}}{2}\,.
\ee
\item Tails:
There are additional nodes in the tropical quiver that are not associated to a color. 
For each edge $E_\alpha$, the tropical quiver contains a sequence of nodes of length $b_\alpha$. The labels of these nodes are given by
\be \label{MultHN}
m_{\alpha, x} = \sum_{y=1}^{x}\left(-\mu_{\alpha, y} + \sum_{c=1}^{n_c} \mu^c_{\alpha, y}\right)\,,
\ee
where $x=1,\dots b_\alpha$.\footnote{Note that one can easily show that all $m_{\alpha,x}$ with $x>b_\alpha$ vanish identically.}
Neighboring nodes pertaining to the same edge are connected once, i.e. the number of intersections between the tail nodes is 
\be
k_{\alpha,x;\beta, y} = \delta_{\alpha,\beta} \left( \delta_{\left|x-y\right|,1}-2\delta_{x,y} \right)\,.
\ee
We have introduced a term so that the self-intersections are $k_{\alpha,x;\alpha,x}=-2$. 
The tail nodes cannot have loops attached to them.
The number of edges between color and tail nodes is given by
\be \label{kAdd}
k^c_{\alpha, x} = \frac{1}{m^c}\left(\mu^c_{\alpha, x}-\mu^c_{\alpha, x+1}\right)\,.
\ee
\end{enumerate}
}
Since each colored sub-polygon $S^c$, consisting of $m^c$ IMPs, is dual to a sub-web of multiplicity $m^c$, we map it to a node in the tropical quiver with label $m^c$. The two contributions to the edges between color nodes in the tropical quiver correspond to the ``stable intersection number'' and 7-brane contribution in the web. We refer the reader to section \ref{sec:BraneWebs} for details on the GTP-to-brane-web map. The tail nodes are not realized in $(P,\{\lambda^c_\alpha\})$ as a colored sub-polygon. The information needed to construct the tails is nonetheless contained in the colored GTP, and is given above in terms of the data defined in the previous subsections. 

\paragraph{Self-intersection.}
The self-intersection $k^{cc}$ of a colored sub-polygon $S^c$ is a measure of the amount of adjoint matter associated to the corresponding node. Specifically, if the self-intersection number differs from $-2$, the $U(m^c)$ ($m^c \neq 1$) node of the magnetic quiver is associated with $\ell^c$ adjoint matter multiplets, indicated by a loop attached to this node. Moreover, the self-intersection of a colored sub-polygon is related to the rank of the Higgsed sub-sector of the SCFT that this sub-polygon corresponds to. In particular, we find
\be 
k^{cc}=2(r(T^c)-1)\,,
\ee
where $T^c$ is the IMP in \eqref{ColorMultiplicity}. Thus, the self-intersection is bounded from below by $-2$.
The tail nodes have self-intersection $k_{\alpha,x;\alpha,x}=-2$ so they cannot have adjoint matter associated to them, i.e. $l_{\alpha,x}=1+\frac{k_{\alpha,x;\alpha,x}}{2}=0$. For theories with a high number of flavors the appearance of colored sub-polygons with $k^{cc} \neq -2$ is very rare \cite{BBESII}. E.g. this phenomenon will not show up in the examples of section \ref{sec:Examples}, and therefore we do not explicitly give the self-intersection numbers there. This {\it is} a feature of our main example, in a trivial sense however, as we show below.
\\

The tropical quiver will be interpreted as defining a 3d $\mathcal{N}=4$ quiver gauge theory:
vertices $U(m_I)$ with $\ell_I$ adjoint matter multiplets, connected by $k_{I J}$ hypermultiplets.  
The key relation of this tropical quiver to the original 5d QFT, is via its Coulomb branch, which is a hyper-K\"ahler manifold, and is identified with the Higgs branch of the 5d QFT $\mathcal{T}_P$.

{\conjecture[Tropical and Magnetic Quivers]
Let $P$ be a GTP, associated to a 5d SCFT $\mathcal{T}_P$. There is a bijection between the inequivalent colorings of $P$ and the cones of the Higgs branch of $\mathcal{T}_P$; moreover, the tropical quiver associated to a coloring is identified with a magnetic quiver for the corresponding cone in the Higgs branch. 
}\\

We obtain this conjecture by utilizing the 1-1 map to the brane-webs, that describe the 5d $\mathcal{N}=1$ theories, in section \ref{sec:BraneWebs}.

\example
Now we are in the position to compute the tropical quiver of $P$ in \eqref{MainExStart} using its colorings. First, consider \eqref{MainExColor1}. There are two colored sub-polygons
\be 
\begin{tikzpicture}[x=.5cm,y=.5cm]

\draw[step=.5cm,gray,very thin] (0,-1) grid (4,2);

\draw[ligne,blue] (0,0)--(1,-1)--(3,-1)--(4,0)--(2,2)--(0,0);

\node[] at (-2,0) {$S^b=$};
 
\node[bd] at (0,0) {}; 
\node[bd] at (1,-1) {}; 
\node[wd] at (2,-1) {}; 
\node[bd] at (3,-1) {}; 
\node[bd] at (4,0) {}; 
\node[wd] at (3,1) {}; 
\node[bd] at (2,2) {}; 
\node[wd] at (1,1) {}; 

\node[] at (6,0) {$S^g=$};

\draw[step=.5cm,gray,very thin] (8,-1) grid (10,0);

\draw[ligne,green] (8,-1)--(10,-1)--(9,0)--(8,-1);

\node[bd] at (8,-1) {}; 
\node[bd] at (9,-1) {}; 
\node[bd] at (10,-1) {}; 
\node[bd] at (9,0) {};

\end{tikzpicture}
\ee
with multiplicities $m^b=m^g=1$. We compute the intersection of the corresponding color nodes explicitly
\be
k^{bg} =\left(6-0-2-0-2-2\right)=0\,,
\ee
where the area was given in \eqref{MainExAreas} and the partitions are given in \eqref{MainExSb} and \eqref{MainExSg} respectively. The self-intersection numbers of the two colored sub-polygons are
\be 
k^{bb}=2 \cdot 7-3 \cdot 2^2-1-1=0\,, \qquad k^{gg}=2\cdot 1-4\cdot 1=-2\,.
\ee
However, given that $m^b$=1 the adjoint representation is trivial, so we drop the loop in the magnetic quiver.
Now, consider the tails. The only edge of $P'$ with $b_\alpha\neq0$ is $E_2$ with $b_2=3$. Thus we can compute
\be
m_{2,1}=-1+2+1=2\,, \quad m_{2,2}=-2+2+2=2\,, \quad m_{2,3}=-3+2+2=1\,,
\ee
Finally, the intersections between the color and tail nodes are given by
\be
k^b_{2,1}=2\,,\qquad k^g_{2,2}=1\,,
\ee
with all others vanishing. Thus, the tropical quiver for $P'$ with this coloring is
\be
\begin{tikzpicture}[x=.8cm,y=.8cm]
\node (g1) at (0,0) [gauge,label=below:\large{1}] {};
\node (g2) at (1,0) [gauge,label=below:\large{2}] {};
\node (g3) at (2,0) [gauge,label=below:\large{2}] {};
\node (g4) at (3,0) [gaugeblue,label=below:\large{1}] {};
\node (g5) at (1,1) [gaugegreen,label=right:\large{1}] {};
\draw (g1)--(g2)--(g3);
\draw (g2)--(g5);
\draw[double distance=3pt] (g3)--(g4);
\end{tikzpicture}\,.
\ee
We can repeat the analysis above for the second coloring of $P'$ in \eqref{MainExColor2}, and find that the tropical quiver associated to this coloring is
\be
\begin{tikzpicture}[x=.8cm,y=.8cm]
\node (g1) at (0,0) [gauge,label=below:\large{1}] {};
\node (g2) at (1,0) [gauge,label=below:\large{2}] {};
\node (g3) at (2,0) [gaugegreen,label=below:\large{2}] {};
\node (g4) at (3,0) [gaugecyan,label=below:\large{1}] {};
\node (g5) at (3,1) [gaugeblue,label=right:\large{1}] {};
\node (g6) at (1,1) [gauge,label=right:\large{1}] {};
\draw (g1)--(g2)--(g3);
\draw (g2)--(g6);
\draw (g5)--(g3)--(g4);
\draw[double distance=3pt] (g4)--(g5);
\end{tikzpicture}\,.
\ee
We identify each of these tropical quivers with a magnetic quiver, giving the two cones on the Higgs branch of the 5d SCFT $\mathcal{T}_P= \mathcal{T}_{P'}$. This theory can be shown to represent the strongly coupled $SU(4)_4 + 4\bm{F}$ \cite{Cabrera:2018jxt}. 

\section{Symplectic Leaves and Colorings with Internal Edges}
\label{sec:Internal}

Our discussion so far required the GTP to have no internal lines. In M-theory, such geometries correspond to 5d SCFTs -- which are the main focus of our attention. 
In this section we generalize our approach to include GTPs with internal lines, which in the geometry are partial resolutions, and in 5d language correspond to opening up partial (extended) Coulomb branch directions. 
The motivation is two-fold: obviously if one would like to study weakly coupled 5d gauge theories, these have rulings, i.e. internal edges. Secondly, and perhaps more importantly for the current endeavor of mapping out the Higgs branch, we require these to construct the Hasse diagram, i.e. the partially ordered set of symplectic leaves that comprise the Higgs branch, as a hyper-K\"ahler cone. 
The Higgsing is constructed by successively opening partial Coulomb branch directions. 
To construct the associated magnetic quivers after each Higgsing, we require to be able to generalize the algorithm to GTPs with internal lines.

\subsection{Colorings with Internal Edges}
\label{sec:IntEdges}

For a polygon with internal edges, the edge coloring has to be extended to these. The first step is to extend our algorithm \ref{def:Coloring}. For polygons with internal edges, the s-rule only has to be obeyed on external edges (it is irrelevant to apply it to internal edges, given that these are dual to internal 5-branes (which do not end on any 7-branes)). 
A colored sub-polygon $S^c$ needs to be divided completely by internal edges, so that all extreme vertices in $S^c$ are part of at least two edges.  We argue for this approach in the brane picture in section \ref{sec:sub-webs}.

{\algorithm[Coloring for Polygons with Internal Edges]

Consider a GTP $\TD$ with internal edges and denote the sub-polygons by $P_A$, where $A \subset \alpha = \{1, \cdots, n_E\}$, 
 \be
 \label{eq:sub-poly}
 E \cap P_A = \cup_{\alpha \in A} E_\alpha\,,
 \ee 
so the $P_A$ do not have any internal edges. 

A coloring for a GTP $P$ with internal edges $E^{\text{in}} \neq \emptyset$ is a partition
\be
\sum_{c=1}^{n_c} \lambda^c_\alpha = \lambda_\alpha\,.
\ee
For all $c$, the $\{\lambda^c_\alpha\}$ defines a refined Minkowski summand $S^c$ of $P\leq S^{1} \oplus \dots \oplus S^{n_c}$ obeying the following rules:
\begin{enumerate}
\item The $S^c$ can be divided into sub-polygons
\be \label{Subpolygons}
S^c = \bigcup S^c_A\,,
\ee
without internal lines.

\item
For all $c$ and $A$
\be
S^c_A = \underbrace{T^c_A \oplus \dots \oplus T^c_A}_{m^c_A}\,, \qquad T^c_A = \bigoplus_i T^i_A\,,
\ee
where each $T^i_A$ is irreducible and obeys the s-rule minimally, for all external edges of $P$. We require that for $c\not= d$  
\be
T^c_A \neq T^d_A\,, \quad \forall A\,.
\ee
The multiplicity of $S^c$ is given by
\be
m^c = \gcd_{A} m^c_A\,.
\ee
\item The partition is maximal, i.e. there is no sub-partition
\be
S^{c_1} \leq S^{c_2} \oplus S^{c_3}\,,
\ee
such that the resulting coloring is valid.

\end{enumerate}
}

Essentially, we can understand this algorithm as a generalization of the algorithm \ref{def:Coloring} by extending the number of constraints imposed by the internal edges. 

{\example
Let us exemplify this by finding the coloring for the weakly coupled $SU(4)$ description of $P'$ in \eqref{MainExPruned}. Turning on the $SU(4)$ gauge coupling corresponds to adding an internal edge, so that the GTP becomes
\be \label{MainExWeakly}
\begin{tikzpicture}[x=.5cm,y=.5cm]

\draw[step=.5cm,gray,very thin] (0,-1) grid (6,3);

\draw[ligne,black] (0,0)--(1,-1)--(5,-1)--(6,0)--(5,1)--(4,2)--(3,3)--(2,2)--(1,1)--(0,0);
\draw[ligne,black] (3,-1)--(3,3);

\node[] at (-2,0) {$P^{\text{w.c.}}=$};
 
\node[bd] at (0,0) {}; 
\node[bd] at (1,-1) {}; 
\node[bd] at (2,-1) {}; 
\node[bd] at (3,-1) {}; 
\node[bd] at (4,-1) {}; 
\node[bd] at (5,-1) {}; 
\node[bd] at (6,0) {}; 
\node[wd] at (5,1) {}; 
\node[wd] at (4,2) {}; 
\node[bd] at (3,3) {}; 
\node[wd] at (2,2) {}; 
\node[wd] at (1,1) {}; 

\end{tikzpicture}
\ee
At this point, we do not distinguish between black and white vertices along the internal edge. By including the internal edge the GTP data changes to
\be
\ba
&L^\p_\alpha=((1,-1),(1,0),(1,0),(1,1),(-1,1),(-1,-1)\,, \qquad L^{\text{in}}=(0,1)\\
&\lambda^\p_\alpha=(1,2,2,1,3,3)\,, \qquad \{\mu_\alpha\}=(\{1\},\{1^2\},\{1^2\},\{1\},\{3\},\{3\}) \,, \qquad \lambda^{\text{in}}=4\,.
\ea
\ee
Furthermore, the internal edge splits $P$ into two with the edges divided as
\be
A_1 = \{1,2,6,\text{in}\}\,, \qquad A_2=\{3,4,5,\text{in}\}\,.
\ee
We need to find compatible colorings of $P^{\text{w.c.}}_{A_1}$ and $P^{\text{w.c.}}_{A_2}$, i.e. the coloring on the internal edge agrees. The only possible coloring is
\be \label{MainExWeaklyColor}
\begin{tikzpicture}[x=.5cm,y=.5cm]

\draw[step=.5cm,gray,very thin] (0,-1) grid (6,3);

\draw[ligne,blue] (1,-1)--(5,-1);
\draw[ligne,blue] (5,1)--(3,3)--(1,1);
\draw[ligne,blue] (3,1)--(3,3);
\draw[ligne,green] (1,1)--(0,0)--(1,-1);
\draw[ligne,green] (5,-1)--(6,0)--(5,1);
\draw[ligne,green] (3,-1)--(3,1);
 
\node[bd] at (0,0) {}; 
\node[bd] at (1,-1) {}; 
\node[bd] at (2,-1) {}; 
\node[bd] at (3,-1) {}; 
\node[bd] at (4,-1) {}; 
\node[bd] at (5,-1) {}; 
\node[bd] at (6,0) {}; 
\node[wd] at (5,1) {}; 
\node[wd] at (4,2) {}; 
\node[bd] at (3,3) {}; 
\node[wd] at (2,2) {}; 
\node[wd] at (1,1) {}; 
\node[] at (3,0) {};
\node[] at (3,1) {};
\node[] at (3,2) {};

\end{tikzpicture}
\ee
From here, we can again compute the magnetic quiver, noting that the lower line is now divided into two distinct edges. The magnetic quiver turns out to be
\be
\begin{tikzpicture}[x=.8cm,y=.8cm]
\node (g1) at (0,0) [gauge,label=below:\large{1}] {};
\node (g2) at (1,0) [gaugeblue,label=below:\large{2}] {};
\node (g3) at (2,0) [gauge,label=below:\large{1}] {};
\node (g4) at (1,1) [gaugegreen,label=right:\large{1}] {};
\draw (g1)--(g2)--(g3);
\draw[double distance=3pt] (g2)--(g4);
\end{tikzpicture}
\ee
as expected for a weakly coupled $SU(4) + 4\bm{F}$.
}

\subsection{Hasse Diagram and Symplectic Leaves}

\label{sec:Hasse}

The Higgs Branch of a 5d SCFT or gauge theory, has a foliation in terms of symplectic leaves, as reviewed in section \ref{sec:Review}. 
Starting from the magnetic quiver associated to the 5d QFT, the Hasse diagram can be obtained by successive quiver subtractions \cite{Cabrera:2018ann, Bourget:2019aer, Bourget:2019rtl}. 

Here we provide a derivation in terms of the data of the colored polygon $(P, \{\lambda^c_\alpha\})$, which characterize each cone of the Higgs branch of the theory $\mathcal{T}_P$. The full foliation structure is however in general interconnected.

The Hasse diagram is obtained by introducing successive partial resolutions into the GTP. In each step we start with a polygon $P$ and find subtractions $\Delta_i$ that result in the next layer $Q_i$ of the Hasse diagram -- there can in general be multiple such subtractions:
\be
\begin{tikzpicture}
\node (1) [hasse,label=above:$P$] at (0,0) {};
\node (2) [hasse,label=below:$Q_1$] at (-1,-1) {};
\node (3) [hasse,label=below:$Q_n$] at (1,-1) {};
\draw (1) edge [] node[label=left:$\Delta_1$] {} (2);
\draw (1) edge [] node[label=right:$\Delta_n$] {} (3);

\node (4) at (0,-.75) {\large$\cdots$};

\end{tikzpicture}
\ee
If at some point the GTPs $Q_j$ along different branches agree, these nodes should be identified, leading to interconnections. This idea was first introduced in \cite{Bourget:2019aer}, where it was argued in the brane picture that the Higgs branch at specific points along the Coulomb branch reproduces the different layers in the Hasse diagram.

{\definition[Hasse diagram of a GTP] 

Let $P$ be a GTP, characterized by edges $E_\alpha$ with $\lambda^{(P)}_\alpha$ and $\mu^{(P)}_{\alpha,x}$.
A transition in the Hasse diagram  $P \xrightarrow{\Delta} Q$ is a set of internal edges $E_\beta$, which correspond to a partial resolution, with minimal line segments $L_\beta$ and number of line segments $\nu_\beta$, such that:
\begin{enumerate}
\item There is a GTP, $Q$, characterized by
\be
E_{(Q)}^{\p}=E_{(P)}^{\p}\,, \qquad E_{(Q)}^{\text{in}}=E_{(P)}^{\text{in}}\cup E_\beta\,, \qquad
\lambda^{(Q)}_\alpha=\{\lambda^{(P)}_\alpha,\nu_\beta\}\,, \qquad \mu^{(Q)}_{\alpha,x}=\mu^{(P)}_{\alpha,x}\,.
\ee
\item $Q$ admits a coloring, such that there is a color $c^\ast$ with
\be
\nu_\beta^{c^\ast} = \nu_\beta \quad \forall \beta\,,
\ee
with associated $\lambda_\alpha^{c^\ast}$ and $ \mu_{\alpha,x}^{c^\ast}$.
\item There is a GTP $\Delta$ characterized by
\be
E_{(\Delta)} = \{E_\alpha^{(P)} | \lambda_\alpha^{c^\ast} \neq 0\}\,, \qquad \lambda_\alpha^{(\Delta)} = \lambda_\alpha^{c^\ast} \,, \qquad \mu_{\alpha,x}^{(\Delta)} = \mu_{\alpha,x}^{c^\ast}\,,
\ee
such that the magnetic quiver of $\Delta$ is a either the magnetic quiver of a Kleinian singularity or the closure of a minimal nilpotent orbit. 

\end{enumerate}
} 

Note that the leaves that we allow here are those discussed in section \ref{sec:Review}. In case that a given GTP has symplectic leaves that go outside of this class, it would interesting to see how these are realized in the GTP in terms of introducing internal lines. 
We are confident that any effect that occurs in the webs has a counterpart in our formulation in terms of the GTPs, including more general symplectic leaves.

A physical way to understand the transitions in the Hasse diagram is in terms of a Higgsing, or combinatorially in terms of quiver subtractions in the magnetic quiver \cite{Cabrera:2018ann}.  
This concept is reviewed in appendix \ref{app:Webs}. Essentially, we replace the affine Dynkin diagram of an ADE algebra (or a Kleinian singularity) by a single $U(1)$ rebalancing node. The idea is that the newly introduced color $c^\ast$ represents the rebalancing node. Including the $\nu_\beta$, $S^{c^\ast}$ has magnetic quiver $U(1)$ but excluding them it would represent an ADE singularity.
We argue about the details of this process in section \ref{sec:HasseWeb}, where we explain how it relates to a partial opening of a Coulomb branch and subsequent Higgsing.

{\proposition

If $P \xrightarrow{\Delta} Q$ is a transition in a Hasse diagram, we can write $P$ as a refined Minkowski sum
\be
P \leq \Delta \oplus S\,,
\ee
where $S$ is another GTP, which need not be irreducible and could be empty. Conversely, such a Minkowski sum decomposition induces a transition in the Hasse diagram if
\begin{enumerate}
\item The magnetic quiver of $\Delta$ is a symplectic singularity \item There is an extension of $\Delta$ by internal lines $\nu_\beta$ to $\Delta_\nu$ with $\lambda_\alpha^{(\Delta_\nu)}=\{\lambda_\alpha^{(\Delta)},\nu_\beta\}$ such that the s-rule for $\Delta_\nu$ is obeyed minimally on each external edge.
\end{enumerate}
Then, $Q\leq \Delta_\nu \oplus S$ with $\mu^{(Q)}_{\alpha,x}=\mu^{(P)}_{\alpha,x}$.

}
This implies that, should we find a $\Delta$ representing a symplectic singularity as a refined Minkowski summand of $P$, we can subtract the corresponding magnetic quiver, provided that $\Delta$ can be condensed to a $U(1)$ by the inclusion of internal lines.

\section{Examples: SQCDs,  Non-Lagrangian and Toric Models}
\label{sec:Examples}

We now provide a large class of diverse examples to show the workings of our conjecture. We will study 5d SCFTs, which have IR descriptions as SQCD, i.e. $SU(N_c)_k + N_F \bm{F}$, or with additional antisymmetric matter. 
We extend our analysis to models such as $T_N$ as well as descendants of $T_N$ that are non-Lagrangian. 
Another class of known theories are the strictly convex toric theories, which have a direct connection to the work of Altmann \cite{Altmann,Altmann2}.

\subsection{SQCD-like Theories}
\subsubsection{$SU(3)_1+ 6 {\bf F}$ UV SCFT}

Our first example is the strongly coupled SCFT, with IR description given by $SU(3)_1 + 6\bm{F}$. 
The GTP $P$ is given by\footnote{Whichever presentation the reader would want to use for this, the only requirement is that, after applying Hanany-Witten moves, the GTP is convex.}
\be
\begin{tikzpicture}[x=.5cm,y=.5cm]

\draw[step=.5cm,gray,very thin] (0,0) grid (2,4);
\node[] at (-1.5,2) {$P=$};
\draw[ligne] (0,0)--(0,4)--(1,4)--(2,3)--(2,0)--(1,0)--(0,0); 
\node[bd] at (0,0) {}; 
\node[bd] at (0,1) {}; 
\node[bd] at (0,2) {}; 
\node[bd] at (0,3) {}; 
\node[bd] at (0,4) {}; 
\node[bd] at (1,4) {}; 
\node[bd] at (2,3) {};
\node[bd] at (2,2) {}; 
\node[bd] at (2,1) {}; 
\node[bd] at (2,0) {}; 
\node[wd] at (1,0) {}; 
\end{tikzpicture} \,,
\ee
which is characterized by the following data:
\be \label{Ex1Data}
\ba
&L_\alpha=\left((0,-1),(1,0),(0,1),(-1,1),(-1,0)\right)\\
&\lambda_\alpha = (4,2,3,1,1)\,, \qquad \mu_{\alpha,x} = (\{1^4\},\{2\},\{1^3\},\{1\},\{1\})\,.
\ea
\ee
As usual, we label the five edges of $P$ in counterclockwise order, starting from the top left edge.
The unique consistent coloring (blue, green, cyan) of this diagram is given by
\be
\begin{tikzpicture}[x=.5cm,y=.5cm]

\draw[step=.5cm,gray,very thin] (0,0) grid (2,4);

\draw[ligne,cyan] (0,0)--(0,3); 
\draw[ligne,blue] (0,3)--(0,4); 
\draw[ligne,green] (0,4)--(1,4); 
\draw[ligne,blue] (1,4)--(2,3); 
\draw[ligne,cyan] (2,3)--(2,0); 
\draw[ligne,blue] (2,0)--(1,0); 
\draw[ligne,green] (1,0)--(0,0); 

\node[bd] at (0,0) {}; 
\node[bd] at (0,1) {}; 
\node[bd] at (0,2) {}; 
\node[bd] at (0,3) {}; 
\node[bd] at (0,4) {}; 
\node[bd] at (1,4) {}; 
\node[bd] at (2,3) {};
\node[bd] at (2,2) {}; 
\node[bd] at (2,1) {}; 
\node[bd] at (2,0) {}; 
\node[wd] at (1,0) {}; 
\end{tikzpicture} \,,
\ee
i.e. the refined Minkowski sum decomposition of $P$ is given by
\be \label{Ex1MinkSum}
\begin{tikzpicture}[x=.5cm,y=.5cm] 

\node[] at (-1.5,0) {$P<$};

\draw[ligne,cyan] (0,-1.5)--(0,-.5)--(0,.5)--(0,1.5); 
\node[bd] at (0,-1.5) {}; 
\node[wd] at (0,-.5) {}; 
\node[wd] at (0,.5) {}; 
\node[bd] at (0,1.5) {}; 

\node[] at (1,0) {$\oplus$};

\draw[ligne,green] (2,0)--(3,0); 
\node[bd] at (2,0) {}; 
\node[bd] at (3,0) {}; 

\node[] at (4,0) {$\oplus$};

\draw[ligne,blue] (5,-.5)--(5,.5)--(6,-.5)--(5,-.5); 
\node[bd] at (5,-.5) {}; 
\node[bd] at (5,.5) {}; 
\node[bd] at (6,-.5) {}; 
\end{tikzpicture}
\ee
with the partitions and multiplicities of these colorings being
\be \label{Ex1lambdamu}
\ba
&\lambda_\alpha^c = (3,0,3,0,0)\,, \quad \mu_{\alpha x}^c = \left(\{3\},-,\{3\},-,-\right)\,, \quad &&m^c=3\\
&\lambda_\alpha^g = (0,1,0,0,1)\,, \quad \mu_{\alpha x}^g = \left(-,\{1\},-,-,\{1\}\right)\,, \quad &&m^g=1\\
&\lambda_\alpha^b = (1,1,0,1,0)\,, \quad \mu_{\alpha x}^b = \left(\{1\},\{1\},-,\{1\},-\right)\,, \quad &&m^b=1\,.
\ea
\ee
Clearly, all the summands in \eqref{Ex1MinkSum} are irreducible and obey the s-rule minimally (up to multiplicity).
A choice of color sub-division is given by
\be
\begin{tikzpicture}[x=.5cm,y=.5cm]

\draw[step=.5cm,gray,very thin] (0,0) grid (2,4);

\draw[ligne,cyan] (0,0)--(0,3); 
\draw[ligne,blue] (0,3)--(0,4); 
\draw[ligne,green] (0,4)--(1,4); 
\draw[ligne,blue] (1,4)--(2,3); 
\draw[ligne,cyan] (2,3)--(2,0); 
\draw[ligne,blue] (2,0)--(1,0); 
\draw[ligne,green] (1,0)--(0,0); 

\draw[densely dotted,thick,cyan] (1,0)--(1,3);
\draw[densely dotted,thick,green] (0,3)--(1,3);
\draw[densely dotted,thick,blue] (1,3)--(1,4);
\draw[densely dotted,thick,blue] (1,3)--(2,3);

\node[bd] at (0,0) {}; 
\node[bd] at (0,1) {}; 
\node[bd] at (0,2) {}; 
\node[bd] at (0,3) {}; 
\node[bd] at (0,4) {}; 
\node[bd] at (1,4) {}; 
\node[bd] at (2,3) {};
\node[bd] at (2,2) {}; 
\node[bd] at (2,1) {}; 
\node[bd] at (2,0) {}; 
\node[wd] at (1,0) {}; 
\end{tikzpicture} \,,
\ee
so that the areas of the parallelograms are
\be\label{Ex1Areas}
A^{cg} = 3\,, \quad A^{cb} = 3\,, \quad A^{gb}=1\,.
\ee
From \eqref{Ex1lambdamu} and \eqref{Ex1Areas} we can compute the intersections between the color nodes using 
\eqref{kColor} to be
\be
k^{cg} = 1\,, \quad k^{cb} = 0\,, \quad k^{gb}=0\,.
\ee
Now, let's turn to the additional nodes, which appear at edges $E_1$ and $E_3$. The node labels and mutual edges can be read off from the $\mu_{\alpha,x}$ and $\mu^c_{\alpha,x}$ in \eqref{Ex1Data} and \eqref{Ex1lambdamu}:
\be
\ba
&m_{1,x}=(3,2,1)\,,\quad &&k^c_{1,1}=1\,, \quad k^b_{1,1}=1\\
&m_{3,x}=(2,1)\,,\quad &&k^c_{3,1}=1\,,
\ea
\ee
and all others vanishing. Putting all this together the magnetic quiver is given by
\be
\begin{tikzpicture}[x=.8cm,y=.8cm]
\node (g1) at (0,0) [gauge,label=below:\large{1}] {};
\node (g2) at (1,0) [gauge,label=below:\large{2}] {};
\node (g3) at (2,0) [gauge,label=below:\large{3}] {};
\node (g4) at (3,0) [gaugecyan,label=below:\large{3}] {};
\node (g5) at (4,0) [gauge,label=below:\large{2}] {};
\node (g6) at (5,0) [gauge,label=below:\large{1}] {};
\node (g7) at (2,1) [gaugeblue,label=left:\large{1}] {};
\node (g8) at (3,1) [gaugegreen,label=right:\large{1}] {};
\draw (g1)--(g2)--(g3)--(g4)--(g5)--(g6);
\draw (g3)--(g7);
\draw (g4)--(g8);
\end{tikzpicture}
\,.
\ee

Now, let us look at the Hasse diagram of this theory. We can check that $P$ can be written as a refined Minkowski sum
\be
\begin{tikzpicture}[x=.5cm,y=.5cm]

\draw[step=.5cm,gray,very thin] (1,0) grid (3,4);
\draw[step=.5cm,gray,very thin] (5,0) grid (7,3);
\draw[ligne,black] (1,0)--(1,2);
\draw[ligne,black] (1,3)--(1,4)--(2,4)--(3,3);
\draw[ligne,black] (3,2)--(3,0)--(2,0)--(1,0); 
\draw[ligne,black] (1,3)--(1,2);
\draw[ligne,black] (3,2)--(3,3);
\node[bd] at (1,0) {}; 
\node[bd] at (1,1) {}; 
\node[bd] at (1,2) {}; 
\node[bd] at (1,3) {}; 
\node[bd] at (1,4) {}; 
\node[bd] at (2,4) {}; 
\node[bd] at (3,3) {}; 
\node[bd] at (3,2) {};
\node[bd] at (3,1) {}; 
\node[bd] at (3,0) {}; 
\node[wd] at (2,0) {};

\node[] at (2,-2) {$P$};

\node[] at (4,1) {$<$};
\node[] at (4,-2) {$<$};

\draw[ligne,black] (5,0)--(5,3)--(6,3)--(7,2)--(7,0)--(6,0)--(5,0); 
\node[bd] at (5,0) {}; 
\node[bd] at (5,1) {}; 
\node[bd] at (5,2) {}; 
\node[bd] at (5,3) {}; 
\node[bd] at (6,3) {}; 
\node[bd] at (7,2) {};
\node[bd] at (7,1) {}; 
\node[bd] at (7,0) {}; 
\node[wd] at (6,0) {}; 

\node[] at (6,-2) {$\Delta$};

\node[] at (8,1) {$\oplus$};
\node[] at (8,-2) {$\oplus$};

\draw[ligne,black] (9,0)--(9,1); 
\node[bd] at (9,0) {}; 
\node[bd] at (9,1) {};

\node[] at (9,-2) {$S$};

\end{tikzpicture} \,,
\ee
We can now check that this decomposition induces a transition in the Hasse diagram. It is straightforward to see that the magnetic quiver of $\Delta$ is the affine Dynkin diagram of $\mathfrak{d}_5$.
Furthermore, there is an extension of $\Delta$ with internal lines such that the s-rule is minimally obeyed on each external edge, namely
\be
\begin{tikzpicture}[x=.5cm,y=.5cm]

\draw[step=.5cm,gray,very thin] (5,0) grid (7,3);

\draw[ligne,black] (5,0)--(5,3)--(6,3)--(7,2)--(7,0)--(6,0)--(5,0); 
\draw[ligne,black] (5,0)--(6,2)--(7,0);
\draw[ligne,black] (6,2)--(7,2);
\draw[ligne,black] (6,2)--(5,3);
\node[bd] at (5,0) {}; 
\node[bd] at (5,1) {}; 
\node[bd] at (5,2) {}; 
\node[bd] at (5,3) {}; 
\node[bd] at (6,3) {}; 
\node[bd] at (7,2) {};
\node[bd] at (7,1) {}; 
\node[bd] at (7,0) {}; 
\node[wd] at (6,0) {}; 

\end{tikzpicture} \,,
\ee
which is $\Delta_\nu$.
Thus, we can deduce $Q$, and its unique coloring, to be
\be
\begin{tikzpicture}[x=.5cm,y=.5cm]

\draw[step=.5cm,gray,very thin] (1,0) grid (3,4);

\draw[ligne,blue] (1,0)--(1,2);
\draw[ligne,blue] (1,3)--(1,4)--(2,4)--(3,3);
\draw[ligne,blue] (3,2)--(3,0)--(2,0)--(1,0); 
\draw[ligne,blue] (1,0)--(2,2);
\draw[ligne,blue] (3,0)--(2,2);
\draw[ligne,blue] (2,3)--(1,4);
\draw[ligne,blue] (2,3)--(3,3);
\draw[ligne,cyan] (1,3)--(1,2);
\draw[ligne,cyan] (3,2)--(3,3);
\node[bd] at (1,0) {}; 
\node[bd] at (1,1) {}; 
\node[bd] at (1,2) {}; 
\node[bd] at (1,3) {}; 
\node[bd] at (1,4) {}; 
\node[bd] at (2,4) {}; 
\node[bd] at (3,3) {}; 
\node[bd] at (3,2) {};
\node[bd] at (3,1) {}; 
\node[bd] at (3,0) {}; 
\node[wd] at (2,0) {};

\end{tikzpicture} \,,
\ee
where we denote the color $c^\ast$ by blue and the other color in the coloring of $Q$ in cyan.
We can read off the data of the external lines
\be
\ba
&\lambda_\alpha^c = (1,0,1,0,0)\,, \quad \mu_{\alpha, x}^c = \left(\{1\},-,\{1\},-,-\right)\,, \quad &&m^c=1\\
&\lambda_\alpha^b = (3,2,2,1,1)\,, \quad \mu_{\alpha, x}^b = \left(\{1^3\},\{2\},\{1^2\},\{1\},\{1\}\right)\,, \quad &&m^b=1\,.
\ea
\ee
Together with $A^{cb}=2$ we can deduce
\be
\ba
&k^{cb}=0\\
&m_{1,x}=(1,1,1)\,,\quad &&k^c_{1,1}=1\,, \quad k^b_{1,3}=1\\
&m_{3,x}=(1,1)\,,\quad &&k^c_{3,1}=1\,, \quad k^b_{3,2}=1\,,
\ea
\ee
with all others vanishing. Altogether, we arrive at the affine Dynkin diagram of $a_7$. Since the magnetic quiver of $Q$ itself is a symplectic singularity, there is a trivial transition with $\Delta=Q$. Thus, the Hasse diagram of the strongly coupled $SU(3)_1 + 6 \bm{F}$ is
\be
\begin{tikzpicture}
\node (1) [hasse] at (0,0) {};
\node (2) [hasse] at (0,-1) {};
\node (3) [hasse] at (0,-2) {};
\draw (1) edge [] node[label=left:$\mathfrak{d}_5$] {} (2);
\draw (2) edge [] node[label=left:$\mathfrak{a}_7$] {} (3);
\end{tikzpicture} \,.
\ee

\subsubsection{$SU(3)_1+ 6 {\bf F}$ IR Theory}

Now, we turn to the weakly coupled version of the theory, i.e. the gauge theory $SU(3)_1+ 6\bm{F}$, which has a description as a GTP $P$, where we have a ruling  
\be
\begin{tikzpicture}[x=.5cm,y=.5cm]

\draw[step=.5cm,gray,very thin] (0,0) grid (2,4);

\draw[ligne] (0,0)--(0,4)--(1,4)--(2,3)--(2,0)--(1,0)--(0,0); 
\draw[ligne] (1,4)--(1,1)--(0,1);
\draw[ligne] (1,1)--(2,0);
\node[bd] at (0,0) {}; 
\node[bd] at (0,1) {}; 
\node[bd] at (0,2) {}; 
\node[bd] at (0,3) {}; 
\node[bd] at (0,4) {}; 
\node[bd] at (1,4) {}; 
\node[bd] at (2,3) {};
\node[bd] at (2,2) {}; 
\node[bd] at (2,1) {}; 
\node[bd] at (2,0) {}; 
\node[wd] at (1,0) {}; 
\end{tikzpicture} \,,
\ee
There is one valid coloring (with color sub-division), which is given by
\be
\begin{tikzpicture}[x=.5cm,y=.5cm]

\draw[step=.5cm,gray,very thin] (0,0) grid (2,4);

\draw[ligne,blue] (0,0)--(0,1); 
\draw[ligne,cyan] (0,1)--(0,4); 
\draw[ligne,green] (0,4)--(1,4); 
\draw[ligne,blue] (1,4)--(2,3); 
\draw[ligne,cyan] (2,3)--(2,0); 
\draw[ligne,blue] (2,0)--(1,0); 
\draw[ligne,green] (1,0)--(0,0); 
\draw[ligne,cyan] (1,4)--(1,1); 
\draw[ligne,blue] (2,0)--(1,1); 
\draw[ligne,green] (1,1)--(0,1); 
\draw[densely dotted,thick,blue] (1,0)--(1,1);

\node[bd] at (0,0) {}; 
\node[bd] at (0,1) {}; 
\node[bd] at (0,2) {}; 
\node[bd] at (0,3) {}; 
\node[bd] at (0,4) {}; 
\node[bd] at (1,4) {}; 
\node[bd] at (2,3) {};
\node[bd] at (2,2) {}; 
\node[bd] at (2,1) {}; 
\node[bd] at (2,0) {}; 
\node[wd] at (1,0) {}; 
\end{tikzpicture} \,,
\ee
i.e. the coloring data for the external lines is
\be
\ba
&\lambda_\alpha^c = (3,0,0,3,0,0)\,, \quad \mu_{\alpha, x}^c = \left(\{3\},-,-,\{3\},-,-\right)\,, \quad &&m^c=3\\
&\lambda_\alpha^g = (0,0,1,0,0,1)\,, \quad \mu_{\alpha, x}^g = \left(-,-,\{1\},-,-,\{1\}\right)\,, \quad &&m^g=1\\
&\lambda_\alpha^b = (0,1,1,0,1,0)\,, \quad \mu_{\alpha, x}^b = \left(-,\{1\},\{1\},-,\{1\},-\right)\,, \quad &&m^b=1\,.
\ea
\ee
Note, that now we have an additional edge, since the edge on the left is subdivided by an internal line. The intersections between the color nodes are
\be
A^{cg}= A^{cb}=3\,,\quad A^{gb}=1\,,\qquad k^{cg}= k^{cb}=1\,, \quad k^{gb}=0\,.
\ee
The additional nodes appear at edges $E_1$ and $E_4$ with
\be
\ba
&m_{1,x}=(2,1)\,,\quad &&k^c_{1,1}=1\\
&m_{4,x}=(2,1)\,,\quad &&k^c_{4,1}=1\,,
\ea
\ee
with all others vanishing. Putting all together the magnetic quiver of the weakly coupled theory is 
\be
\begin{tikzpicture}[x=.8cm,y=.8cm]
\node (g1) at (0,0) [gauge,label=below:\large{1}] {};
\node (g2) at (1,0) [gauge,label=below:\large{2}] {};
\node (g3) at (2,0) [gaugecyan,label=below:\large{3}] {};
\node (g4) at (3,0) [gauge,label=below:\large{2}] {};
\node (g5) at (4,0) [gauge,label=below:\large{1}] {};
\node (g6) at (1,1) [gaugeblue,label=left:\large{1}] {};
\node (g7) at (3,1) [gaugegreen,label=right:\large{1}] {};
\draw (g1)--(g2)--(g3)--(g4)--(g5);
\draw (g6)--(g3)--(g7);
\end{tikzpicture}
\,.
\ee

Again, we can look at the Hasse diagram. We can show that the following Minkowski sum decomposition induces a transition in the Hasse diagram
\be
\begin{tikzpicture}[x=.5cm,y=.5cm]

\draw[step=.5cm,gray,very thin] (1,0) grid (3,4);
\draw[step=.5cm,gray,very thin] (5,0) grid (7,3);
\draw[ligne,black] (1,0)--(1,2);
\draw[ligne,black] (1,3)--(1,4)--(2,4)--(3,3);
\draw[ligne,black] (3,2)--(3,0)--(2,0)--(1,0); 
\draw[ligne,black] (1,3)--(1,2);
\draw[ligne,black] (3,2)--(3,3);
\draw[ligne,black] (1,0)--(2,1);
\draw[ligne,black] (3,1)--(2,1);
\draw[ligne,black] (2,4)--(2,1);
\node[bd] at (1,0) {}; 
\node[bd] at (1,1) {}; 
\node[bd] at (1,2) {}; 
\node[bd] at (1,3) {}; 
\node[bd] at (1,4) {}; 
\node[bd] at (2,4) {}; 
\node[bd] at (3,3) {}; 
\node[bd] at (3,2) {};
\node[bd] at (3,1) {}; 
\node[bd] at (3,0) {}; 
\node[wd] at (2,0) {};

\node[] at (2,-2) {$P$};

\node[] at (4,1) {$<$};
\node[] at (4,-2) {$<$};

\draw[ligne,black] (5,0)--(5,3)--(6,3)--(7,2)--(7,0)--(6,0)--(5,0); 
\draw[ligne,black] (5,0)--(6,1);
\draw[ligne,black] (7,1)--(6,1);
\draw[ligne,black] (6,3)--(6,1);
\node[bd] at (5,0) {}; 
\node[bd] at (5,1) {}; 
\node[bd] at (5,2) {}; 
\node[bd] at (5,3) {}; 
\node[bd] at (6,3) {}; 
\node[bd] at (7,2) {};
\node[bd] at (7,1) {}; 
\node[bd] at (7,0) {}; 
\node[wd] at (6,0) {}; 

\node[] at (6,-2) {$\Delta_{\phantom{\nu}}$};

\node[] at (8,1) {$\oplus$};
\node[] at (8,-2) {$\oplus$};

\draw[ligne,black] (9,0)--(9,1); 
\node[bd] at (9,0) {}; 
\node[bd] at (9,1) {};

\node[] at (9,-2) {$S$};

\end{tikzpicture}
\,, \qquad \qquad
\begin{tikzpicture}[x=.5cm,y=.5cm]

\draw[step=.5cm,gray,very thin] (1,0) grid (3,4);
\draw[step=.5cm,gray,very thin] (5,0) grid (7,3);
\draw[ligne,blue] (1,0)--(1,1);
\draw[ligne,blue] (1,2)--(1,4)--(2,4)--(3,3)--(3,2);
\draw[ligne,blue] (3,1)--(3,0)--(2,0)--(1,0); 
\draw[ligne,black] (1,1)--(1,2);
\draw[ligne,black] (3,1)--(3,2);
\draw[ligne,blue] (1,0)--(2,1);
\draw[ligne,blue] (3,1)--(2,1);
\draw[ligne,blue] (2,4)--(2,2);
\draw[ligne,blue] (1,4)--(2,2);
\draw[ligne,black] (2,1)--(2,2);

\node[bd] at (1,0) {}; 
\node[bd] at (1,1) {}; 
\node[bd] at (1,2) {}; 
\node[bd] at (1,3) {}; 
\node[bd] at (1,4) {}; 
\node[bd] at (2,4) {}; 
\node[bd] at (3,3) {}; 
\node[bd] at (3,2) {};
\node[bd] at (3,1) {}; 
\node[bd] at (3,0) {}; 
\node[wd] at (2,0) {};

\node[] at (2,-2) {$Q$};

\node[] at (4,1) {$<$};
\node[] at (4,-2) {$<$};

\draw[ligne,blue] (5,0)--(5,3)--(6,3)--(7,2)--(7,0)--(6,0)--(5,0); 
\draw[ligne,blue] (5,0)--(6,1);
\draw[ligne,blue] (7,1)--(6,1);
\draw[ligne,blue] (6,3)--(6,1);
\draw[ligne,blue] (5,3)--(6,1);
\node[bd] at (5,0) {}; 
\node[bd] at (5,1) {}; 
\node[bd] at (5,2) {}; 
\node[bd] at (5,3) {}; 
\node[bd] at (6,3) {}; 
\node[bd] at (7,2) {};
\node[bd] at (7,1) {}; 
\node[bd] at (7,0) {}; 
\node[wd] at (6,0) {}; 

\node[] at (6,-2) {$\Delta_\nu$};

\node[] at (8,1) {$\oplus$};
\node[] at (8,-2) {$\oplus$};

\draw[ligne,black] (9,0)--(9,1); 
\node[bd] at (9,0) {}; 
\node[bd] at (9,1) {};

\node[] at (9,-2) {$S$};

\end{tikzpicture} \,,
\ee
where the partitions of $Q$ and $P$ agree.
We can again compute the magnetic quivers of $\Delta_\nu$ and $Q$ and find the affine Dynkin diagrams of $\mathfrak{d}_4$ and $\mathfrak{a}_5$ respectively. Thus, the Higgs branch of the weakly coupled $SU(3)_1 + 6\bm{F}$ is
\be
\begin{tikzpicture}
\node (1) [hasse] at (0,0) {};
\node (2) [hasse] at (0,-1) {};
\node (3) [hasse] at (0,-2) {};
\draw (1) edge [] node[label=left:$\mathfrak{d}_4$] {} (2);
\draw (2) edge [] node[label=left:$\mathfrak{a}_5$] {} (3);
\end{tikzpicture} \,.
\ee

\subsection{$T_4$}

The trinions $T_N$ with extremal vertices $(0,0), \ (N, 0), \ (0,N)$ were studied in the context of 5d SCFTs in \cite{Benini:2009gi}.
Their (and their descendants') magnetic quivers and Hasse diagrams were derived in \cite{Eckhard:2020jyr}. We will derive these from the colored polygon method for $T_4$. 

The toric polygon for $T_4$ is
\be
\begin{tikzpicture}[x=.5cm,y=.5cm]
\node[] at (-1.5,2) {$P=$};
\draw[step=.5cm,gray,very thin] (0,0) grid (4,4);
\draw[ligne] (0,0)--(0,4)--(4,0)--(0,0); 
\node[bd] at (0,0) {}; 
\node[bd] at (0,1) {}; 
\node[bd] at (0,2) {}; 
\node[bd] at (0,3) {}; 
\node[bd] at (0,4) {}; 
\node[bd] at (1,3) {};
\node[bd] at (2,2) {};
\node[bd] at (3,1) {};
\node[bd] at (4,0) {}; 
\node[bd] at (3,0) {}; 
\node[bd] at (2,0) {}; 
\node[bd] at (1,0) {}; 
\end{tikzpicture} 
\ee
The data characterizing $P$ is
\be \label{T4Data}
\ba
&L_\alpha=\left((0,-1),(1,0),(-1,1) \right)\\
&\lambda_\alpha = (4,4,4)\,, \qquad \mu_{\alpha,x} = (\{1^4\},\{1^4\},\{1^4\})\,.
\ea
\ee
We first compute the magnetic quiver. The only coloring that is consistent is by a single color $n_c=1$:
\be
\begin{tikzpicture}[x=.5cm,y=.5cm]
\draw[step=.5cm,gray,very thin] (0,0) grid (4,4);
\draw[ligne, blue] (0,0)--(0,4)--(4,0)--(0,0); 
\node[bd] at (0,0) {}; 
\node[bd] at (0,1) {}; 
\node[bd] at (0,2) {}; 
\node[bd] at (0,3) {}; 
\node[bd] at (0,4) {}; 
\node[bd] at (1,3) {};
\node[bd] at (2,2) {};
\node[bd] at (3,1) {};
\node[bd] at (4,0) {}; 
\node[bd] at (3,0) {}; 
\node[bd] at (2,0) {}; 
\node[bd] at (1,0) {}; 
\end{tikzpicture}  \,.
\ee
The coloring is specified by the partition data 
\be
\lambda^b_\alpha = (4, 4, 4)\,, \qquad \mu^b_\alpha = (\{4\},\{ 4\}, \{4\})\,, \qquad m^b =4\,.
\ee
We find that there is one vertex in the magnetic quiver from the single color with label 4. 
All three edges give rise to additional (non-color) nodes in the magnetic quiver. Their labels and intersections with the blue node are
\be 
	m_{\alpha,x} = (3,2,1)\,, \qquad k^b_{\alpha,1} = 1\,,
\ee
with all other vanishing.
The complete magnetic quiver is
\be
\begin{tikzpicture}[x=.8cm,y=.8cm]
\node[] at (-1.5, 1) {$\text{MQ}(T_4)= \qquad$}; 

\node (g1) at (0,0) [gauge,label=below:\large{1}] {};
\node (g2) at (1,0) [gauge,label=below:\large{2}] {};
\node (g3) at (2,0) [gauge,label=below:\large{3}] {};
\node (g4) at (3,0) [gaugeblue,label=below:\large{4}] {};
\node (g5) at (4,0) [gauge,label=below:\large{3}] {};
\node (g6) at (5,0) [gauge,label=below:\large{2}] {};
\node (g7) at (6,0) [gauge,label=below:\large{1}] {};
\node (g8) at (3,1) [gauge,label=right:\large{3}] {};
\node (g9) at (3,2) [gauge,label=right:\large{2}] {};
\node (g10) at (3,3) [gauge,label=right:\large{1}] {};
\draw (g1)--(g2)--(g3)--(g4)--(g5)--(g6)--(g7);
\draw (g4)--(g8)--(g9)--(g10);
\end{tikzpicture}
\,.
\ee

To determine the Hasse diagram, we determine all the subdiagrams that correspond to rank 1 theories: There is one $\mathfrak{e}_6$ theory and three ways to embed the $\mathfrak{e}_7$:
\be
\begin{tikzpicture}[x=.5cm,y=.5cm]
\node[] at (-1.5,1) {$\mathfrak{e}_6: \qquad $};
\draw[step=.5cm,gray,very thin] (0,0) grid (3,3);
\draw[ligne] (0,0)--(0,3)--(3,0)--(0,0); 
\node[bd] at (0,0) {}; 
\node[bd] at (0,1) {}; 
\node[bd] at (0,2) {}; 
\node[bd] at (0,3) {}; 
\node[bd] at (1,2) {};
\node[bd] at (2,1) {};
\node[bd] at (3,0) {}; 
\node[bd] at (2,0) {}; 
\node[bd] at (1,0) {}; 
\end{tikzpicture} 
\qquad 
\begin{tikzpicture}[x=.5cm,y=.5cm]
\node[] at (-1.5,1) {$\mathfrak{e}_7: \qquad $};
\draw[step=.5cm,gray,very thin] (0,0) grid (4,4);
\draw[ligne] (0,0)--(0,4)--(4,0)--(0,0); 
\node[bd] at (0,0) {}; 
\node[bd] at (0,1) {}; 
\node[bd] at (0,2) {}; 
\node[bd] at (0,3) {}; 
\node[bd] at (0,4) {}; 
\node[wd] at (1,3) {};
\node[bd] at (2,2) {};
\node[wd] at (3,1) {};
\node[bd] at (4,0) {}; 
\node[bd] at (3,0) {}; 
\node[bd] at (2,0) {}; 
\node[bd] at (1,0) {}; 
\end{tikzpicture} 
\qquad 
\begin{tikzpicture}[x=.5cm,y=.5cm]
\draw[step=.5cm,gray,very thin] (0,0) grid (4,4);
\draw[ligne] (0,0)--(0,4)--(4,0)--(0,0);
\node[bd] at (0,0) {}; 
\node[bd] at (0,1) {}; 
\node[bd] at (0,2) {}; 
\node[bd] at (0,3) {}; 
\node[bd] at (0,4) {}; 
\node[bd] at (1,3) {};
\node[bd] at (2,2) {};
\node[bd] at (3,1) {};
\node[bd] at (4,0) {}; 
\node[wd] at (3,0) {}; 
\node[bd] at (2,0) {}; 
\node[wd] at (1,0) {}; 
\end{tikzpicture} 
\qquad 
\begin{tikzpicture}[x=.5cm,y=.5cm]
\draw[step=.5cm,gray,very thin] (0,0) grid (4,4);
\draw[ligne] (0,0)--(0,4)--(4,0)--(0,0);
\node[bd] at (0,0) {}; 
\node[wd] at (0,1) {}; 
\node[bd] at (0,2) {}; 
\node[wd] at (0,3) {}; 
\node[bd] at (0,4) {}; 
\node[bd] at (1,3) {};
\node[bd] at (2,2) {};
\node[bd] at (3,1) {};
\node[bd] at (4,0) {}; 
\node[bd] at (3,0) {}; 
\node[bd] at (2,0) {}; 
\node[bd] at (1,0) {}; 
\end{tikzpicture} 
\ee
These are the first four transverse slices of the Hasse diagram. The magnetic quiver of the above diagrams can readily be checked to give the affine Dynkin diagrams of the corresponding groups.
Clearly, the refined Minkowski sum $P\leq \Delta \oplus S$ can be realised in two ways
\be 
\begin{tikzpicture}[x=.5cm,y=.5cm]

\node[] at (5,1) {$P<$};

\draw[step=.5cm,gray,very thin] (7,0) grid (10,3);
\draw[ligne] (7,0)--(7,3)--(10,0)--(7,0); 
\node[bd] at (7,0) {}; 
\node[bd] at (7,1) {}; 
\node[bd] at (7,2) {}; 
\node[bd] at (7,3) {}; 
\node[bd] at (8,2) {};
\node[bd] at (9,1) {};
\node[bd] at (10,0) {}; 
\node[bd] at (9,0) {}; 
\node[bd] at (8,0) {}; 

\node[] at (11,1) {$\oplus$};

\node[bd] at (12,0) {};
\node[bd] at (13,0) {};
\node[bd] at (12,1) {};
\draw[ligne] (12,0)--(13,0)--(12,1)--(12,0); 

\node[] at (16,1) {$P<$};

\draw[step=.5cm,gray,very thin] (18,0) grid (22,4);
\draw[ligne] (18,0)--(18,4)--(22,0)--(18,0); 
\node[bd] at (18,0) {}; 
\node[bd] at (18,1) {}; 
\node[bd] at (18,2) {}; 
\node[bd] at (18,3) {}; 
\node[bd] at (18,4) {}; 
\node[wd] at (19,3) {};
\node[bd] at (20,2) {};
\node[wd] at (21,1) {};
\node[bd] at (22,0) {}; 
\node[bd] at (21,0) {}; 
\node[bd] at (20,0) {}; 
\node[bd] at (19,0) {}; 
\end{tikzpicture}
\ee
and likewise for the other two representations of $\mathfrak{e}_7$. Note that in the second case $S$ is empty. There is an extension of each of the $\Delta$ with internal lines such that the s-rule is minimally obeyed on each external edge ($\Delta_\nu$):
\be
\begin{tikzpicture}[x=.5cm,y=.5cm]
\node[] at (-1.5,1) {$\mathfrak{e}_6: \qquad $};
\draw[step=.5cm,gray,very thin] (0,0) grid (3,3);
\draw[ligne] (0,0)--(0,3)--(3,0)--(0,0); 
\draw[ligne] (0,0)--(1,1)--(3,0); 
\draw[ligne] (1,1)--(0,3);  
\node[bd] at (0,0) {}; 
\node[bd] at (0,1) {}; 
\node[bd] at (0,2) {}; 
\node[bd] at (0,3) {}; 
\node[bd] at (1,2) {};
\node[bd] at (2,1) {};
\node[bd] at (3,0) {}; 
\node[bd] at (2,0) {}; 
\node[bd] at (1,0) {}; 
\end{tikzpicture} 
\qquad 
\begin{tikzpicture}[x=.5cm,y=.5cm]
\node[] at (-1.5,1) {$\mathfrak{e}_7: \qquad $};
\draw[step=.5cm,gray,very thin] (0,0) grid (4,4);
\draw[ligne] (0,0)--(0,4)--(4,0)--(0,0); 
\draw[ligne] (0,0)--(1,1)--(4,0);
\draw[ligne] (1,1)--(0,4); 
\node[bd] at (0,0) {}; 
\node[bd] at (0,1) {}; 
\node[bd] at (0,2) {}; 
\node[bd] at (0,3) {}; 
\node[bd] at (0,4) {}; 
\node[wd] at (1,3) {};
\node[bd] at (2,2) {};
\node[wd] at (3,1) {};
\node[bd] at (4,0) {}; 
\node[bd] at (3,0) {}; 
\node[bd] at (2,0) {}; 
\node[bd] at (1,0) {}; 
\end{tikzpicture} 
\qquad 
\begin{tikzpicture}[x=.5cm,y=.5cm]
\draw[step=.5cm,gray,very thin] (0,0) grid (4,4);
\draw[ligne] (0,0)--(0,4)--(4,0)--(0,0); 
\draw[ligne] (0,0)--(1,2)--(4,0);
\draw[ligne] (1,2)--(0,4); 
\node[bd] at (0,0) {}; 
\node[bd] at (0,1) {}; 
\node[bd] at (0,2) {}; 
\node[bd] at (0,3) {}; 
\node[bd] at (0,4) {}; 
\node[bd] at (1,3) {};
\node[bd] at (2,2) {};
\node[bd] at (3,1) {};
\node[bd] at (4,0) {}; 
\node[wd] at (3,0) {}; 
\node[bd] at (2,0) {}; 
\node[wd] at (1,0) {}; 
\end{tikzpicture} 
\qquad 
\begin{tikzpicture}[x=.5cm,y=.5cm]
\draw[step=.5cm,gray,very thin] (0,0) grid (4,4);
\draw[ligne] (0,0)--(0,4)--(4,0)--(0,0); 
\draw[ligne] (0,0)--(2,1)--(4,0);
\draw[ligne] (2,1)--(0,4); 
\node[bd] at (0,0) {}; 
\node[wd] at (0,1) {}; 
\node[bd] at (0,2) {}; 
\node[wd] at (0,3) {}; 
\node[bd] at (0,4) {}; 
\node[bd] at (1,3) {};
\node[bd] at (2,2) {};
\node[bd] at (3,1) {};
\node[bd] at (4,0) {}; 
\node[bd] at (3,0) {}; 
\node[bd] at (2,0) {}; 
\node[bd] at (1,0) {}; 
\end{tikzpicture} 
\ee
The $\mathfrak{e}_6$ induces a transition in the Hasse diagram with $Q$ given by
\be
 \begin{tikzpicture}[x=.5cm,y=.5cm]
\node[] at (-2,2) {$T_4-\mathfrak{e}_6= $};
\draw[step=.5cm,gray,very thin] (0,0) grid (4,4);
\draw[ligne, black] (0,0)--(0,4)--(4,0)--(0,0); 
\draw[ligne, black] (0,0)--(1,1);
\draw[ligne, black] (4,0)--(2,1);
\draw[ligne, black] (0,4)--(1,2);
\node[bd] at (0,0) {}; 
\node[bd] at (0,1) {}; 
\node[bd] at (0,2) {}; 
\node[bd] at (0,3) {}; 
\node[bd] at (0,4) {}; 
\node[bd] at (1,3) {};
\node[bd] at (2,2) {};
\node[bd] at (3,1) {};
\node[bd] at (4,0) {}; 
\node[bd] at (3,0) {}; 
\node[bd] at (2,0) {}; 
\node[bd] at (1,0) {}; 
\end{tikzpicture} 
\ee
To compute the magnetic quiver for this diagram we find the unique color sub-division 
\be
\begin{tikzpicture}[x=.5cm,y=.5cm]
\draw[step=.5cm,gray,very thin] (0,0) grid (4,4);
\draw[ligne, blue] (0,0)--(0,1);
\draw[ligne, blue] (0,2)--(0,4);
\draw[ligne, blue] (0,4)--(1,3);
\draw[ligne, blue] (2,2)--(4,0)--(2,0);
\draw[ligne, blue] (0,0)--(1,0);
\draw[ligne, blue] (0,0)--(1,1);
\draw[ligne, blue] (4,0)--(2,1);
\draw[ligne, blue] (0,4)--(1,2);
\draw[densely dotted,thick, blue] (1,0)--(1,1)--(0,1);
\draw[densely dotted,thick, blue] (2,0)--(2,2);
\draw[densely dotted,thick, blue] (0,2)--(1,2)--(1,3);
\draw[ligne, cyan] (0,1)--(0,2);
\draw[ligne, cyan] (1,0)--(2,0);
\draw[ligne, cyan] (1,3)--(2,2);
\draw[densely dotted,thick, cyan] (1,1)--(1,2)--(2,1)--(1,1);
\node[bd] at (0,0) {}; 
\node[bd] at (0,1) {}; 
\node[bd] at (0,2) {}; 
\node[bd] at (0,3) {}; 
\node[bd] at (0,4) {}; 
\node[bd] at (1,3) {};
\node[bd] at (2,2) {};
\node[bd] at (3,1) {};
\node[bd] at (4,0) {}; 
\node[bd] at (3,0) {}; 
\node[bd] at (2,0) {}; 
\node[bd] at (1,0) {}; 
\end{tikzpicture} 
\ee
The data are  
\be 
\ba
\lambda^b_\alpha &= (3,3,3)\cr
\lambda^c_\alpha &= (1,1,1)
\ea \qquad 
\ba
\mu^b_\alpha & = (\{1^3\},\{1^3\},\{1^3\}) \cr 
\mu^c_\alpha &= (\{1\},\{1\},\{1\})
\ea\qquad 
\ba
m^b &=1\cr 
m^c &=1\cr 
\ea
\ee
from which we deduce
\be
\ba
A^{cb} &=3\cr
m_{\alpha,x} &=(1,1,1)
\ea \qquad
\ba
k^{cb} &=0\cr
k^b_{\alpha,1} &=1
\ea \qquad
\ba 
\cr
k^c_{\alpha,3} &=1
\ea \qquad
\ba 
\cr
\forall \alpha
\ea
\ee
with all others vanishing.
Each of the three edges therefore contributes a chain three of $U(1)$ nodes. 
The magnetic quiver is thus 
\be
\begin{tikzpicture}[x=.8cm,y=.8cm]
\node[] at (-1.5, 1) {$\text{MQ}(T_4-\mathfrak{e}_6)= \qquad$}; 

\node (g1) at (0,0) [gauge,label=below:\large{1}] {};
\node (g2) at (1,0) [gauge,label=below:\large{1}] {};
\node (g3) at (2,0) [gauge,label=below:\large{1}] {};
\node (g4) at (3,0) [gaugeblue,label=below:\large{1}] {};
\node (g5) at (4,0) [gauge,label=below:\large{1}] {};
\node (g6) at (5,0) [gauge,label=below:\large{1}] {};
\node (g7) at (6,0) [gauge,label=below:\large{1}] {};
\node (g8) at (3,1) [gauge,label=right:\large{1}] {};
\node (g9) at (3,2) [gauge,label=right:\large{1}] {};
\node (g10) at (3,3) [gauge,label=right:\large{1}] {};
\node (g11) at (3,4) [gaugecyan,label=right:\large{1}] {};
\draw (g1)--(g2)--(g3)--(g4)--(g5)--(g6)--(g7);
\draw (g4)--(g8)--(g9)--(g10)--(g11);
\draw (g1)--(g11)--(g7);
\end{tikzpicture}
\,.
\ee

Next we note that there are three embeddings of $\mathfrak{a}_7$ singularities into the partially resolved diagram $T_4-\mathfrak{e}_6$:
\be
\begin{tikzpicture}[x=.5cm,y=.5cm]
\node[] at (-1.5,2) {$\mathfrak{a}_7: \qquad  $};
\draw[step=.5cm,gray,very thin] (0,0) grid (4,4);
\draw[ligne] (0,0)--(0,4)--(4,0)--(0,0); 
\draw[ligne] (0,0)--(1,1);
\draw[ligne] (0,4)--(1,2);
\draw[ligne] (2,1)--(4,0);
\node[bd] at (0,0) {}; 
\node[bd] at (0,1) {}; 
\node[bd] at (0,2) {}; 
\node[bd] at (0,3) {}; 
\node[bd] at (0,4) {}; 
\node[bd] at (1,3) {};
\node[wd] at (2,2) {};
\node[bd] at (3,1) {};
\node[bd] at (4,0) {}; 
\node[bd] at (3,0) {}; 
\node[bd] at (2,0) {}; 
\node[bd] at (1,0) {}; 
\end{tikzpicture} 
\qquad 
\begin{tikzpicture}[x=.5cm,y=.5cm]
\draw[step=.5cm,gray,very thin] (0,0) grid (4,4);
\draw[ligne] (0,0)--(0,4)--(4,0)--(0,0); 
\draw[ligne] (1,2)--(0,4);
\draw[ligne] (0,0)--(1,1);
\draw[ligne] (2,1)--(4,0);
\node[bd] at (0,0) {}; 
\node[bd] at (0,1) {}; 
\node[bd] at (0,2) {}; 
\node[bd] at (0,3) {}; 
\node[bd] at (0,4) {}; 
\node[bd] at (1,3) {};
\node[bd] at (2,2) {};
\node[bd] at (3,1) {};
\node[bd] at (4,0) {}; 
\node[bd] at (3,0) {}; 
\node[wd] at (2,0) {}; 
\node[bd] at (1,0) {}; 
\end{tikzpicture} 
\qquad 
\begin{tikzpicture}[x=.5cm,y=.5cm]
\draw[step=.5cm,gray,very thin] (0,0) grid (4,4);
\draw[ligne] (0,0)--(0,4)--(4,0)--(0,0); 
\draw[ligne] (2,1)--(4,0); 
\draw[ligne] (0,4)--(1,2);
\draw[ligne] (1,1)--(0,0);
\node[bd] at (0,0) {}; 
\node[bd] at (0,1) {}; 
\node[wd] at (0,2) {}; 
\node[bd] at (0,3) {}; 
\node[bd] at (0,4) {}; 
\node[bd] at (1,3) {};
\node[bd] at (2,2) {};
\node[bd] at (3,1) {};
\node[bd] at (4,0) {}; 
\node[bd] at (3,0) {}; 
\node[bd] at (2,0) {}; 
\node[bd] at (1,0) {}; 
\end{tikzpicture} 
\ee
It can again be checked that the magnetic quiver of all these three diagrams is an affine Dynkin diagram for $\mathfrak{a}_7$. As before, there exists an extension of these diagrams ($\Delta_\nu$) by internal lines such that the s-rule is minimally obeyed on each external edge. The embedding of these extended diagrams in $T_4-\mathfrak{e}_6$ is
\be\label{T4e6a7} 
 \begin{tikzpicture}[x=.5cm,y=.5cm]
 \node[] at (-2.5,2) {$T_4 - {\mathfrak{e}_6}- \mathfrak{a}_7
=\qquad  $};
\draw[step=.5cm,gray,very thin] (0,0) grid (4,4);
\draw[ligne] (0,0)--(0,4)--(4,0)--(0,0); 
\draw[ligne] (0,0)--(1,1)--(4,0);
\draw[ligne] (1,1)--(0,4); 
\draw[ligne] (0,4)--(1,2)--(2,1)--(4,0);
\node[bd] at (0,0) {}; 
\node[bd] at (0,1) {}; 
\node[bd] at (0,2) {}; 
\node[bd] at (0,3) {}; 
\node[bd] at (0,4) {}; 
\node[bd] at (1,3) {};
\node[bd] at (2,2) {};
\node[bd] at (3,1) {};
\node[bd] at (4,0) {}; 
\node[bd] at (3,0) {}; 
\node[bd] at (2,0) {}; 
\node[bd] at (1,0) {}; 
\end{tikzpicture} 
\qquad 
\begin{tikzpicture}[x=.5cm,y=.5cm]
\draw[step=.5cm,gray,very thin] (0,0) grid (4,4);
\draw[ligne] (0,0)--(0,4)--(4,0)--(0,0); 
\draw[ligne] (0,0)--(1,2)--(4,0);
\draw[ligne] (1,2)--(0,4); 
\draw[ligne] (0,0)--(1,1)--(2,1)--(4,0);
\node[bd] at (0,0) {}; 
\node[bd] at (0,1) {}; 
\node[bd] at (0,2) {}; 
\node[bd] at (0,3) {}; 
\node[bd] at (0,4) {}; 
\node[bd] at (1,3) {};
\node[bd] at (2,2) {};
\node[bd] at (3,1) {};
\node[bd] at (4,0) {}; 
\node[bd] at (3,0) {}; 
\node[bd] at (2,0) {}; 
\node[bd] at (1,0) {}; 
\end{tikzpicture} 
\qquad 
\begin{tikzpicture}[x=.5cm,y=.5cm]
\draw[step=.5cm,gray,very thin] (0,0) grid (4,4);
\draw[ligne] (0,0)--(0,4)--(4,0)--(0,0); 
\draw[ligne] (0,0)--(2,1)--(4,0);
\draw[ligne] (2,1)--(0,4); 
\draw[ligne] (0,0)--(1,1)--(1,2)--(0,4);
\node[bd] at (0,0) {}; 
\node[bd] at (0,1) {}; 
\node[bd] at (0,2) {}; 
\node[bd] at (0,3) {}; 
\node[bd] at (0,4) {}; 
\node[bd] at (1,3) {};
\node[bd] at (2,2) {};
\node[bd] at (3,1) {};
\node[bd] at (4,0) {}; 
\node[bd] at (3,0) {}; 
\node[bd] at (2,0) {}; 
\node[bd] at (1,0) {}; 
\end{tikzpicture} 
\ee
The magnetic quiver of each of these diagrams is 
\be
MQ (T_4 - \mathfrak{e}_6 - \mathfrak{a}_7) = \mathfrak{a}_3 \,,
\ee
which is the final symplectic leave in the $\mathfrak{e}_6$ branch of the Hasse diagram.

Likewise along the $\mathfrak{e}_7$ branches, the first step (the subtraction of the $\mathfrak{e}_7$ slice) is obtained by the following partial triangulation of the $T_4$ diagram 
\be
\begin{tikzpicture}[x=.5cm,y=.5cm]
\draw[step=.5cm,gray,very thin] (0,0) grid (4,4);
 \node[] at (-2.5,2) {$T_4 - {\mathfrak{e}_7}
=\qquad  $};
\draw[ligne] (0,0)--(0,4)--(4,0)--(0,0); 
\draw[ligne] (0,0)--(1,1)--(4,0);
\draw[ligne] (1,1)--(0,4); 
\node[bd] at (0,0) {}; 
\node[bd] at (0,1) {}; 
\node[bd] at (0,2) {}; 
\node[bd] at (0,3) {}; 
\node[bd] at (0,4) {}; 
\node[bd] at (1,3) {};
\node[bd] at (2,2) {};
\node[bd] at (3,1) {};
\node[bd] at (4,0) {}; 
\node[bd] at (3,0) {}; 
\node[bd] at (2,0) {}; 
\node[bd] at (1,0) {}; 
\end{tikzpicture} 
\qquad 
\begin{tikzpicture}[x=.5cm,y=.5cm]
\draw[step=.5cm,gray,very thin] (0,0) grid (4,4);
\draw[ligne] (0,0)--(0,4)--(4,0)--(0,0); 
\draw[ligne] (0,0)--(1,2)--(4,0);
\draw[ligne] (1,2)--(0,4); 
\node[bd] at (0,0) {}; 
\node[bd] at (0,1) {}; 
\node[bd] at (0,2) {}; 
\node[bd] at (0,3) {}; 
\node[bd] at (0,4) {}; 
\node[bd] at (1,3) {};
\node[bd] at (2,2) {};
\node[bd] at (3,1) {};
\node[bd] at (4,0) {}; 
\node[bd] at (3,0) {}; 
\node[bd] at (2,0) {}; 
\node[bd] at (1,0) {}; 
\end{tikzpicture} 
\qquad 
\begin{tikzpicture}[x=.5cm,y=.5cm]
\draw[step=.5cm,gray,very thin] (0,0) grid (4,4);
\draw[ligne] (0,0)--(0,4)--(4,0)--(0,0); 
\draw[ligne] (0,0)--(2,1)--(4,0);
\draw[ligne] (2,1)--(0,4); 
\node[bd] at (0,0) {}; 
\node[bd] at (0,1) {}; 
\node[bd] at (0,2) {}; 
\node[bd] at (0,3) {}; 
\node[bd] at (0,4) {}; 
\node[bd] at (1,3) {};
\node[bd] at (2,2) {};
\node[bd] at (3,1) {};
\node[bd] at (4,0) {}; 
\node[bd] at (3,0) {}; 
\node[bd] at (2,0) {}; 
\node[bd] at (1,0) {}; 
\end{tikzpicture} 
\ee
The magnetic quiver of this theory is readily obtained to be 
\be
\begin{tikzpicture}[x=.8cm,y=.8cm]
\node[] at (-2, .5) {$\text{MQ}(T_4-\mathfrak{e}_7)= \qquad$}; 

\node (g1) at (0,0) [gauge,label=below:\large{1}] {};
\node (g2) at (1,0) [gauge,label=below:\large{2}] {};
\node (g3) at (2,0) [gauge,label=below:\large{1}] {};
\node (g4) at (1,1) [gauge,label=right:\large{1}] {};
\draw (g1)--(g2)--(g3);
\draw[double distance = 3pt] (g2)--(g4);
\end{tikzpicture}
\,.
\ee
Finally, we note that there are $\mathfrak{a}_1$ singularities realizable as 
\be
\begin{tikzpicture}[x=.5cm,y=.5cm]
\draw[step=.5cm,gray,very thin] (0,0) grid (4,4);
 \node[] at (-1.5,2) {$\mathfrak{a}_1=  \qquad  $};
\draw[ligne] (0,0)--(0,4)--(4,0)--(0,0); 
\draw[ligne] (0,0)--(1,1)--(4,0);
\draw[ligne] (1,1)--(0,4); 
\node[bd] at (0,0) {}; 
\node[bd] at (0,1) {}; 
\node[bd] at (0,2) {}; 
\node[bd] at (0,3) {}; 
\node[bd] at (0,4) {}; 
\node[bd] at (1,3) {};
\node[wd] at (2,2) {};
\node[bd] at (3,1) {};
\node[bd] at (4,0) {}; 
\node[bd] at (3,0) {}; 
\node[bd] at (2,0) {}; 
\node[bd] at (1,0) {}; 
\end{tikzpicture} 
\qquad 
\begin{tikzpicture}[x=.5cm,y=.5cm]
\draw[step=.5cm,gray,very thin] (0,0) grid (4,4);
\draw[ligne] (0,0)--(0,4)--(4,0)--(0,0); 
\draw[ligne] (0,0)--(1,2)--(4,0);
\draw[ligne] (1,2)--(0,4); 
\node[bd] at (0,0) {}; 
\node[bd] at (0,1) {}; 
\node[wd] at (0,2) {}; 
\node[bd] at (0,3) {}; 
\node[bd] at (0,4) {}; 
\node[bd] at (1,3) {};
\node[bd] at (2,2) {};
\node[bd] at (3,1) {};
\node[bd] at (4,0) {}; 
\node[bd] at (3,0) {}; 
\node[bd] at (2,0) {}; 
\node[bd] at (1,0) {}; 
\end{tikzpicture} 
\qquad 
\begin{tikzpicture}[x=.5cm,y=.5cm]
\draw[step=.5cm,gray,very thin] (0,0) grid (4,4);
\draw[ligne] (0,0)--(0,4)--(4,0)--(0,0); 
\draw[ligne] (0,0)--(2,1)--(4,0);
\draw[ligne] (2,1)--(0,4); 
\node[bd] at (0,0) {}; 
\node[bd] at (0,1) {}; 
\node[bd] at (0,2) {}; 
\node[bd] at (0,3) {}; 
\node[bd] at (0,4) {}; 
\node[bd] at (1,3) {};
\node[bd] at (2,2) {};
\node[bd] at (3,1) {};
\node[bd] at (4,0) {}; 
\node[bd] at (3,0) {}; 
\node[wd] at (2,0) {}; 
\node[bd] at (1,0) {}; 
\end{tikzpicture} 
\ee
These are precisely subtracted by partially triangulating further as in (\ref{T4e6a7}), which is where the branches meet. 
This completes the Hasse diagram of $T_4$, in agreement with \cite{Eckhard:2020jyr}:

\be
\begin{tikzpicture}
\node (1) [hasse] at (0,0) {};
\node (2) [hasse] at (-1,-2) {};
\node (3) [hasse] at (1,-2) {};
\node (4) [hasse] at (2,-2) {};
\node (5) [hasse] at (3,-2) {};
\node (6) [hasse] at (1,-4) {};
\node (7) [hasse] at (2,-4) {};
\node (8) [hasse] at (3,-4) {};
\node (9) [hasse] at (2,-5) {};
\draw (1) edge [] node[label=left:$\mathfrak{e}_6$] {} (2);
\draw (1) edge [] node[label=right:$\mathfrak{e}_7$, yshift=-.3cm] {} (3);
\draw (1) edge [] node[label=right:$\mathfrak{e}_7$, yshift=-.3cm,xshift=.1cm] {} (4);
\draw (1) edge [] node[label=right:$\mathfrak{e}_7$, yshift=-.3cm,xshift=.3cm] {} (5);
\draw (3) edge [] node[near start, label=right:$\mathfrak{a}_1$] {} (6);
\draw (4) edge [] node[near start, label=right:$\mathfrak{a}_1$] {} (7);
\draw (5) edge [] node[near start, label=right:$\mathfrak{a}_1$] {} (8);
\draw (2) edge [] node[label=left:$\mathfrak{a}_7$,yshift=-.4cm,xshift=.4cm] {} (6);
\draw (2) edge [] node[label=left:$\mathfrak{a}_7$,yshift=-.4cm,xshift=.7cm] {} (7);
\draw (2) edge [] node[label=left:$\mathfrak{a}_7$,yshift=-.4cm,xshift=.9cm] {} (8);
\draw (6) edge [] node[label=left:$\mathfrak{a}_3$] {} (9);
\draw (7) edge [] node[label=right:$\mathfrak{a}_3$,xshift=-.2cm] {} (9);
\draw (8) edge [] node[label=right:$\mathfrak{a}_3$] {} (9);
\end{tikzpicture}\,.
\ee


\subsection{$SU(4)_{3/2} + 1{\bf AS} + 7 {\bf F}$}

\label{sec:ExAS}

Let us now look at a more complicated example, the $SU(4)_{3/2} + 1{\bf AS} + 7 {\bf F}$. The GTP $P$ with a consistent coloring is given by
\be \label{Ex2Coloring}
\begin{tikzpicture}[x=.5cm,y=.5cm]
\draw[step=.5cm,gray,very thin] (0,0) grid (4,7);
\draw[ligne, blue](0,0)--(0,1)--(0,3);
\draw[ligne, blue](4,3)--(4,0);
\draw[ligne, cyan](0,3)--(1,4)--(3,6)--(4,7)--(4,3);
\draw[ligne, cyan](4,0)--(3,0)--(2,0)--(1,0)--(0,0);
\draw[densely dotted, thick, cyan](0,3)--(4,3);
\node[bd] at (0,0) {}; 
\node[wd] at (0,1) {}; 
\node[bd] at (0,2) {}; 
\node[bd] at (0,3) {}; 
\node[wd] at (1,4) {}; 
\node[bd] at (2,5) {}; 
\node[wd] at (3,6) {}; 
\node[bd] at (4,7) {}; 
\node[bd] at (4,6) {}; 
\node[bd] at (4,5) {}; 
\node[bd] at (4,4) {}; 
\node[bd] at (4,3) {}; 
\node[bd] at (4,2) {}; 
\node[bd] at (4,1) {}; 
\node[bd] at (4,0) {}; 
\node[wd] at (3,0) {}; 
\node[wd] at (2,0) {}; 
\node[wd] at (1,0) {}; 
\end{tikzpicture}
\ee
We can summarise the data by
\be
\ba
&L_\alpha=((-1,-1),(0,-1),(1,0),(0,1))\,, \\
&\lambda_\alpha=(4,3,4,7)\,, \qquad &&\mu_{\alpha,x}=\left(\{2^2\},\{2,1\},\{4\},\{1^7\}\right)\\
&\lambda^c_\alpha=(4,0,4,4)\,, \qquad &&\mu^c_{\alpha,x}=\left(\{4\},-,\{4\},\{4\}\right)\\
&\lambda^b_\alpha=(0,3,0,3)\,, \qquad &&\mu^b_{\alpha,x}=\left(-,\{3\},-,\{3\}\right)\,.
\ea
\ee
With $A^{cb}=12$ we can deduce for the color nodes
\be
m^c=4\,,\quad m^b=3\,, \qquad k^{rb}=0\,.
\ee
Furthermore, there are additional nodes on edges $E_1$, $E_2$ and $E_4$ with
\be
\ba
&m_{1,1} = 2\,, \qquad &&k^c_{1,1}=1\\
&m_{2,1} = 1\,, \qquad &&k^b_{2,1}=1\\
&m_{4,x} = (6,5,4,3,2,1)\,, \qquad &&k^c_{4,1}=1\,, \quad k^b_{4,1}=1\,,
\ea
\ee
with all others vanishing. The magnetic quiver is thus given by 
\be \label{Ex2MQ}
\begin{tikzpicture}[x=.8cm,y=.8cm]
\node (g1) at (0,0) [gauge,label=below:\large{1}] {};
\node (g2) at (1,0) [gauge,label=below:\large{2}] {};
\node (g3) at (2,0) [gauge,label=below:\large{3}] {};
\node (g4) at (3,0) [gauge,label=below:\large{4}] {};
\node (g5) at (4,0) [gauge,label=below:\large{5}] {};
\node (g6) at (5,0) [gauge,label=below:\large{6}] {};
\node (g7) at (6,0) [gaugecyan,label=below:\large{4}] {};
\node (g8) at (7,0) [gauge,label=below:\large{2}] {};
\node (g9) at (5,1) [gaugeblue,label=left:\large{3}] {};
\node (g10) at (6,1) [gauge,label=right:\large{1}] {};
\draw (g1)--(g2)--(g3)--(g4)--(g5)--(g6)--(g7)--(g8);
\draw (g6)--(g9)--(g10);
\end{tikzpicture}
\,.
\ee

We can now compute the Hasse diagram of this theory. It turns out that there are two distinct branches which we will discuss in turn. First, we can write
\be \label{Ex2HasseE6}
\begin{tikzpicture}[x=.5cm,y=.5cm]

\draw[step=.5cm,gray,very thin] (8,0) grid (12,7);
\draw[step=.5cm,gray,very thin] (14,0) grid (16,4);
\draw[step=.5cm,gray,very thin] (18,0) grid (20,3);

\draw[ligne, black](8,0)--(8,1)--(8,2);
\draw[ligne, black](12,2)--(12,0);
\draw[ligne, black](10,5)--(11,6)--(12,7)--(12,5);
\draw[ligne, black](12,0)--(11,0)--(10,0);
\draw[ligne, black](8,2)--(8,3)--(9,4)--(10,5);
\draw[ligne, black](12,5)--(12,2);
\draw[ligne, black](10,0)--(9,0)--(8,0);
\node[bd] at (8,0) {}; 
\node[wd] at (8,1) {}; 
\node[bd] at (8,2) {}; 
\node[bd] at (8,3) {}; 
\node[wd] at (9,4) {}; 
\node[bd] at (10,5) {}; 
\node[wd] at (11,6) {}; 
\node[bd] at (12,7) {}; 
\node[bd] at (12,6) {}; 
\node[bd] at (12,5) {}; 
\node[bd] at (12,4) {}; 
\node[bd] at (12,3) {}; 
\node[bd] at (12,2) {}; 
\node[bd] at (12,1) {}; 
\node[bd] at (12,0) {}; 
\node[wd] at (11,0) {}; 
\node[wd] at (10,0) {}; 
\node[wd] at (9,0) {}; 

\node[] at (10,-2) {$P$};

\node[] at (13,2) {$<$};
\node[] at (13,-2) {$<$};

\draw[ligne,black] (14,0)--(14,2)--(16,4)--(16,0)--(15,0)--(14,0); 
\node[bd] at (14,0) {}; 
\node[bd] at (14,1) {}; 
\node[bd] at (14,2) {}; 
\node[bd] at (15,3) {}; 
\node[bd] at (16,4) {}; 
\node[bd] at (16,3) {};
\node[bd] at (16,2) {}; 
\node[bd] at (16,1) {}; 
\node[bd] at (16,0) {};
\node[wd] at (15,0) {};
\node[bd] at (14,0) {};

\node[] at (15,-2) {$\Delta_{\phantom{\nu}}$};

\node[] at (17,2) {$\oplus$};
\node[] at (17,-2) {$\oplus$};

\draw[ligne,black] (18,0)--(18,1)--(20,3)--(20,0)--(18,0); 
\node[bd] at (18,0) {}; 
\node[bd] at (18,1) {};
\node[bd] at (19,2) {};
\node[bd] at (20,3) {};
\node[bd] at (20,2) {};
\node[bd] at (20,1) {};
\node[bd] at (20,0) {};
\node[bd] at (19,0) {};

\node[] at (19,-2) {$S$};
\end{tikzpicture} \,,
\qquad \qquad
\begin{tikzpicture}[x=.5cm,y=.5cm]

\draw[step=.5cm,gray,very thin] (14,0) grid (16,4);
\draw[step=.5cm,gray,very thin] (18,0) grid (20,3);
\draw[step=.5cm,gray,very thin] (8,0) grid (12,7);

\draw[ligne, blue](8,0)--(8,1)--(8,2);
\draw[ligne, blue](12,2)--(12,0);
\draw[ligne, blue](10,5)--(11,6)--(12,7)--(12,5);
\draw[ligne, blue](12,0)--(11,0)--(10,0);
\draw[ligne, blue](8,0)--(9,2);
\draw[ligne, blue](8,3)--(9,3);
\draw[ligne, blue](12,0)--(11,2);
\draw[ligne, blue](12,7)--(11,5);
\draw[ligne, black](8,2)--(8,3)--(9,4)--(10,5);
\draw[ligne, black](12,5)--(12,2);
\draw[ligne, black](10,0)--(9,0)--(8,0);
\node[bd] at (8,0) {}; 
\node[wd] at (8,1) {}; 
\node[bd] at (8,2) {}; 
\node[bd] at (8,3) {}; 
\node[wd] at (9,4) {}; 
\node[bd] at (10,5) {}; 
\node[wd] at (11,6) {}; 
\node[bd] at (12,7) {}; 
\node[bd] at (12,6) {}; 
\node[bd] at (12,5) {}; 
\node[bd] at (12,4) {}; 
\node[bd] at (12,3) {}; 
\node[bd] at (12,2) {}; 
\node[bd] at (12,1) {}; 
\node[bd] at (12,0) {}; 
\node[wd] at (11,0) {}; 
\node[wd] at (10,0) {}; 
\node[wd] at (9,0) {}; 

\node[] at (10,-2) {$Q$};

\node[] at (13,2) {$<$};
\node[] at (13,-2) {$<$};

\draw[ligne,blue] (14,0)--(14,2)--(16,4)--(16,0)--(15,0)--(14,0); 
\draw[ligne,blue] (14,0)--(15,2);
\draw[ligne,blue] (14,2)--(15,2);
\draw[ligne,blue] (16,4)--(15,2);
\draw[ligne,blue] (16,0)--(15,2);
\node[bd] at (14,0) {}; 
\node[bd] at (14,1) {}; 
\node[bd] at (14,2) {}; 
\node[bd] at (15,3) {}; 
\node[bd] at (16,4) {}; 
\node[bd] at (16,3) {};
\node[bd] at (16,2) {}; 
\node[bd] at (16,1) {}; 
\node[bd] at (16,0) {};
\node[wd] at (15,0) {};
\node[bd] at (14,0) {};

\node[] at (15,-2) {$\Delta_\nu$};

\node[] at (17,2) {$\oplus$};
\node[] at (17,-2) {$\oplus$};

\draw[ligne,black] (18,0)--(18,1)--(20,3)--(20,0)--(18,0); 
\node[bd] at (18,0) {}; 
\node[bd] at (18,1) {};
\node[bd] at (19,2) {};
\node[bd] at (20,3) {};
\node[bd] at (20,2) {};
\node[bd] at (20,1) {};
\node[bd] at (20,0) {};
\node[bd] at (19,0) {};

\node[] at (19,-2) {$S$};

\end{tikzpicture} \,,
\ee
where $\Delta$ corresponds to an $\mathfrak{e}_6$.

Now, let us compute the magnetic quiver of $Q$. We start with the coloring, now respecting the internal lines
\be
\begin{tikzpicture}[x=.5cm,y=.5cm]

\draw[step=.5cm,gray,very thin] (8,0) grid (12,7);

\draw[ligne, blue](8,0)--(8,1)--(8,2);
\draw[ligne, blue](12,2)--(12,0);
\draw[ligne, blue](10,5)--(11,6)--(12,7)--(12,5);
\draw[ligne, blue](12,0)--(11,0)--(10,0);
\draw[ligne, blue](8,0)--(9,2);
\draw[ligne, blue](8,3)--(9,3);
\draw[ligne, blue](12,0)--(11,2);
\draw[ligne, blue](12,7)--(11,5);
\draw[ligne, green](8,2)--(8,3);
\draw[ligne, cyan](8,3)--(9,4)--(10,5);
\draw[ligne, cyan](12,5)--(12,3);
\draw[ligne, green](12,3)--(12,2);
\draw[ligne, cyan](10,0)--(9,0)--(8,0);

\draw[densely dotted, thick, cyan](9,2)--(11,2);
\draw[densely dotted, thick, blue](10,0)--(11,2)--(12,2);
\draw[densely dotted, thick, blue](8,2)--(9,2);
\draw[densely dotted, thick, blue](10,5)--(12,5);
\draw[densely dotted, thick, cyan](9,3)--(11,5)--(11,3)--(9,3);
\draw[densely dotted, thick, green](9,2)--(9,3);
\draw[densely dotted, thick, green](11,2)--(11,3);
\draw[densely dotted, thick, blue](11,3)--(12,3);

\node[bd] at (8,0) {}; 
\node[wd] at (8,1) {}; 
\node[bd] at (8,2) {}; 
\node[bd] at (8,3) {}; 
\node[wd] at (9,4) {}; 
\node[bd] at (10,5) {}; 
\node[wd] at (11,6) {}; 
\node[bd] at (12,7) {}; 
\node[bd] at (12,6) {}; 
\node[bd] at (12,5) {}; 
\node[bd] at (12,4) {}; 
\node[bd] at (12,3) {}; 
\node[bd] at (12,2) {}; 
\node[bd] at (12,1) {}; 
\node[bd] at (12,0) {}; 
\node[wd] at (11,0) {}; 
\node[wd] at (10,0) {}; 
\node[wd] at (9,0) {}; 

\end{tikzpicture} \,,
\ee

Again, we first compute the data of the coloring to be
\be
\ba
&\lambda^c_\alpha=(2,0,2,2)\,, \qquad &&\mu^c_{\alpha,x}=\left(\{2\},-,\{2\},\{2\}\right)\\
&\lambda^g_\alpha=(0,1,0,1)\,, \qquad &&\mu^g_{\alpha,x}=\left(-,\{1\},-,\{1\}\right)\\
&\lambda^b_\alpha=(2,2,2,4)\,, \qquad &&\mu^b_{\alpha,x}=\left(\{1^2\},\{1^2\},\{2\},\{1^4\}\right)\,.
\ea
\ee
The color nodes are characterized by
\be
\ba
&m^c = 2\,, \qquad &&m^g=1\,, \qquad m^b=1\\
&k^{cg}=0\,, \qquad &&k^{cb}=0\,, \qquad k^{gb}=0\,.
\ea
\ee
Note that the multiplicity of the blue node is one, due to the internal lines. The additional nodes appear on $E_1$ and $E_4$ and are
\be
\ba
&m_{1,1} = 1\,, \qquad &&k^c_{1,1}=1\\
&m_{4,x} = (3,3,3,3,2,1)\,, \qquad &&k^c_{4,1}=1\,, \quad k^g_{4,1}=1 \,, \quad k^b_{4,4}=1\,,
\ea
\ee
with all others vanishing, whereas for $E_2$ we see that the multiplicity of the possible single additional node $m_{2,1}$ vanishes. From this we see the magnetic quiver is
\be
\begin{tikzpicture}[x=.8cm,y=.8cm]
\node (g1) at (0,0) [gauge,label=below:\large{1}] {};
\node (g2) at (1,0) [gaugecyan,label=below:\large{2}] {};
\node (g3) at (2,0) [gauge,label=below:\large{3}] {};
\node (g4) at (3,0) [gauge,label=below:\large{3}] {};
\node (g5) at (4,0) [gauge,label=below:\large{3}] {};
\node (g6) at (5,0) [gauge,label=below:\large{3}] {};
\node (g7) at (6,0) [gauge,label=below:\large{2}] {};
\node (g8) at (7,0) [gauge,label=below:\large{1}] {};
\node (g9) at (2,1) [gaugegreen,label=left:\large{1}] {};
\node (g10) at (5,1) [gaugeblue,label=right:\large{1}] {};
\draw (g1)--(g2)--(g3)--(g4)--(g5)--(g6)--(g7)--(g8);
\draw (g3)--(g9);
\draw (g6)--(g10);
\end{tikzpicture}\,.
\ee

To see the next step in the Hasse diagram we write
\be
\begin{tikzpicture}[x=.5cm,y=.5cm]

\draw[step=.5cm,gray,very thin] (14,0) grid (17,6);
\draw[step=.5cm,gray,very thin] (8,0) grid (12,7);

\draw[ligne, black](8,0)--(8,1)--(8,2);
\draw[ligne, black](12,2)--(12,0);
\draw[ligne, black](10,5)--(11,6)--(12,7)--(12,5);
\draw[ligne, black](12,0)--(11,0)--(10,0);
\draw[ligne, black](8,0)--(9,2);
\draw[ligne, black](8,3)--(9,3);
\draw[ligne, black](12,0)--(11,2);
\draw[ligne, black](12,7)--(11,5);
\draw[ligne, black](8,2)--(8,3)--(9,4)--(10,5);
\draw[ligne, black](12,5)--(12,2);
\draw[ligne, black](10,0)--(9,0)--(8,0);
\node[bd] at (8,0) {}; 
\node[wd] at (8,1) {}; 
\node[bd] at (8,2) {}; 
\node[bd] at (8,3) {}; 
\node[wd] at (9,4) {}; 
\node[bd] at (10,5) {}; 
\node[wd] at (11,6) {}; 
\node[bd] at (12,7) {}; 
\node[bd] at (12,6) {}; 
\node[bd] at (12,5) {}; 
\node[bd] at (12,4) {}; 
\node[bd] at (12,3) {}; 
\node[bd] at (12,2) {}; 
\node[bd] at (12,1) {}; 
\node[bd] at (12,0) {}; 
\node[wd] at (11,0) {}; 
\node[wd] at (10,0) {}; 
\node[wd] at (9,0) {}; 

\node[] at (13,2) {$<$};

\draw[ligne,black] (14,0)--(14,1)--(14,3)--(15,4)--(17,6)--(17,5)--(17,4)--(17,0)--(16,0)--(15,0)--(14,0); 
\draw[ligne,black] (14,0)--(15,2);
\draw[ligne,black] (14,3)--(15,3);
\draw[ligne,black] (17,6)--(16,4);
\draw[ligne,black] (17,0)--(16,2);
\node[bd] at (14,0) {}; 
\node[wd] at (14,1) {}; 
\node[bd] at (14,2) {}; 
\node[bd] at (14,3) {}; 
\node[wd] at (15,4) {}; 
\node[bd] at (16,5) {}; 
\node[bd] at (17,6) {}; 
\node[wd] at (17,5) {}; 
\node[wd] at (17,4) {}; 
\node[bd] at (17,3) {};
\node[bd] at (17,2) {}; 
\node[bd] at (17,1) {}; 
\node[bd] at (17,0) {};
\node[wd] at (16,0) {};
\node[wd] at (15,0) {};
\node[bd] at (14,0) {};

\node[] at (18,2) {$\oplus$};

\draw[ligne,black] (19,2)--(20,3)--(20,2)--(19,2); 
\node[bd] at (19,2) {}; 
\node[bd] at (20,3) {};
\node[bd] at (20,2) {};
\end{tikzpicture}\,,
\qquad\qquad
\begin{tikzpicture}[x=.5cm,y=.5cm]

\draw[step=.5cm,gray,very thin] (14,0) grid (17,6);
\draw[step=.5cm,gray,very thin] (8,0) grid (12,7);

\draw[ligne, blue](8,0)--(8,1)--(8,3);
\draw[ligne, blue](12,2)--(12,0);
\draw[ligne, blue](10,5)--(11,6)--(12,7)--(12,5);
\draw[ligne, blue](12,0)--(11,0);
\draw[ligne, blue](8,0)--(9,2);
\draw[ligne, blue](8,3)--(9,3);
\draw[ligne, blue](12,0)--(11,2);
\draw[ligne, blue](12,7)--(11,5);
\draw[ligne, blue](8,3)--(9,4);
\draw[ligne, blue](12,4)--(12,2);
\draw[ligne, blue](10,0)--(9,0)--(8,0);
\draw[ligne, blue](9,2)--(9,3)--(10,4);
\draw[ligne, blue](10,2)--(11,2)--(11,4);

\draw[ligne, black](9,4)--(10,5);
\draw[ligne, black](12,4)--(12,5);
\draw[ligne, black](10,0)--(11,0);

\node[bd] at (8,0) {}; 
\node[wd] at (8,1) {}; 
\node[bd] at (8,2) {}; 
\node[bd] at (8,3) {}; 
\node[wd] at (9,4) {}; 
\node[bd] at (10,5) {}; 
\node[wd] at (11,6) {}; 
\node[bd] at (12,7) {}; 
\node[bd] at (12,6) {}; 
\node[bd] at (12,5) {}; 
\node[bd] at (12,4) {}; 
\node[bd] at (12,3) {}; 
\node[bd] at (12,2) {}; 
\node[bd] at (12,1) {}; 
\node[bd] at (12,0) {}; 
\node[wd] at (11,0) {}; 
\node[wd] at (10,0) {}; 
\node[wd] at (9,0) {}; 

\node[] at (13,2) {$<$};

\draw[ligne,blue] (14,0)--(14,1)--(14,3)--(15,4)--(17,6)--(17,5)--(17,4)--(17,0)--(16,0)--(15,0)--(14,0); 
\draw[ligne,blue] (14,0)--(15,2);
\draw[ligne,blue] (14,3)--(15,3);
\draw[ligne,blue] (15,2)--(15,3)--(16,4);
\draw[ligne,blue] (15,2)--(16,2);
\draw[ligne,blue] (17,6)--(16,4);
\draw[ligne,blue] (17,0)--(16,2)--(16,4);
\node[bd] at (14,0) {}; 
\node[wd] at (14,1) {}; 
\node[bd] at (14,2) {}; 
\node[bd] at (14,3) {}; 
\node[wd] at (15,4) {}; 
\node[bd] at (16,5) {}; 
\node[bd] at (17,6) {}; 
\node[wd] at (17,5) {}; 
\node[wd] at (17,4) {}; 
\node[bd] at (17,3) {};
\node[bd] at (17,2) {}; 
\node[bd] at (17,1) {}; 
\node[bd] at (17,0) {};
\node[wd] at (16,0) {};
\node[wd] at (15,0) {};
\node[bd] at (14,0) {};

\node[] at (18,2) {$\oplus$};

\draw[ligne,black] (19,2)--(20,3)--(20,2)--(19,2); 
\node[bd] at (19,2) {}; 
\node[bd] at (20,3) {};
\node[bd] at (20,2) {};
\end{tikzpicture}\,,
\ee
and compute that the magnetic quiver of $\Delta$ and $Q$ to be $\mathfrak{d}_7$ and $\mathfrak{a}_8$ respectively.

Let us now turn to the other branch of the Hasse diagram. We can actually embed a different Minkowski summand into $P$, namely a diagram corresponding to $\mathfrak{e}_8$
\be \label{Ex2HasseE8}
\begin{tikzpicture}[x=.5cm,y=.5cm]

\draw[step=.5cm,gray,very thin] (8,0) grid (12,7);
\draw[step=.5cm,gray,very thin] (14,0) grid (18,7);

\draw[ligne, black](8,0)--(8,1)--(8,2);
\draw[ligne, black](12,2)--(12,0);
\draw[ligne, black](10,5)--(11,6)--(12,7)--(12,5);
\draw[ligne, black](12,0)--(11,0)--(10,0);
\draw[ligne, black](8,2)--(8,3)--(9,4)--(10,5);
\draw[ligne, black](12,5)--(12,2);
\draw[ligne, black](10,0)--(9,0)--(8,0);
\node[bd] at (8,0) {}; 
\node[wd] at (8,1) {}; 
\node[bd] at (8,2) {}; 
\node[bd] at (8,3) {}; 
\node[wd] at (9,4) {}; 
\node[bd] at (10,5) {}; 
\node[wd] at (11,6) {}; 
\node[bd] at (12,7) {}; 
\node[bd] at (12,6) {}; 
\node[bd] at (12,5) {}; 
\node[bd] at (12,4) {}; 
\node[bd] at (12,3) {}; 
\node[bd] at (12,2) {}; 
\node[bd] at (12,1) {}; 
\node[bd] at (12,0) {}; 
\node[wd] at (11,0) {}; 
\node[wd] at (10,0) {}; 
\node[wd] at (9,0) {};

\node[] at (13,2) {$<$};

\draw[ligne, black](14,0)--(14,1)--(14,2);
\draw[ligne, black](18,2)--(18,0);
\draw[ligne, black](16,5)--(17,6)--(18,7)--(18,5);
\draw[ligne, black](18,0)--(17,0)--(16,0);
\draw[ligne, black](14,2)--(14,3)--(15,4)--(16,5);
\draw[ligne, black](18,5)--(18,2);
\draw[ligne, black](16,0)--(15,0)--(14,0);
\node[bd] at (14,0) {}; 
\node[wd] at (14,1) {}; 
\node[wd] at (14,2) {}; 
\node[bd] at (14,3) {}; 
\node[wd] at (15,4) {}; 
\node[bd] at (16,5) {}; 
\node[wd] at (17,6) {}; 
\node[bd] at (18,7) {}; 
\node[bd] at (18,6) {}; 
\node[bd] at (18,5) {}; 
\node[bd] at (18,4) {}; 
\node[bd] at (18,3) {}; 
\node[bd] at (18,2) {}; 
\node[bd] at (18,1) {}; 
\node[bd] at (18,0) {}; 
\node[wd] at (17,0) {}; 
\node[wd] at (16,0) {}; 
\node[wd] at (15,0) {};

\end{tikzpicture} \,,
\qquad \qquad
\begin{tikzpicture}[x=.5cm,y=.5cm]

\draw[step=.5cm,gray,very thin] (8,0) grid (12,7);
\draw[step=.5cm,gray,very thin] (14,0) grid (18,7);

\draw[ligne, blue](8,0)--(8,1)--(8,2);
\draw[ligne, blue](12,2)--(12,0);
\draw[ligne, blue](10,5)--(11,6)--(12,7)--(12,5);
\draw[ligne, blue](12,0)--(11,0)--(10,0);
\draw[ligne, blue](8,2)--(8,3)--(9,4)--(10,5);
\draw[ligne, blue](12,5)--(12,2);
\draw[ligne, blue](10,0)--(9,0)--(8,0);
\draw[ligne, blue](8,0)--(11,4);
\draw[ligne, blue](8,3)--(11,4);
\draw[ligne, blue](12,0)--(11,4);
\draw[ligne, blue](12,7)--(11,4);
\node[bd] at (8,0) {}; 
\node[wd] at (8,1) {}; 
\node[bd] at (8,2) {}; 
\node[bd] at (8,3) {}; 
\node[wd] at (9,4) {}; 
\node[bd] at (10,5) {}; 
\node[wd] at (11,6) {}; 
\node[bd] at (12,7) {}; 
\node[bd] at (12,6) {}; 
\node[bd] at (12,5) {}; 
\node[bd] at (12,4) {}; 
\node[bd] at (12,3) {}; 
\node[bd] at (12,2) {}; 
\node[bd] at (12,1) {}; 
\node[bd] at (12,0) {}; 
\node[wd] at (11,0) {}; 
\node[wd] at (10,0) {}; 
\node[wd] at (9,0) {};

\node[] at (13,2) {$<$};

\draw[ligne, blue](14,0)--(14,1)--(14,2);
\draw[ligne, blue](18,2)--(18,0);
\draw[ligne, blue](16,5)--(17,6)--(18,7)--(18,5);
\draw[ligne, blue](18,0)--(17,0)--(16,0);
\draw[ligne, blue](14,2)--(14,3)--(15,4)--(16,5);
\draw[ligne, blue](18,5)--(18,2);
\draw[ligne, blue](16,0)--(15,0)--(14,0);
\draw[ligne, blue](14,0)--(17,4);
\draw[ligne, blue](14,3)--(17,4);
\draw[ligne, blue](18,0)--(17,4);
\draw[ligne, blue](18,7)--(17,4);
\node[bd] at (14,0) {}; 
\node[wd] at (14,1) {}; 
\node[wd] at (14,2) {}; 
\node[bd] at (14,3) {}; 
\node[wd] at (15,4) {}; 
\node[bd] at (16,5) {}; 
\node[wd] at (17,6) {}; 
\node[bd] at (18,7) {}; 
\node[bd] at (18,6) {}; 
\node[bd] at (18,5) {}; 
\node[bd] at (18,4) {}; 
\node[bd] at (18,3) {}; 
\node[bd] at (18,2) {}; 
\node[bd] at (18,1) {}; 
\node[bd] at (18,0) {}; 
\node[wd] at (17,0) {}; 
\node[wd] at (16,0) {}; 
\node[wd] at (15,0) {};

\end{tikzpicture} \,,
\ee
It is easy to compute the magnetic quiver of the rightmost diagram to be the affine diagram of $\mathfrak{a}_1$. This means that in total the Hasse diagram of the strongly coupled $SU(4)_{3/2} + 1\bm{AS} + 7\bm{F}$ is
\be
\begin{tikzpicture}
\node (1) [hasse] at (0,0) {};
\node (2) [hasse] at (-1,-1) {};
\node (3) [hasse] at (-1,-2) {};
\node (4) [hasse] at (0,-3) {};
\node (5) [hasse] at (1,-2.5) {};
\draw (1) edge [] node[label=left:$\mathfrak{e}_6$] {} (2);
\draw (2) edge [] node[label=left:$\mathfrak{d}_7$] {} (3);
\draw (3) edge [] node[label=left:$\mathfrak{a}_8$] {} (4);
\draw (1) edge [] node[label=right:$\mathfrak{e}_8$] {} (5);
\draw (5) edge [] node[label=right:$\mathfrak{a}_1$] {} (4);
\end{tikzpicture}\,.
\ee

Let us quickly comment on why there is no $e_7$ leaf. Given the shape of $P$ the possible $e_7$ GTP is given by
\be \label{Ex2HasseE7}
\begin{tikzpicture}[x=.5cm,y=.5cm]

\draw[step=.5cm,gray,very thin] (0,0) grid (3,5);

\draw[ligne,blue] (0,0)--(0,1)--(0,2)--(3,5)--(3,0)--(2,0)--(1,0)--(0,0); 
\draw[ligne,blue] (0,0)--(2,3);
\draw[ligne,blue] (3,0)--(2,3);
\draw[ligne,blue] (0,2)--(2,3);
\draw[ligne,blue] (3,5)--(2,3);
\node[bd] at (0,0) {}; 
\node[wd] at (0,1) {}; 
\node[bd] at (0,2) {}; 
\node[bd] at (1,3) {}; 
\node[bd] at (2,4) {}; 
\node[bd] at (3,5) {}; 
\node[bd] at (3,4) {}; 
\node[bd] at (3,3) {};
\node[bd] at (3,2) {}; 
\node[bd] at (3,1) {}; 
\node[bd] at (3,0) {};
\node[wd] at (2,0) {};
\node[wd] at (1,0) {};
\node[bd] at (0,0) {};
\end{tikzpicture}\,.
\ee
However, on edge $E_1$, the S-rule would demand $\mu^b_{1,x}=\{1^3\}$, which is incompatible with $P$, which has $\mu_{1,x}=\{2^2\}$. Similar incompatibilities hold for all other possible leaves.

\subsection{Pruning}

\label{sec:ExPruning}

As we emphasized, some GTPs benefit from pruning, before the algorithm is applied. We now provide another example for this.
Take the GTP $P$
\be \label{ExPruning}
\begin{tikzpicture}[x=.5cm,y=.5cm]
\draw[step=.5cm,gray,very thin] (0,0) grid (4,8);
\draw[ligne](0,0)--(1,2)--(3,6)--(4,8);
\draw[ligne](4,8)--(4,0);
\draw[ligne](4,0)--(3,0)--(2,0)--(1,0)--(0,0);
\node[bd] at (0,0) {}; 
\node[wd] at (1,2) {}; 
\node[bd] at (2,4) {}; 
\node[wd] at (3,6) {}; 
\node[bd] at (4,8) {}; 
\node[bd] at (4,7) {}; 
\node[bd] at (4,6) {}; 
\node[bd] at (4,5) {}; 
\node[bd] at (4,4) {}; 
\node[bd] at (4,3) {}; 
\node[bd] at (4,2) {}; 
\node[bd] at (4,1) {}; 
\node[bd] at (4,0) {}; 
\node[wd] at (3,0) {}; 
\node[wd] at (2,0) {}; 
\node[wd] at (1,0) {}; 
\end{tikzpicture}
\ee
which has
\be
\ba
L_1 = (-1,-2)\,,\quad L_2 = (1,0)\,, \quad L_3=(0,1)\\
\lambda_{\alpha}=(4,4,8)\,,\qquad \mu_{\alpha,i}=(\{2^2\},\{4\},\{1^8\})\,,
\ea
\ee
and consider the pruning $P'=(P,\mu^3_1,-)$, i.e. along the edge $E_3$. First we need to compute
\be
\det(L_3,L_1)=+1 \rightarrow L_1 \in E_+\,, \qquad \det(L_3,L_2)=-1 \rightarrow L_2 \in E_-\,.
\ee
From this we can completely build $P'$.
\begin{enumerate}
\item The unaltered edges are $E_-= 4 L_2$ with $\mu_2=\{4\}$
\item Removing the edge segment we prune along we have $\{1^7\} L_3$
\item The slope of $E_+$ changes, i.e. we have $L_1\to (-1,-2) + (0,1)=(-1,-1)$ with unchanged $\mu_1=\{2^2\}$
\item The final slope is along $-L_{3}=(0,-1)$ with $\mu_{-3}=\{-1+ 4\} = \{3\}$.
\end{enumerate}
Thus, $P'$ is given by
\be
\begin{tikzpicture}[x=.5cm,y=.5cm]
\draw[step=.5cm,gray,very thin] (14,0) grid (18,7);
\draw[ligne, black](14,0)--(14,1)--(14,2);
\draw[ligne, black](18,2)--(18,0);
\draw[ligne, black](16,5)--(17,6)--(18,7)--(18,5);
\draw[ligne, black](18,0)--(17,0)--(16,0);
\draw[ligne, black](14,2)--(14,3)--(15,4)--(16,5);
\draw[ligne, black](18,5)--(18,2);
\draw[ligne, black](16,0)--(15,0)--(14,0);
\node[bd] at (14,0) {}; 
\node[wd] at (14,1) {}; 
\node[wd] at (14,2) {}; 
\node[bd] at (14,3) {}; 
\node[wd] at (15,4) {}; 
\node[bd] at (16,5) {}; 
\node[wd] at (17,6) {}; 
\node[bd] at (18,7) {}; 
\node[bd] at (18,6) {}; 
\node[bd] at (18,5) {}; 
\node[bd] at (18,4) {}; 
\node[bd] at (18,3) {}; 
\node[bd] at (18,2) {}; 
\node[bd] at (18,1) {}; 
\node[bd] at (18,0) {}; 
\node[wd] at (17,0) {}; 
\node[wd] at (16,0) {}; 
\node[wd] at (15,0) {}; 

\node[] at (16,-1) {1.};
\node[] at (19,3) {2.};
\node[] at (15.5,5.5) {3.};
\node[] at (13,1.5) {4.};

\end{tikzpicture}
\ee
where we labeled the four sides corresponding to their origin in the enumeration. From the discussion in the previous example we know that the magnetic quiver of $P'$ is the affine Dynkin diagram of $\mathfrak{e}_8$. Consequently, both $P'$ and $P$ represent the rank one $E_8$-theory.

\subsection{Isolated Toric Singularities}

For isolated toric singularities, i.e. which are strictly convex, with $\lambda_\alpha=1 \ \forall \alpha$, the derivation of the MQ and Hasse diagram substantially simplifies. As this is an interesting class of theories we will here discuss this simplified setting. 
Furthermore, the polygons for each of these have a Minkowski sum decomposition, which in this case is well-known to map to the deformations of the singularities by the work of Altmann \cite{Altmann,Altmann2}.  We can consider the setup discussed in this paper as a generalization of this to not strictly convex toric and generalized toric polygons. 

As the edge lengths are all $1$ with no vertices along the edges, the multiplicities of all the nodes in the magnetic quiver is always $1$. Each color furthermore contributes precisely one vertex and there are no tails, i.e. the number of colors determines the Higgs branch dimension plus 1. 

The Hasse diagram is obtained by simply collapsing along one Minkowski summand: if 
\be
P= P_1  + \cdots +  P_{n_c} \,,
\ee
then the Hasse diagram has $n_c$ branches. Note that in this strictly convex case the partition sum and Minkowski sum agree. Each branch correspond to the quiver subtraction of the theory 
\be
\Delta_i = P_1 + \cdots+ P_{i-1} + P_{i+1} +\cdots + P_{n_c} \,.
\ee
If $\Delta_i$ is a trivial theory, then this branch is empty. 
The next level in the Hasse diagram is obtained by taking the sub-polygon $\Delta_i$ in $P$ and assigning it a single color.

The simplest example is the `beetle' \cite{Closset:2018bjz} 
\be
\begin{tikzpicture}[x=.5cm,y=.5cm]
\node[] at (-2,0) {$P_{\text{beetle}}= \qquad $};
\draw[step=.5cm,gray,very thin] (0,-1) grid (3,1);
\draw[ligne] (0,0)--(1,1)--(2,1)--(3,0)--(2,-1)--(1,-1)--(0,0); 
\node[bd] at (0,0) {}; 
\node[bd] at (1,1) {}; 
\node[bd] at (2,1) {}; 
\node[bd] at (3,0) {}; 
\node[bd] at (2,-1) {}; 
\node[bd] at (1,-1) {}; 
\end{tikzpicture} 
\ee
There is precisely one coloring and an associated Minkowski sum decomposition 
\be
 \qquad 
\begin{tikzpicture}[x=.5cm,y=.5cm]
\draw[step=.5cm,gray,very thin] (0,-1) grid (3,1);
\draw[ligne, cyan] (0,0)--(1,1);
\draw[ligne, cyan] (1,0)--(2,1);
\draw[ligne, cyan] (2,-1)--(3,0);
\draw[ligne, blue] (1,1)--(2,1);
\draw[ligne, blue] (0,0)--(1,0);
\draw[ligne, blue] (1,-1)--(2,-1);
\draw[ligne, green] (1,-1)--(0,0) ;
\draw[ligne, green] (1,0)--(2,-1) ;
\draw[ligne, green] (2,1)--(3,0) ;
\node[bd] at (0,0) {}; 
\node[bd] at (1,1) {}; 
\node[bd] at (2,1) {}; 
\node[bd] at (3,0) {}; 
\node[bd] at (2,-1) {}; 
\node[bd] at (1,-1) {}; 
\end{tikzpicture} 
=
\begin{tikzpicture}[x=.5cm,y=.5cm]
\draw[ligne, cyan] (0,0)--(1,1);
\end{tikzpicture} 
+
\begin{tikzpicture}[x=.5cm,y=.5cm]
\draw[ligne, blue] (0,0)--(1,0);
\end{tikzpicture} 
+
\begin{tikzpicture}[x=.5cm,y=.5cm]
\draw[ligne, green] (0,1)--(1,0) ;
\end{tikzpicture} 
\ee
The magnetic quiver is read off simply by computing the areas between the various bi-colored parallelograms (there is only the stable intersection in this case) 
\be
\begin{tikzpicture}[x=.8cm,y=.8cm]
\node[] at (-3,0.5) {$\text{MQ}(P_{\text{beetle}})= \qquad $};
\node (g1) at (0,0) [gaugeblue,label=below:\large{1}] {};
\node (g2) at (-1,1) [gaugecyan,label=above:\large{1}] {};
\node (g3) at (1,1) [gaugegreen,label=above:\large{1}] {};
\draw (g2)--(g1)--(g3);
\draw[doublearrow] (g2)--(g3);
\end{tikzpicture}
\ee
The Hasse diagram is obtained by taking the Minkowski sum of two colors only. This is nontrivial (i.e. not a rank 0 theory) only for red + green, which is the $A_1$ theory, which results in 
\be
\begin{tikzpicture}[x=.5cm,y=.5cm]
\node[] at (-4, 0) {$P_{\text{beetle}} -  A_1 = P_{\text{beetle}} - $};
\draw[ligne, cyan] (1,0)--(2,1);
\draw[ligne, cyan] (2,-1)--(3,0);
\draw[ligne,green] (1,0)--(2,-1) ;
\draw[ligne, green] (2,1)--(3,0) ;
\node[bd] at (2,1) {}; 
\node[bd] at (3,0) {}; 
\node[bd] at (2,-1) {}; 
\node[bd] at (1,0) {}; 
\end{tikzpicture} \qquad 
\begin{tikzpicture}[x=.5cm,y=.5cm]
\node[] at (-1,0) {$= \qquad $};
\draw[step=.5cm,gray,very thin] (0,-1) grid (3,1);
\draw[ligne, cyan] (0,0)--(1,1);
\draw[ligne, cyan] (1,0)--(2,1);
\draw[ligne, cyan] (2,-1)--(3,0);
\draw[ligne, blue] (1,1)--(2,1);
\draw[ligne, blue] (0,0)--(1,0);
\draw[ligne, blue] (1,-1)--(2,-1);
\draw[ligne, cyan] (1,-1)--(0,0) ;
\draw[ligne, cyan] (1,0)--(2,-1) ;
\draw[ligne, cyan] (2,1)--(3,0) ;
\node[bd] at (0,0) {}; 
\node[bd] at (1,1) {}; 
\node[bd] at (2,1) {}; 
\node[bd] at (3,0) {}; 
\node[bd] at (2,-1) {}; 
\node[bd] at (1,-1) {}; 
\node[] at (4,0){$\qquad = A_1 $};
\end{tikzpicture} 
\,.
\ee

A slightly more complicated example is the octagon, which has multiple branches. The toric polygon is 
\be
\begin{tikzpicture}[x=.5cm,y=.5cm]
\node[] at (-3, 0) {$P_{\text{octagon}} = \qquad $} ;
\draw[step=.5cm,gray,very thin] (0,-1) grid (3,2);
\draw[ligne] (0,0)--(0,1)--(1,2)--(2,2)--(3,1)--(3,0)--(2,-1)--(1,-1)--(0,0); 
\node[bd] at (0,0) {}; 
\node[bd] at (0,1) {}; 
\node[bd] at (1,2) {};
\node[bd] at (2,2) {}; 
\node[bd] at (3,1) {}; 
\node[bd] at (3,0) {}; 
\node[bd] at (2,-1) {};
\node[bd] at (1,-1) {};
\end{tikzpicture} 
\ee
There are three colorings, and associated Minkowski sum decompositions
\be
\ba
(P_{\text{octagon}}, \{\lambda^c_{(1)}\}) & = \qquad 
\begin{tikzpicture}[x=.5cm,y=.5cm]
\draw[step=.5cm,gray,very thin] (0,-1) grid (3,2);
\draw[ligne, cyan] (0,0)--(0,1); 
\draw[ligne, blue] (0,1)--(1,2); 
\draw[ligne, cyan] (1,2)--(2,2); 
\draw[ligne, green] (2,2)--(3,1); 
\draw[ligne, blue] (3,1)--(3,0); 
\draw[ligne, cyan] (3,0)--(2,-1);
\draw[ligne, blue] (2,-1)--(1,-1);  
\draw[ligne, green] (1,-1)--(0,0); 
\node[bd] at (0,0) {}; 
\node[bd] at (0,1) {}; 
\node[bd] at (1,2) {};
\node[bd] at (2,2) {}; 
\node[bd] at (3,1) {}; 
\node[bd] at (3,0) {}; 
\node[bd] at (2,-1) {};
\node[bd] at (1,-1) {};
\end{tikzpicture} 
= 
\begin{tikzpicture}[x=.5cm,y=.5cm]
\draw[ligne, cyan] (0,0)--(1,1)--(0,1)--(0,0);
\end{tikzpicture} 
+
\begin{tikzpicture}[x=.5cm,y=.5cm]
\draw[ligne, blue] (0,0)--(1,0)--(1,1) -- (0,0);
\end{tikzpicture} 
+
\begin{tikzpicture}[x=.5cm,y=.5cm]
\draw[ligne, green] (0,1)--(1,0) ;
\end{tikzpicture} 
\cr 
(P_{\text{octagon}}, \{\lambda^c_{(2)}\}) &= \qquad 
\begin{tikzpicture}[x=.5cm,y=.5cm]
\draw[step=.5cm,gray,very thin] (0,-1) grid (3,2);
\draw[ligne, blue] (0,0)--(0,1); 
\draw[ligne, cyan] (0,1)--(1,2); 
\draw[ligne, green] (1,2)--(2,2); 
\draw[ligne, orange] (2,2)--(3,1); 
\draw[ligne, blue] (3,1)--(3,0); 
\draw[ligne, cyan] (3,0)--(2,-1);
\draw[ligne, green] (2,-1)--(1,-1);  
\draw[ligne, orange] (1,-1)--(0,0); 
\node[bd] at (0,0) {}; 
\node[bd] at (0,1) {}; 
\node[bd] at (1,2) {};
\node[bd] at (2,2) {}; 
\node[bd] at (3,1) {}; 
\node[bd] at (3,0) {}; 
\node[bd] at (2,-1) {};
\node[bd] at (1,-1) {};
\end{tikzpicture} 
= 
\begin{tikzpicture}[x=.5cm,y=.5cm]
\draw[ligne, cyan] (0,0)--(1,1);
\end{tikzpicture} 
+
\begin{tikzpicture}[x=.5cm,y=.5cm]
\draw[ligne, green] (0,0)--(1,0);
\end{tikzpicture} 
+
\begin{tikzpicture}[x=.5cm,y=.5cm]
\draw[ligne, orange] (0,1)--(1,0) ;
\end{tikzpicture} 
+
\begin{tikzpicture}[x=.5cm,y=.5cm]
\draw[ligne, blue] (0,0)--(0,1) ;
\end{tikzpicture} 
\cr 
(P_{\text{octagon}}, \{\lambda^c_{(3)}\}) &= \qquad 
\begin{tikzpicture}[x=.5cm,y=.5cm]
\draw[step=.5cm,gray,very thin] (0,-1) grid (3,2);
\draw[ligne, cyan] (0,0)--(0,1); 
\draw[ligne, green] (0,1)--(1,2); 
\draw[ligne, blue] (1,2)--(2,2); 
\draw[ligne, cyan] (2,2)--(3,1); 
\draw[ligne, blue] (3,1)--(3,0); 
\draw[ligne, green] (3,0)--(2,-1);
\draw[ligne, cyan] (2,-1)--(1,-1);  
\draw[ligne, blue] (1,-1)--(0,0); 
\node[bd] at (0,0) {}; 
\node[bd] at (0,1) {}; 
\node[bd] at (1,2) {};
\node[bd] at (2,2) {}; 
\node[bd] at (3,1) {}; 
\node[bd] at (3,0) {}; 
\node[bd] at (2,-1) {};
\node[bd] at (1,-1) {};
\end{tikzpicture} 
= 
\begin{tikzpicture}[x=.5cm,y=.5cm]
\draw[ligne, cyan] (0,0)--(1,0)--(0,1)--(0,0);
\end{tikzpicture} 
+
\begin{tikzpicture}[x=.5cm,y=.5cm]
\draw[ligne, blue] (0,1)--(1,1)--(1,0) -- (0,1);
\end{tikzpicture} 
+
\begin{tikzpicture}[x=.5cm,y=.5cm]
\draw[ligne, green] (0,0)--(1,1) ;
\end{tikzpicture} 
\ea
\ee
The magnetic quivers are computed by filling the coloring 
\be
\begin{tikzpicture}[x=.8cm,y=.8cm]
\node[] at (-6,0.5) {$MQ(P_{\text{octagon}}, \{\lambda^c_{(1)}\})= MQ(P_{\text{octagon}}, \{\lambda^c_{(3)}\}) =$};
\node (g1) at (0,0) [gaugeblue,label=below:\large{1}] {};
\node (g2) at (2,0) [gaugegreen,label=below:\large{1}] {};
\node (g3) at (1,1) [gaugecyan,label=above:\large{1}] {};
\draw[doublearrow] (g1)--(g2)--(g3)--(g1);
\end{tikzpicture}
\ee
and for the dimension three cone of the Higgs branch 
\be
\begin{tikzpicture}[x=.8cm,y=.8cm]
\node[] at (-3,1){$MQ(P_{\text{octagon}}, \{\lambda^c_{(2)}\}) =$};
\node (g1) at (0,0) [gaugecyan,label=below:\large{1}] {};
\node (g2) at (2,0) [gaugeorange,label=below:\large{1}] {};
\node (g3) at (1,1) [gaugeblue,label=below:\large{1}] {};
\node (g4) at (1,2) [gaugegreen,label=above:\large{1}] {};
\draw (g1)--(g3)--(g2)--(g4)--(g1);
\draw (g3)--(g4);
\draw[doublearrow] (g1)--(g2);
\end{tikzpicture}
\ee
To compute the Hasse diagram, we consider both branches: for $\lambda_{(1)}$ and $\lambda_{(3)}$
there are three slices
\be
A_1= \begin{tikzpicture}[x=.5cm,y=.5cm]
\draw[ligne, blue] (0,0)--(1,0);
\draw[ligne, blue] (0,1)--(1,1)--(1,2)--(0,1);
\draw[ligne, blue] (2,2)--(2,1);
\draw[ligne, cyan] (1,2)--(2,2);
\draw[ligne, cyan] (0,0)--(0,1);
\draw[ligne, cyan] (1,0)--(1,1)--(2,1)--(1,0);
\end{tikzpicture}\,,\qquad 
A_1' =  
\begin{tikzpicture}[x=.5cm,y=.5cm]
\draw[ligne, cyan] (0,1)--(0,2)--(1,2)--(0,1);
\draw[ligne, cyan] (2,1)--(1,0);
\draw[ligne, green] (0,1)--(1,0);
\draw[ligne, green] (1,2)--(2,1);
\end{tikzpicture}\,,\qquad 
A_1''= 
\begin{tikzpicture}[x=.5cm,y=.5cm]
\draw[ligne, blue] (1,0)--(2,0)--(2,1)--(1,0);
\draw[ligne, blue] (0,1)--(1,2);
\draw[ligne, green] (0,1)--(1,0);
\draw[ligne, green] (1,2)--(2,1);
\end{tikzpicture}\,,\qquad 
\ee
Subtracting this, results in the following three diagrams 
\be
\ba
(P_{\text{octagon}}, \{\lambda^c_{(1)}\}) - A_1 &=  
\begin{tikzpicture}[x=.5cm,y=.5cm]
\draw[step=.5cm,gray,very thin] (0,-1) grid (3,2);
\draw[ligne, purple] (0,0)--(0,1); 
\draw[ligne, purple] (0,1)--(1,2); 
\draw[ligne, purple] (1,2)--(2,2); 
\draw[ligne, green] (2,2)--(3,1); 
\draw[ligne, purple] (3,1)--(3,0); 
\draw[ligne, purple] (3,0)--(2,-1);
\draw[ligne, purple] (2,-1)--(1,-1);  
\draw[ligne, green] (1,-1)--(0,0); 
\node[bd] at (0,0) {}; 
\node[bd] at (0,1) {}; 
\node[bd] at (1,2) {};
\node[bd] at (2,2) {}; 
\node[bd] at (3,1) {}; 
\node[bd] at (3,0) {}; 
\node[bd] at (2,-1) {};
\node[bd] at (1,-1) {};
\end{tikzpicture} 
\cr 
(P_{\text{octagon}}, \{\lambda^c_{(1)}\}) - A_1' 
&=
\begin{tikzpicture}[x=.5cm,y=.5cm]
\draw[step=.5cm,gray,very thin] (0,-1) grid (3,2);
\draw[ligne, purple] (0,0)--(0,1); 
\draw[ligne, blue] (0,1)--(1,2); 
\draw[ligne, purple] (1,2)--(2,2); 
\draw[ligne, purple] (2,2)--(3,1); 
\draw[ligne, blue] (3,1)--(3,0); 
\draw[ligne, purple] (3,0)--(2,-1);
\draw[ligne, blue] (2,-1)--(1,-1);  
\draw[ligne, purple] (1,-1)--(0,0); 
\node[bd] at (0,0) {}; 
\node[bd] at (0,1) {}; 
\node[bd] at (1,2) {};
\node[bd] at (2,2) {}; 
\node[bd] at (3,1) {}; 
\node[bd] at (3,0) {}; 
\node[bd] at (2,-1) {};
\node[bd] at (1,-1) {};
\end{tikzpicture} 
\cr
(P_{\text{octagon}}, \{\lambda^c_{(1)}\}) - A_1'' 
&=
\begin{tikzpicture}[x=.5cm,y=.5cm]
\draw[step=.5cm,gray,very thin] (0,-1) grid (3,2);
\draw[ligne, cyan] (0,0)--(0,1); 
\draw[ligne, purple] (0,1)--(1,2); 
\draw[ligne, cyan] (1,2)--(2,2); 
\draw[ligne, purple] (2,2)--(3,1); 
\draw[ligne, purple] (3,1)--(3,0); 
\draw[ligne, cyan] (3,0)--(2,-1);
\draw[ligne, purple] (2,-1)--(1,-1);  
\draw[ligne, purple] (1,-1)--(0,0); 
\node[bd] at (0,0) {}; 
\node[bd] at (0,1) {}; 
\node[bd] at (1,2) {};
\node[bd] at (2,2) {}; 
\node[bd] at (3,1) {}; 
\node[bd] at (3,0) {}; 
\node[bd] at (2,-1) {};
\node[bd] at (1,-1) {};
\end{tikzpicture} 
\ea
\ee
All these three colored GTPs have MQ given by the Kleinian singularity $A_3$ (two vertices with four lines connecting them).
Along the other branch with the $\lambda_{(2)}$ coloring there are three singularities 
\be
\mathfrak{a}_2^{c}= \begin{tikzpicture}[x=.5cm,y=.5cm]
\draw[ligne, blue] (0,0)--(0,1);
\draw[ligne, green] (0,0)--(1,0);
\draw[ligne, blue] (1,1)--(1,2);
\draw[ligne, blue] (2,1)--(2,2);
\draw[ligne, green] (1,1)--(2,1);
\draw[ligne, green] (1,2)--(2,2);
\draw[ligne, cyan] (0,0)--(1,1);
\draw[ligne, cyan] (0,1)--(1,2);
\draw[ligne, cyan] (1,0)--(2,1);
\end{tikzpicture}\,,\qquad
\mathfrak{a}_2^{r}= \begin{tikzpicture}[x=.5cm,y=.5cm]
\draw[ligne, blue] (0,1)--(0,2);
\draw[ligne, green] (0,2)--(1,2);
\draw[ligne, blue] (1,0)--(1,1);
\draw[ligne, blue] (2,0)--(2,1);
\draw[ligne, green] (1,1)--(2,1);
\draw[ligne, green] (1,0)--(2,0);
\draw[ligne, orange] (0,1)--(1,0);
\draw[ligne, orange] (0,2)--(1,1);
\draw[ligne, orange] (1,2)--(2,1);
\end{tikzpicture}\,,\qquad
\mathfrak{a}_2^{g}= \begin{tikzpicture}[x=.5cm,y=.5cm]
\draw[ligne, blue] (0,0)--(0,1);
\draw[ligne, blue] (1,1)--(1,2);
\draw[ligne, blue] (2,0)--(2,1);
\draw[ligne, cyan] (0,0)--(1,1);
\draw[ligne, cyan] (0,1)--(1,2);
\draw[ligne, cyan] (1,-1)--(2,0);
\draw[ligne, orange] (0,0)--(1,-1);
\draw[ligne, orange] (1,1)--(2,0);
\draw[ligne, orange] (1,2)--(2,1);
\end{tikzpicture}\,,\qquad
\mathfrak{a}_2^{b}= \begin{tikzpicture}[x=.5cm,y=.5cm]
\draw[ligne, cyan] (0,1)--(1,2);
\draw[ligne, cyan] (1,0)--(2,1);
\draw[ligne, cyan] (2,0)--(3,1);
\draw[ligne, orange] (0,1)--(1,0);
\draw[ligne, orange] (1,2)--(2,1);
\draw[ligne, orange] (2,2)--(3,1);
\draw[ligne, green] (1,2)--(2,2);
\draw[ligne, green] (2,1)--(3,1);
\draw[ligne, green] (1,0)--(2,0);
\end{tikzpicture}\,.
\ee
The singularity after subtraction of these $\mathfrak{a}_2$, which is achieved by identifying the colors appearing in these sub-polygons we find 
\be
\ba
(P_{\text{octagon}}, \{\lambda^c_{(2)}\}) - \mathfrak{a}_2^c
&=
\begin{tikzpicture}[x=.5cm,y=.5cm]
\draw[step=.5cm,gray,very thin] (0,-1) grid (3,2);
\draw[ligne, purple] (0,0)--(0,1); 
\draw[ligne, purple] (0,1)--(1,2); 
\draw[ligne, purple] (1,2)--(2,2); 
\draw[ligne, orange] (2,2)--(3,1); 
\draw[ligne, purple] (3,1)--(3,0); 
\draw[ligne, purple] (3,0)--(2,-1);
\draw[ligne, purple] (2,-1)--(1,-1);  
\draw[ligne, orange] (1,-1)--(0,0); 
\node[bd] at (0,0) {}; 
\node[bd] at (0,1) {}; 
\node[bd] at (1,2) {};
\node[bd] at (2,2) {}; 
\node[bd] at (3,1) {}; 
\node[bd] at (3,0) {}; 
\node[bd] at (2,-1) {};
\node[bd] at (1,-1) {};
\end{tikzpicture}   = A_3 \cr 
(P_{\text{octagon}}, \{\lambda^c_{(2)}\}) - \mathfrak{a}_2^r
&=
\begin{tikzpicture}[x=.5cm,y=.5cm]
\draw[step=.5cm,gray,very thin] (0,-1) grid (3,2);
\draw[ligne, purple] (0,0)--(0,1); 
\draw[ligne, cyan] (0,1)--(1,2); 
\draw[ligne, purple] (1,2)--(2,2); 
\draw[ligne, purple] (2,2)--(3,1); 
\draw[ligne, purple] (3,1)--(3,0); 
\draw[ligne, cyan] (3,0)--(2,-1);
\draw[ligne, purple] (2,-1)--(1,-1);  
\draw[ligne, purple] (1,-1)--(0,0); 
\node[bd] at (0,0) {}; 
\node[bd] at (0,1) {}; 
\node[bd] at (1,2) {};
\node[bd] at (2,2) {}; 
\node[bd] at (3,1) {}; 
\node[bd] at (3,0) {}; 
\node[bd] at (2,-1) {};
\node[bd] at (1,-1) {};
\end{tikzpicture}   = A_3 \cr 
(P_{\text{octagon}}, \{\lambda^c_{(2)}\}) - \mathfrak{a}_2^g
&=
\begin{tikzpicture}[x=.5cm,y=.5cm]
\draw[step=.5cm,gray,very thin] (0,-1) grid (3,2);
\draw[ligne, purple] (0,0)--(0,1); 
\draw[ligne, purple] (0,1)--(1,2); 
\draw[ligne, green] (1,2)--(2,2); 
\draw[ligne, purple] (2,2)--(3,1); 
\draw[ligne, purple] (3,1)--(3,0); 
\draw[ligne, purple] (3,0)--(2,-1);
\draw[ligne, green] (2,-1)--(1,-1);  
\draw[ligne, purple] (1,-1)--(0,0); 
\node[bd] at (0,0) {}; 
\node[bd] at (0,1) {}; 
\node[bd] at (1,2) {};
\node[bd] at (2,2) {}; 
\node[bd] at (3,1) {}; 
\node[bd] at (3,0) {}; 
\node[bd] at (2,-1) {};
\node[bd] at (1,-1) {};
\end{tikzpicture}   = A_2 \cr 
(P_{\text{octagon}}, \{\lambda^c_{(2)}\}) - \mathfrak{a}_2^b
&=
\begin{tikzpicture}[x=.5cm,y=.5cm]
\draw[step=.5cm,gray,very thin] (0,-1) grid (3,2);
\draw[ligne, blue] (0,0)--(0,1); 
\draw[ligne, purple] (0,1)--(1,2); 
\draw[ligne, purple] (1,2)--(2,2); 
\draw[ligne, purple] (2,2)--(3,1); 
\draw[ligne, blue] (3,1)--(3,0); 
\draw[ligne, purple] (3,0)--(2,-1);
\draw[ligne, green] (2,-1)--(1,-1);  
\draw[ligne, purple] (1,-1)--(0,0); 
\node[bd] at (0,0) {}; 
\node[bd] at (0,1) {}; 
\node[bd] at (1,2) {};
\node[bd] at (2,2) {}; 
\node[bd] at (3,1) {}; 
\node[bd] at (3,0) {}; 
\node[bd] at (2,-1) {};
\node[bd] at (1,-1) {};
\end{tikzpicture}   = A_2 \,.
\ea
\ee
The combined Hasse diagram is then as follows: 
\be
\begin{tikzpicture}
\node (1) [hasse] at (0,0) {};
\node (2) [hasse] at (-8,2) {};
\node (3) [hasse] at (-6,2) {};
\node (4) [hasse] at (-4,2) {};
\node (5) [hasse] at (-2,2) {};
\node (6) [hasse] at (2,2) {};
\node (7) [hasse] at (4,2) {};
\node (8) [hasse] at (6,2) {};
\node (9) [hasse] at (8,2) {};

\node (10) [hasse] at (-6,4) {};
\node (11) [hasse] at (6,4) {};
\node (12) [hasse] at (0,4) {};

\node (13) [hasse] at (0,6) {};

\draw (1) edge [] node[label=left:$A_3$] {} (2);
\draw (1) edge [] node[label=left:$A_3$] {} (3);
\draw (1) edge [] node[label=left:$A_3$] {} (4);
\draw (1) edge [] node[label=left:$A_2$] {} (5);
\draw (1) edge [] node[label=right:$A_2$] {} (6);
\draw (1) edge [] node[label=right:$A_3$] {} (7);
\draw (1) edge [] node[label=right:$A_3$] {} (8);
\draw (1) edge [] node[label=right:$A_3$] {} (9);

\draw (2) edge [] node[label=right:$A_1$] {} (10);
\draw (3) edge [] node[label=right:$A_1$] {} (10);
\draw (4) edge [] node[label=right:$A_1$] {} (10);

\draw (7) edge [] node[label=right:$A_1$] {} (11);
\draw (8) edge [] node[label=right:$A_1$] {} (11);
\draw (9) edge [] node[label=right:$A_1$] {} (11);

\draw (5) edge [] node[label=right:$A_1$] {} (12);
\draw (6) edge [] node[label=right:$A_1$] {} (12);

\draw (4) edge [] node[label=right:$A_2$] {} (13);
\draw (12) edge [] node[label=right:$A_1$] {} (13);
\draw (7) edge [] node[label=right:$A_2$] {} (13);

\end{tikzpicture}\,.
\ee

\subsection{Non-Lagrangian Theories}

With the algorithm presented in this paper we can also study non-Lagrangian theories. A prime example are the theories $B_N^{(i)}$ introduced in \cite{Eckhard:2020jyr}. Here, we will study the magnetic quiver for $B_N^{(1)}$, for which the colored GTP is given by (drawn for $N=7$)
\be
\begin{tikzpicture}[x=.5cm,y=.5cm]
\draw[step=.5cm,gray,very thin] (0,0) grid (6,6);
\draw[ligne,green] (0,1)--(0,2); 
\draw[ligne,green] (6,1)--(6,0); 
\draw[ligne,blue] (0,2)--(0,6)--(6,1); 
\draw[ligne,blue] (6,0)--(0,1); 
\draw[thick, densely dotted,blue] (6,1)--(0,2); 
\node[bd,label=left:{(0,1)}] at (0,1) {}; 
\node[bd] at (0,2) {}; 
\node[bd] at (0,3) {}; 
\node[bd] at (0,4) {}; 
\node[bd] at (0,5) {}; 
\node[bd,label=left:{(0,$N$-1)}] at (0,6) {}; 
\node[bd,label=right:{($N$-1,1)}] at (6,1) {};
\node[bd,label=right:{($N$-1,0)}] at (6,0) {};
\end{tikzpicture} 
\ee
i.e. the GTP agrees with the toric diagram and there are two colors, independent of $N$. The data for the GTP and its coloring are
\be
\ba
&L_\alpha=((0,-1),(N-1,-1),(0,1),(-(N-1),N-2))  \\
&\lambda_\alpha=(N-2,1,1,1)\,, \qquad \mu_{\alpha,x}=\left(\{1^{N-2}\},\{1\},\{1\},\{1\}\right)\\
&\lambda^b_\alpha=(N-3,1,0,1)\,, \qquad \mu^b_{\alpha,x}=\left(\{N-3\},\{1\},-,\{1\}\right)\\
&\lambda^g_\alpha=(1,0,1,0)\,, \qquad \qquad \ \mu^g_{\alpha,x}=\left(\{1\},-,\{1\},-\right)\,,
\ea
\ee
so both color nodes have multiplicity one.
From this we can compute
\be
A^{bg}=N-1 \,,\qquad k^{bg}=2\,.
\ee
The additional nodes come only from $L_1$ with
\be
m_{1,x} = \left(N-3,N-4,\dots,1\right)\,, \qquad k_{1,1}^b = N-3\,, \quad k_{1,1}^g=1\,.
\ee
We can put this together to compute the magnetic quiver
\be
\begin{tikzpicture}[x=.8cm,y=.8cm]
\node (g1) at (0,0) [gauge,label=below:\large{1}] {};
\node (g2) at (1,0) [] {$\cdots$};
\node (g3) at (2,0) [gauge,label=below:\large{$N$-4}] {};
\node (g4) at (3.5,0) [gauge,label=below:\large{$N$-3}] {};
\node (g5) at (5,0) [gaugegreen,label=below:\large{1}] {};
\node (g6) at (3.5,1.5) [gaugeblue,label=above:\large{1}] {};
\draw (g1)--(g2)--(g3)--(g4)--(g5);
\draw[double distance = 3pt] (g5)--(g6);
\draw (g4)--(g6) node[midway, left] {\large$N$-3};
\end{tikzpicture}
\ee

\section{Derivation from the Tropics}
\label{sec:BraneWebs}

We now turn to deriving the rules that we propose in section \ref{sec:ColoringRules}. The origin is in tropical geometry, or in physics language, the $(p,q)$ 5-brane-webs -- the precise relation to tropical geometry is only strictly known in the case when the GTP is toric, and generalizing this to non-toric GTPs would be very interesting indeed. 
These are dual to the generalized toric polygons, and in this context some progress was recently made in extracting the Higgs branch magnetic quivers and Hasse diagrams \cite{Cabrera:2018ann,Cabrera:2018jxt,Cabrera:2019izd,Bourget:2019aer,Bourget:2019rtl,Cabrera:2019dob,Grimminger:2020dmg, Eckhard:2020jyr, Bourget:2020asf}.
For self-consistency, we provide a brief summary of the webs, the concept of sub-webs and the derivation of magnetic quivers in appendix \ref{app:Webs}.
In this section, we give a precise dictionary between the concepts in the webs, and the data that we introduced in the GTP, and thereby provide a derivation of the rules set out in section \ref{sec:ColoringRules}. 

\subsection{Sub-web Decomposition to Colored Polygon}

\label{sec:sub-webs}

The most important identification is between a GTP and brane-web.
First, recall that a $(p,q)$ 5-brane-web is made up of connecting $(p,q)$ 5-branes, where we distinguish between external branes (which end on corresponding $(p,q)$ 7-branes) and internal branes (which only connect to other $(p,q)$ 5-branes).
Given a web, the corresponding GTP is its dual graph. As such, a set of $n_5$ $(p,q)$ 5-branes maps to a line in the GTP, that is perpendicular to the branes,  i.e. to an edge $E_\alpha$ with the properties
\be \label{WebEalpha}
L_\alpha = (-q,p)\,, \qquad \lambda_\alpha = n_5\,.
\ee
This holds both for internal and external branes. Note that charge conservation in the web ensures closedness of the GTP. A stack of $n_5$ external 5-branes, which end on $n_7$ 7-branes gives rise to an external edge in the GTP with $n_7+1$ vertices along it, i.e.
\be 
	b_\alpha=n_7-1\,.
\ee 
Such a web configuration naturally defines a partition of $n_5$ into the number of 5-branes ending on each 7-brane, which is exactly given by the $\mu_{\alpha,x}$.
Note that from the $L_\alpha$ and $\mu_{\alpha,x}$ we can infer the positions of the vertices $\bm{v}_i$, up to equivalent configurations, as follows. Choosing an initial position for $\bm{v}_{\alpha, 0}^{\text{ex}}$ we take any reordered set of $\mu_{\alpha,x}$, which we call $\mu_{\alpha,i}$. All such reorderings are equivalent. The positions of the black vertices are then defined inductively by
\be
\bm{v}_{\alpha, i+1} = \bm{v}_{\alpha, i} + \mu_{\alpha,i+1} L_\alpha\,.
\ee
Furthermore, for neighboring edges $E_\alpha$ and $E_{\alpha+1}$ we have
\be
\p E_\alpha \cap \p E_{\alpha +1}=  \bm{v}_{\alpha, b_\alpha+1}^{\text{ex}} = \bm{v}_{\alpha+1, 0}^{\text{ex}}\,.
\ee
The brane-web thus induces the vertices $V_b$ of the corresponding GTP.

Next, we discuss the idea of pruning. We can take a $(p,q)$ 7-brane in the brane-web, on which $n_5$ 5-branes end, and push it through the entire web, turning it into a $(-p,-q)$ 7-brane. This operation needs to preserve both the total monodromy and charge conservation. The first condition is ensured by changing the branch cut of each of the passed 7-branes. The second condition fixes the number of $(-p,-q)$ 5-branes ending on the moved $(-p,-q)$ 7-brane.
This can be done in either direction around the web. By explicit computation we arrive at the results in section \ref{sec:Pruning}.

To extract the magnetic quiver, the brane-web is divided into sub-webs that are themselves consistent brane-webs. There can be multiple (maximal) divisions into sub-webs, which are identified with different cones of the Higgs branch. Here, we will only consider a single division. We distinguish between two kinds of sub-webs, which are those that are associated to colors in the GTP, and those that are not but which give rise to tails in the tropical quiver.

For now we will focus on the first class of sub-webs $W^c$, which contain 5-branes that end on two distinct kinds of 7-branes. Each $W^c$ is mapped to a colored sub-polygon $S^c$, as specified in definition \ref{def:Coloring}. An edge segment is part of $S^c$ if the corresponding 5-brane is contained in $W^c$, defining the partition of $\lambda_\alpha$ into $\lambda^c_\alpha$ for each edge. The refined Minkowski sum of two sub-polygons $S^{c_1} \oplus S^{c_2}$ is understood as combining two sub-webs $W^{c_1}$ and $W^{c_2}$, where we identify the 7-branes involved in the $W^c$.

Definition \ref{def:Coloring} contains certain conditions for a coloring of a GTP to be valid, which have their origin in the brane-web: 
\begin{enumerate}
\item Each sub-web $W^c$ gives rise to a sub-polygon $S^c$ because charge conservation in the brane-web ensures closed-ness in the GTP.
\item A sub-web $W^c$ has a multiplicity, i.e. it consists of $m^c$ identical minimal webs $W^c_{\text{min}}$. 
\item {\bf s-rule.} The s-rule states that the number of D5-branes that may connect a single D7-brane and a stack of NS5-branes is less or equal to the number of NS5-branes in the stack. For a given sub-web $W^c$, the $\mu^c_{\alpha,1}$ essentially count the number of NS5-branes in such a configuration, or any $SL(2,\Z)$-transformed junction of this kind.  
Let us consider a minimal web $W^c_{\text{min}}$. It obeys the s-rule if there is a full resolution such that each intersection obeys the s-rule locally. This idea was advanced in \cite{Benini:2009gi} from which the introduction of tiles follows. In earlier work, 
an $SL(2, \Z)$-invariant formulation has already been discussed in \cite{Bergman:1998ej}. Here we will provide a general formula for this, which to our understanding has not appeared thus far. 
We propose that the intersection of three types of branes $(p_i,q_i)$ obeys the s-rule, if 
\be \label{sruleweb}
\left|\det\begin{pmatrix} p_i & q_i \\ p_j & q_j \end{pmatrix}\right| \geq \text{gcd}(p_k,q_k)^2\,,
\ee
for every permutation $(i,j,k)$ of $(1,2,3)$. This is represented by \eqref{sruletiles}. We provide a proof of the formula (\ref{sruleweb}) in the next paragraph. 
\item The division into sub-webs is maximal, i.e. no sub-web can be further subdivided while still satisfying the s-rule. The same must be true for the partitions $\{\lambda^c_\alpha \}$.
\end{enumerate}

\paragraph{Derivation of the s-rule for simple intersections of 5-branes.}
Consider a junction of three 5-branes (possibly with multiplicity) of type $(p_i , q_i)$, for $i=1,2,3$. Let $\delta_i = \mathrm{gcd}(p_i,q_i)$ be the multiplicities of the branes. We want to find a condition on the charges $(p_i , q_i)$ for this intersection to be supersymmetric when all the 5-branes on each of the three legs end on a single 7-brane, as illustrated below:
\be 
\begin{tikzpicture}[
roundnode/.style={circle, draw=black, thick, fill=white, minimum size=4mm},
]
\node[roundnode] at (-3,1) (1) {};
\node[roundnode] at (2,2) (2) {};
\node[roundnode] at (1,-2) (3) {};
\node at (-4,1) {$(p_1,q_1)$};
\node at (3,2) {$(p_2,q_2)$};
\node at (2,-2) {$(p_3,q_3)$};
\draw[thick] (1) -- (0,0) {};
\draw[thick] (2) -- (0,0) {};
\draw[thick] (3) -- (0,0) {};
\end{tikzpicture}\,.
\ee
By B{\'e}zout's identity, there exist two integers $u$ and $v$ such that $u p_1 + v q_1 = \delta_1$. One can then consider the matrix 
\begin{equation}
\left(
\begin{array}{cc}
u & v \\ - \frac{q_1}{\delta_1} &  \frac{p_1}{\delta_1}
\end{array}
 \right) \in \mathrm{SL}(2,\mathbb{Z}) \, . 
\end{equation}
Using the $\mathrm{SL}(2,\mathbb{Z})$ transformation given by multiplying on the right the charges $(p_i , q_i)$ by that matrix, one transforms $(p_1,q_1)$ to $(\delta_1 , 0)$. The other charges $(p_i,q_i)$ (for $i=2,3$) are transformed to \begin{equation}
\label{eqNS5}
\left( u p_i + v q_i , \frac{1}{\delta_1}  \det\begin{pmatrix} p_1 & q_1 \\ p_i & q_i \end{pmatrix} \right)  \,.
\end{equation}
After this transformation, the branes labeled by index $1$ are $\delta_1$ D5 branes which all end on the same D7 brane:  
\be 
\begin{tikzpicture}[
roundnode/.style={circle, draw=black, thick, fill=white, minimum size=4mm},
]
\node[roundnode] at (-2,0) (1) {};
\node[roundnode] at (1,2) (2) {};
\node[roundnode] at (1,-2) (3) {};
\node at (-3,0) {$(\delta_1,0)$};
\node at (4,2) {$\left( u p_2 + v q_2 , \frac{1}{\delta_1}  \det\begin{pmatrix} p_1 & q_1 \\ p_2 & q_2 \end{pmatrix} \right)$};
\node at (4,-2) {$\left( u p_3 + v q_3 , \frac{1}{\delta_1}  \det\begin{pmatrix} p_1 & q_1 \\ p_3 & q_3 \end{pmatrix} \right)$};
\draw[thick] (1) -- (0,0) {};
\draw[thick] (2) -- (0,0) {};
\draw[thick] (3) -- (0,0) {};
\end{tikzpicture}\,.
\ee
Therefore the s-rule says that the number of NS5-branes, given by the absolute value of the second charge in (\ref{eqNS5}), has to be at least $\delta_1$, i.e.
\be
  \frac{1}{\delta_1} \left| \det\begin{pmatrix} p_1 & q_1 \\ p_i & q_i \end{pmatrix}  \right| \geq \delta_1 \, . 
\ee
Using charge conservation, $p_1 + p_2 + p_3 = 0$ and $q_1 + q_2 + q_3 = 0$, one can replace the determinant on the right by the determinant of $(p_2,q_2)$ and $(p_3,q_3)$, which finally gives 
\begin{equation}
 \left|  \det\begin{pmatrix} p_2 & q_2 \\ p_3 & q_3 \end{pmatrix} \right| \geq (\delta_1)^2 \, . 
\end{equation}
The same reasoning is true for any permutation of the indices $(1,2,3)$, giving the rule (\ref{sruleweb}). This formula is clearly $\mathrm{SL}(2,\mathbb{Z})$-invariant, as wanted.

\paragraph{A comment on internal edges.}
Let us briefly comment on the case where a GTP has internal edges, which is the subject of section \ref{sec:IntEdges}. This corresponds to a brane-web containing internal 5-branes, i.e. 5-branes that connect two 5-brane junctions (rather than ending on a 7-brane). Such a set of internal 5-branes necessarily pertains to a collection of sub-webs $\{W^c\}$. There is no obstruction to sending the length of the internal 5-branes to infinity, i.e. the webs connected by the internal 5-branes are independent of each other, apart from a matching condition along the internal 5-branes, and must obey the s-rule locally. This is the reason the s-rule should be applied separately to each $S^c_A$, as detailed in section \ref{sec:IntEdges}.

\subsection{Magnetic Quivers}

What we have seen so far is that a colored GTP contains the same data as the sub-web decomposition of the dual web. 
We will now argue for the derivation of the tropical quiver from GTP data, as explained in section \ref{sec:MQ}, starting from the brane-web.

Each sub-web $W^c$ corresponds to a node in the magnetic quiver, labeled by the multiplicity $m^c$ of the sub-web. The sub-webs are mapped to colorings, or colored sub-polygons $S^c$, in the GTP. To each color we thus associate a node in the tropical quiver with label $m^c=\text{gcd}_\alpha(\lambda^c_\alpha)$.

The intersections between the color nodes are computed as follows:
Consider the color sub-division of the GTP. The area $A^{c_1c_2}$ corresponds to the stable intersection (SI) between the sub-webs $W^{c_1}$ and $W^{c_2}$ in tropical geometry \cite{Cabrera:2018jxt}. Recall, that the SI is defined by infinitesimally displacing the two minimal sub-webs and computing the intersection
\be \label{SI}
\text{SI}=\sum \left|\det\begin{pmatrix} p^{c_1} & q^{c_1}\\ p^{c_2} & q^{c_2} \end{pmatrix}\right|\,,
\ee
where the sum goes over all intersections and the $(p^c,q^c)$ are normalized by a factor of $m^c$. Now consider the parallelograms $G^{c_1c_2}$ in the color sub-divided polygon, with the two sides characterized by $\lambda^{c_1}_\alpha L_\alpha$ and $\lambda^{c_2}_\beta L_\beta$. The Euclidean area of $G^{c_1c_2}$ is then given by
\be\label{AreaG}
\text{Area}(G^{c_1c_2}) = \lambda^{c_1}_\alpha \lambda^{c_2}_\beta \left|\det\left(L_\alpha,L_\beta\right)\right| = \lambda^{c_1}_\alpha \lambda^{c_2}_\beta  \left|\det\begin{pmatrix} q^{c_1} & -p^{c_1}\\ q^{c_2} & -p^{c_2} \end{pmatrix}\right|\,,
\ee
because of \eqref{WebEalpha}. However, here the $(p^c,q^c)$ are normalized to be coprime, so, by the inclusion of the $\lambda^c$, this agrees with the SI in \eqref{SI}.

There is an additional contribution to the intersection between two sub-webs, $W^{c_1}$ and $W^{c_2}$, which is the effective number of 7-branes on which they both end.
To see this contribution in the GTP consider a single edge $E_\alpha$, corresponding to a tail of $(p,q)$ 5- and 7-branes, all of the same $p$ and $q$, in the brane-web.
Recall that the number of 7-branes that an edge $E_\alpha$ gives rise to is $n_7=b_\alpha+1$, and the intermediate segments (which are stacks of 5-branes in the web) are labeled by $x$. 
Consider the 7-brane between segments $x-1$ and $x$. The number of 5-branes in color $c$ on either side differ by $\mu^c_{\alpha, x}$, by definition. The sum over all 7-branes yields the second contribution to the intersection between $W^{c_1}$ and $W^{c_2}$.
However, we need to divide by the product of the multiplicities of the two sub-webs to obtain the effective intersection in the magnetic quiver, given by
\be
-\frac{1}{m^{c_1}m^{c_2}} \sum_{x=1}^{b_\alpha} \mu^{c_1}_{\alpha,x} \mu^{c_2}_{\alpha,x}\,.
\ee
Together with the SI contribution \eqref{AreaG} this gives the number of edges \eqref{kColor} between nodes in the tropical quiver associated to the two colors $c_1$ and $c_2$.

Furthermore, we can determine the number of 5-branes in a tail, which do not belong to any of the $W^c$. These sub-webs $W_{\alpha,x}$, spanning the segment $x$ between 7-branes in a tail labeled by $\alpha$, do not appear as sub-polygons in the GTP. Nonetheless, they do give rise to additional nodes in the magnetic quiver, and their presence can be detected in the GTP. Their multiplicities $m_{\alpha,x}$ are given by the total number of 5-branes along a segment $x$ in the tail $\alpha$, minus the number of 5-branes in that segment that belong to a $W^c$. The former is the total number of 5-branes protruding from the internal web in this direction, $\lambda_\alpha$, minus the number of 5-branes that have ended on a 7-brane further up the tail, $\sum_{y=1}^x \mu_{\alpha, y}$. The computation of the latter is the same but restricted to a coloring $c$.
\footnote{By some abuse of notation we say $\mu^c_{\alpha, y}=0$ if $y$ is larger than the length of the partition $\mu^c_{\alpha, x}$.} Hence, the multiplicities of the sub-webs $W_{\alpha,x}$, in terms of GTP data, are
\be
m_{\alpha,x} = \sum_{y=1}^x \left(- \mu_{\alpha,y} + \sum_{c=1}^{n_c} \mu^c_{\alpha,y} \right)\,,
\ee
which is \eqref{MultHN}.
These sub-webs give rise to the additional tail nodes in the magnetic quiver with labels $m_{\alpha,x}$. They are connected to their nearest neighbors by a single edge, since the only contribution to the intersection between $W_{\alpha,x}$ and $W_{\alpha,x+1}$ comes from ending on opposite sides of the same 7-brane.

Finally, we determine the number of edges between color nodes and tail nodes in the tropical quiver. 
Consider a tail segment $x$ made up of $m_{\alpha, x}$ 5-branes between two 7-branes. The number of 5-branes in color $c$ ending on these two 7-branes is given by $\mu^c_{\alpha, x}$ and $\mu^c_{\alpha, x+1}$ respectively. Each time the number of $c$-colored 5-branes ending on the 7-branes on either side of a segment differs, the corresponding tail node is connected to the $c$-color node. Thus, the intersection, after accounting for multiplicity, is
\be
\frac{1}{m^c}\left(\mu^c_{\alpha, x} - \mu^c_{\alpha, x+1}\right)\,,
\ee
equivalent to \eqref{kAdd}.

\subsection{Hasse Diagram}

\label{sec:HasseWeb}

Now, let us turn to the computation of the Hasse diagrams in section \ref{sec:Hasse}.
Recall that given a magnetic quiver of a theory, a transition in the Hasse diagram corresponds to the quiver subtraction of a symplectic leaf, i.e. either the affine Dynkin diagram of an $ADE$ Lie algebra $\mathfrak{g}$ corresponding to the magnetic quiver of the closure of a minimal nilpotent orbit, or a Kleinian singularity -- these are summarized in table \ref{tab:SymSing}. 
It was argued in \cite{Bourget:2019aer, Bourget:2019rtl} that this process can be translated into the language of brane-webs as follows. Consider a brane-web $W_P$, corresponding to a theory with some original magnetic quiver. A quiver subtraction by a symplectic leaf \footnote{We assume for notational simplicity that the singularity is of $ADE$ type, the logic holds in the Kleinian case.} $\mathfrak{g}$ corresponds to turning on certain Coulomb branch parameters of the theory, i.e. the introduction of internal 5-branes in the brane-web, resulting in a new brane-web $W_Q$. The Coulomb branch parameters must be turned on in such a way that, in a maximal sub-division of $W_Q$, all the newly introduced 5-branes belong to the same sub-web $W_{\Delta_\nu}$. At the origin of its Coulomb branch (we could call this brane-web $W_{\Delta}$ in parallel with the notation in section \ref{sec:IntEdges}), the magnetic quiver of this sub-web is the affine Dynkin diagram of $\mathfrak{g}$. By going into the Coulomb branch of this sub-theory, the associated Higgs branch reduces to $\mathfrak{g} \to \mathfrak{u}(1)$. In the magnetic quiver this corresponds to a quiver subtraction with subsequent rebalancing with a $\mathfrak{u}(1)$ node.

Now, we can further translate this into the language of GTPs. Turning on Coulomb branch parameters corresponds to a partial resolution of the GTP $P$, i.e. the introduction of internal edges $E_\beta$ of length $\nu_\beta$, dual to the newly introduced internal 5-branes in the web, which results in a new GTP $Q$. The internal 5-branes belong to a sub-web $W_{\Delta_\nu}$, corresponding to a color sub-polygon $\Delta_\nu$ in $Q$, which includes the internal edges. To describe the Higgs branch at a generic point on the Coulomb branch, the magnetic quiver of $\Delta_{\nu}$ should be a single $U(1)$ node. However, at the origin of the Coulomb branch, i.e. in the absence of the $E_\beta$ where the GTP is $\Delta$, the magnetic quiver correspond to $\mathfrak{g}$.

\begin{table}
\centering
\begin{tabular}{| c | c|}
 \hline
$(p,q)$ 5-Brane-web/Tropics & Generalized toric polygon\\
\hline\hline
$n_5$ coincident $(p,q)$ 5-branes & Edge $E_\alpha$ with $L_\alpha=(q,-p)$ and $\lambda_\alpha=n_5$\\
\hline
External and internal 5-branes & External edges $E^\p$ and internal edges $E^{\text{in}}$\\
\hline
$n_5$ 5-branes subsequently ending on $n_7$ 7-branes & Partition $\mu_{\alpha, x}$ of length $b_\alpha=n_7-1$\\
\hline
Compact and non-compact faces & Black vertices $V_b$\\
\hline
Division into sub-webs & Coloring\\
\hline
Internal sub-web $W^c$ & Colored subdiagram $S^c$\\
\hline
Partition of coincident 5-branes into sub-webs & $\lambda_\alpha = \sum \lambda_\alpha^c$\\
\hline
Multiplicity of $W^c$ & $m^c$\\
\hline
S-rule, $\#$ branes of $W^c$ ending on each 7-brane &  $\mu^c_{\alpha, x}$\\
\hline
Stable Intersection & Mixed Volume $MV(S^{c_1},S^{c_2})$\\
\hline
Multiplicities of sub-webs between 7-branes & $m_{\alpha, x}$\\
\hline
5-brane displacement that is quiver subtraction & Transition $P \xrightarrow{\Delta} Q$\\
\hline
Coulomb branch parameters & Partial resolution $\nu_\beta$\\
\hline
\end{tabular}
\label{test}
\caption{Dictionary between brane-webs/tropics and colored GTPs: sub-webs, magnetic quivers and Hasse diagram.}
\end{table}

\subsection{Example}

As an instructive example let us consider $SU(4)_\frac{3}{2} + 1 \bm{AS} + 7 \bm{F}$ at strong coupling, which was discussed in detail in section \ref{sec:ExAS} from the GTP point of view. The brane-web for this theory was discussed in \cite{Bergman:2015dpa} and is given by
\be
\begin{tikzpicture}[
roundnode/.style={circle, draw=black, thick, fill=white, minimum size=4mm},
]

\node (1) {};
\node[roundnode] at ($(1)+(-1.5,0)$) (2) {};
\node[roundnode,label=above:{$(-1,0)$}] at ($(1)+(-3,0)$) (3) {};
\node[roundnode] at ($(1)+(-1,1)$) (4) {};
\node[roundnode,label=left:{$(-1,1)$}] at ($(1)+(-2,2)$) (5) {};
\node[roundnode,label=right:{$(0,-1)$}] at ($(1)+(0,-1.5)$) (6) {};
\node[roundnode] at ($(1)+(1.5,0)$) (7) {};
\node[roundnode] at ($(1)+(3,0)$) (8) {};
\node[roundnode] at ($(1)+(4.5,0)$) (9) {};
\node[roundnode] at ($(1)+(6,0)$) (10) {};
\node[roundnode] at ($(1)+(7.5,0)$) (11) {};
\node[roundnode] at ($(1)+(9,0)$) (12) {};
\node[roundnode,label=below:{$(1,0)$}] at ($(1)+(10.5,0)$) (13) {};

\draw[thick] ($(1)$) -- (2) node[midway,below] {3};
\draw[thick] (2) -- (3) node[midway,below] {1};
\draw[thick] ($(1)$) -- (4) node[midway,above right] {4};
\draw[thick] (4) -- (5) node[midway, above right] {2};
\draw[thick] ($(1)$) -- (6) node[midway,right] {4};
\draw[thick] ($(1)$) -- (7) node[midway,above] {7};
\draw[thick] (7) -- (8) node[midway,above] {6};
\draw[thick] (8) -- (9) node[midway,above] {5};
\draw[thick] (9) -- (10) node[midway,above] {4};
\draw[thick] (10) -- (11) node[midway,above] {3};
\draw[thick] (11) -- (12) node[midway,above] {2};
\draw[thick] (12) -- (13) node[midway,above] {1};

\end{tikzpicture}\,,
\ee
where we state the number of 5-branes and the type of 7-branes explicitly. There are no internal 5-branes, so the dual GTP does not have internal edges. There are four different types of 7-branes in the web, corresponding to four external edges in the GTP. From \eqref{WebEalpha} we can read off that the GTP has edges
\be
L_\alpha = ((-1,-1),(0,-1),(1,0),(0,1))\,, \qquad \lambda_\alpha =(4,3,4,7)\,,
\ee
where $\lambda_\alpha$ is the number of 5-branes protruding from the central junction. From the number of 7-branes in each tail we find
\be 
	b_\alpha=(1,1,0,6)\,.
\ee
Furthermore, we can read off the $\mu_{\alpha,x}$ as the number of 5-branes ending on each of the 7-branes. For example, for the first edge (with the $(-1,1)$-branes), two of the four 5-branes end on each of the two 7-branes, i.e. $\mu_{1,x}=\{2,2\}$. Conversely, for the second edge, of the three 5-branes two and one end on the two 7-branes respectively, so $\mu_{2,x}=\{2,1\}$. Using this we can easily determine the GTP.

Now, let us determine the division of the brane-web into sub-webs. This is a simple exercise and we arrive at
\be 
\begin{tikzpicture}[
roundnode/.style={circle, draw=black, thick, fill=white, minimum size=4mm},
]

\node (1) {};
\node[roundnode] at ($(1)+(-1.5,0)$) (2) {};
\node[roundnode,label=above:{$(-1,0)$}] at ($(1)+(-3,0)$) (3) {};
\node[roundnode] at ($(1)+(-1,1)$) (4) {};
\node[roundnode,label=left:{$(-1,1)$}] at ($(1)+(-2,2)$) (5) {};
\node[roundnode,label=right:{$(0,-1)$}] at ($(1)+(0,-1.5)$) (6) {};
\node[roundnode] at ($(1)+(1.5,0)$) (7) {};
\node[roundnode] at ($(1)+(3,0)$) (8) {};
\node[roundnode] at ($(1)+(4.5,0)$) (9) {};
\node[roundnode] at ($(1)+(6,0)$) (10) {};
\node[roundnode] at ($(1)+(7.5,0)$) (11) {};
\node[roundnode] at ($(1)+(9,0)$) (12) {};
\node[roundnode,label=below:{$(1,0)$}] at ($(1)+(10.5,0)$) (13) {};

\draw[blue,thick] ($(1)+(0,-1pt)$) -- ($(2.east)+(0,-1pt)$) node[midway,below] {3};
\draw[thick] (2) -- (3) node[midway,below] {1};
\draw[cyan,thick] ($(1)+(0,1pt)$) -- ($(4.south east)$) node[midway,above right] {4};
\draw[thick] (4) -- (5) node[midway, above right] {2};
\draw[cyan,thick] ($(1)+(0,1pt)$) -- ($(6.north)$);
\draw[cyan,thick] ($(1)+(0,1pt)$) -- ($(7.west)+(0,1pt)$);
\draw[blue,thick] ($(1)+(0,-1pt)$) -- ($(7.west)+(0,-1pt)$);
\draw[thick] (7) -- (8) node[midway,above] {6};
\draw[thick] (8) -- (9) node[midway,above] {5};
\draw[thick] (9) -- (10) node[midway,above] {4};
\draw[thick] (10) -- (11) node[midway,above] {3};
\draw[thick] (11) -- (12) node[midway,above] {2};
\draw[thick] (12) -- (13) node[midway,above] {1};

\end{tikzpicture}\,,
\ee
where we colored the sub-webs in blue $W_b$, and cyan $W_c$. This is equivalent to the coloring of the GTP shown in \eqref{Ex2Coloring}. The number of 5-branes coming out of the central junction are
\be 
	\lambda^b_\alpha=(0,3,0,3)\,, \qquad \lambda^c_\alpha=(4,0,4,4)\,.
\ee
In this case the application of the s-rule is very simple, since the sub-webs are just (stacks of) a fundamental D5-NS5-brane junction and a D5-brane, so both sub-webs end on the first 7-branes they encounter. For example, we have $\mu^b_{4,x}=\{ 3\}$ and $\mu^c_{4,x}=\{ 4\}$, which contributes exactly 
\be
	-\frac{1}{m^b m^c} \sum_{x=1}^6 \mu^b_{4,x}\mu^c_{4,x}=-\frac{1}{3 \cdot 4} \left( 3 \cdot 4 +0 \cdots +0 \right)=-1
\ee
to the intersection of $W^b$, $W^c$. We can easily see that the stable intersection between the two colored sub-webs is one, after accounting for multiplicity, so they actually do not intersect in the magnetic quiver.
Since $W^b$ and $W^c$ both end on the first 7-branes they encounter, the multiplicities $m_{\alpha,x}$ of the $W_{\alpha,x}$ sub-webs, coming from the 5-branes suspended between two identical kinds of 7-branes, are not reduced, e.g. we have $m_{4,x}=\{6,5,4,3,2,1\}$. Finally, the colored sub-webs intersect $W_{\alpha,x}$ if a different number of colored branes end on the 7-branes on either side of $W_{\alpha,x}$. This is true e.g. for $W^c$ and $W_{4,1}$, since four cyan branes end on the 7-brane to the left of $W_{4,1}$ and none end on the 7-brane to the right. Following the usual rules we can quickly check that the magnetic quiver of this web agrees with \eqref{Ex2MQ}.

Let us now consider the first transition in the Hasse diagram $P \xrightarrow{\Delta} Q$. From the magnetic quiver it is clear that we can subtract an $e_6$ singularity. The corresponding Coulomb branch deformation in the web is given by
\be
\label{BraneHasse}
\begin{tikzpicture}[
roundnode/.style={circle, draw=black, thick, fill=white, minimum size=4mm},
]

\node (1) {};
\node[roundnode] at ($(1)+(-2,0)$) (2) {};
\node[roundnode] at ($(1)+(-3.5,0)$) (3) {};
\node[roundnode] at ($(1)+(-1.5,1.5)$) (4) {};
\node[roundnode] at ($(1)+(-2.5,2.5)$) (5) {};
\node[roundnode] at ($(1)+(0,-1.5)$) (6) {};
\node[roundnode] at ($(1)+(2,0)$) (7) {};
\node[roundnode] at ($(1)+(3.5,0)$) (8) {};
\node[roundnode] at ($(1)+(5,0)$) (9) {};
\node[roundnode] at ($(1)+(6.5,0)$) (10) {};
\node[roundnode] at ($(1)+(8,0)$) (11) {};
\node[roundnode] at ($(1)+(9.5,0)$) (12) {};
\node[roundnode] at ($(1)+(11,0)$) (13) {};

\draw[thick] ($(7.west)+(0,1pt)$) -- ($(2.east)+(0,1pt)$) node[near end,above] {1} node[near start,above] {3};
\draw[thick] ($(4.north west)+(1pt,1pt)$) -- ($(5.south east)+(1pt,1pt)$) node[midway,right] {1};
\draw[thick] ($(7.east)+(0,1pt)$) -- ($(8.west)+(0,1pt)$) node[midway,above] {3};
\draw[thick] ($(8.east)+(0,1pt)$) -- ($(9.west)+(0,1pt)$) node[midway,above] {3};
\draw[thick] ($(9.east)+(0,1pt)$) -- ($(10.west)+(0,1pt)$) node[midway,above] {3};
\draw[thick] ($(10.east)+(0,1pt)$) -- ($(11.west)+(0,1pt)$) node[midway,above] {3};
\draw[thick] ($(11.east)+(0,1pt)$) -- ($(12.west)+(0,1pt)$) node[midway,above] {2};
\draw[thick] ($(12.east)+(0,1pt)$) -- ($(13.west)+(0,1pt)$) node[midway,above] {1};
\draw[thick] ($(4.south east)+(1pt,1pt)$) -- ($(4)+(1.5,-1.5)+(1pt,1pt)$) node[near start,right] {2};
\draw[thick] ($(6.north)+(1pt,0)$) -- ($(4)+(1.5,-1.5)+(1pt,1pt)$) node[near start,right] {2};

\draw[blue,thick] ($(2.west)+(0,-1pt)$) -- ($(3.east)+(0,-1pt)$) node[midway,below] {1};
\draw[blue,thick] ($(7.east)+(0,-1pt)$) -- ($(8.west)+(0,-1pt)$) node[midway,below] {3};
\draw[blue,thick] ($(8.east)+(0,-1pt)$) -- ($(9.west)+(0,-1pt)$) node[midway,below] {2};
\draw[blue,thick] ($(9.east)+(0,-1pt)$) -- ($(10.west)+(0,-1pt)$) node[midway,below] {1};
\draw[blue,thick] ($(2.east)+(0,-1pt)$) -- ($(2)+(1,-1pt)$) node[midway,below] {2};
\draw[blue,thick] ($(2)+(1,-1pt)$) -- ($(6)+(-1pt,1)$) node[midway,below] {1};
\draw[blue,thick] ($(7)+(-1,-1pt)$) -- ($(6)+(-1pt,1)$) node[midway,below] {1};
\draw[blue,thick] ($(7)+(-1,-1pt)$) -- ($(7.west)+(0,-1pt)$) node[midway,below] {4};
\draw[blue,thick] ($(6.north)+(-1pt,0)$) -- ($(6)+(-1pt,1)$) node[midway,left] {2};
\draw[blue,thick] ($(4.south east)+(-1pt,-1pt)$) -- ($(4)+(.5,-.5)+(0,-1pt)$) node[midway,left] {2};
\draw[blue,thick] ($(2)+(1,-1pt)$) -- ($(4)+(.5,-.5)+(0,-1pt)$) node[midway,left] {1};
\draw[blue,thick] ($(7)+(-1,-1pt)$) -- ($(4)+(.5,-.5)+(0,-1pt)$) node[midway,above] {1};
\draw[blue,thick] ($(4.north west)+(-1pt,-1pt)$) -- ($(5.south east)+(-1pt,-1pt)$) node[midway,left] {1};

\end{tikzpicture}\,,
\ee
where the $e_6$ sub-web ($W_{\Delta_\nu}$) is shown in blue. We arrive at this 5-brane deformation by first considering the quiver subtraction in the magnetic quiver, from which we deduce how many of each of the 5-branes should contribute to the $e_6$ sub-web. In a second step we need to find the Coulomb branch deformation that enforces this sub-web. Although there might seem to be a lot of possibilities, the s-rule constrains the setup such that this is only allowed deformation.
We can read off the GTP after the transition, i.e. $Q$ in \eqref{Ex2HasseE6}, which is the dual diagram of \eqref{BraneHasse}. It is obtained from $P$ by adding the internal edges dual to the internal 5-branes in \eqref{BraneHasse}. Similarly, the GTP $\Delta$, indicating the subtracted symplectic leave is dual to the blue $e_6$ sub-web at the origin of the Coulomb branch.

\subsubsection*{Acknowledgements}
We thank Mohammad Akhond, Fabio Apruzzi, Cyril Closset, Eric Fusy, Simone Giacomelli, Julius Grimminger, Amihay Hanany, Hirotaka Hayashi, Dave Morrison, Yuji Tachikawa, Futoshi Yagi, Yi-Nan Wang and Zhenghao Zhong for discussions. The work of SSN and JE, and in part MvB is supported by the ERC
Consolidator Grant number 682608 ``Higgs bundles: Supersymmetric Gauge Theories and
Geometry (HIGGSBNDL)". The work of AB is supported by the STFC Consolidated Grant  ST/P000762/1. SSN acknowledges support also from the Simons Foundation.


\newpage

\appendix

\section{Recap: Webs, Magnetic Quivers and Hasse Diagrams}
\label{app:Webs}

In this section we quickly review the concept of $(p,q)$ 5-brane-webs and how to read off the magnetic quiver and Hasse diagram of the corresponding theory. A $(p,q)$ 5-brane-web is a type IIB configuration consisting of 5-branes, extended along the $(x_0,\dots,x_4)$-directions and at an angle $\theta$ in the $(x_5,x_6)$-direction, and 7-branes, extended along the $(x_0,\dots,x_4,x_7,x_8,x_9)$-directions. To preserve eight supercharges the configuration needs to obey
\begin{enumerate}
\item $(p,q)$ 5-branes lie at angles\footnote{For this we choose $\tau_{\text{IIB}}=i$ for simplicity.}
\be
\tan \theta = \frac{p}{q}\,.
\ee
\item $(p,q)$ 5-branes end on corresponding $(p,q)$ 7-branes.
\item At each junction, where multiple 5-branes meet, charge conservation implies
\be
\sum_i \left(p_i,q_i\right) = (0,0)\,,
\ee
where we take the orientation such that all branes are outgoing.
\item
The web obeys the s-rule. We discuss the implications in detail around \eqref{sruleweb}.
\end{enumerate}
Each such $(p,q)$ 5-brane-web describes a 5d $\mathcal{N}=1$ theory. Reviewing the connection between these two descriptions is beyond the scope of this appendix, however, the brane-webs corresponding to $SU(N)$ quiver gauge theories with fundamental and antisymmetric matter, which are relevant in this paper, are given in \cite{Bergman:2015dpa,Zafrir:2015rga,Cabrera:2018jxt}.

The method to compute the (unitary) magnetic quiver of a brane-web was developed in \cite{Cabrera:2018jxt}. The idea is to divide the web into a (maximal) set of sub-webs that themselves are consistent, supersymmetric brane-webs. In type IIB this is justified by displacing these sub-webs along the $(x_7,x_8,x_9)$-directions, which make up the 5d $\mathcal{N}=1$ R-symmetry. Each different division into sub-webs produces a magnetic quiver and corresponds to a distinct component in the Higgs branch.
Given a division into sub-webs, the magnetic quiver is computed as follows
\begin{enumerate}
\item 
A sub-web corresponds to a $U(1)$ node. If there are $m$ equivalent (i.e. identical and coincident) sub-webs this is enhanced to $U(m)$.
\item
The number of bifundamental matter multiplets, i.e. intersections between the nodes, is determined by two contributions:
\begin{enumerate}
\item The stable intersection number is determined by regarding the 5-branes of the sub-webs as tropical curves, and infinitesimally displacing the curves. Then, in each point where 5-branes from the two sub-webs intersect there is a contribution to the stable intersection
\be
\left|\det\begin{pmatrix} p_1 & q_1 \\ p_2 & q_2 \end{pmatrix}\right|\,,
\ee
where the intersecting branes are $(p_1,q_1)$ and $(p_2,q_2)$. The sum over these contributions is independent of the choice of displacement.
\item The second contribution comes from 7-branes. If two 5-branes belonging to different sub-webs end on opposite sides of a 7-brane, there is a contribution of $+1$, if they end on the same side of a 7-brane, there is a contribution of $-1$.
\end{enumerate}
The sum over all these contributions counts the intersection between two unitary nodes.
\end{enumerate}

Finally, let us look at the Hasse diagram, which can be seen from the repeated quiver subtraction of symplectic leaves in the magnetic quiver. The balance of the $I$'th node in the magnetic quiver is given in terms of the node labels $U(m_J)$ and the edge multiplicities $k_{IJ}$ as
\be
\beta_I = -2m_I+\sum_{J\neq I}k_{IJ}m_J+2\ell_I(m_I-1)\,, 
\ee
where $J$ runs over all nodes in the magnetic quiver.
In a consistent magnetic quiver all balances are non-negative. We can subtract two quivers $Q^+$ and $Q^-$, if all their edge multiplicities $k_{IJ}$ agree \footnote{We can label the quiver nodes in any way. If $Q^-$ has fewer nodes than $Q^+$ this condition is only required for the edges of $Q^-$.} and $m_I^+-m_I^->0$ for all nodes. Then, we compute the resulting quiver $Q=Q^+-Q^-$,
\begin{enumerate}
\item The node multiplicities are $m_I=m_I^+-m_I^-$ and the edge multiplicities are $k_{IJ}=k_{IJ}^+$.
\item To rebalance we add an additional $U(1)$ node and connect it such that the balances of all nodes with non-negative multiplicity retain the balance $\beta_I=\beta_I^+$.
\end{enumerate}
The Hasse diagram is obtained by repeatedly subtracting symplectic leaves. These can be affine Dynkin diagrams of an ADE type Lie algebra $\mathfrak{g}$ or a Kleinian singularity $A_{N-1}$. We summarize them in table \ref{tab:SymSing}. Note that the quaternionic dimension of the Higgs branch (at a given point in the Hasse diagram) is given by
\be
\dim_{\mathbb{H}}(\mathcal{H})=-1+\sum_{I} m_I\,,
\ee
and the flavor symmetry of a theory is given by the lowest-lying leave in the Hasse diagram.

\section{Mathematica Code}
\label{app:MMA}

For the convenience of the reader, we attach to this paper a notebook written with {\tt{Mathematica 11.3}}. The notebook contains a collection of auxiliary routines which implement the various steps of the algorithm outlined in section \ref{sec:ColoringRules}, with one crucial exception regarding the s-rule.

The algorithm presented in the main text is a general prescription to derive the magnetic quiver. However, in general it is not known how to generate all possible resolutions, in order to check the s-rule of a given diagram within a reasonable time. To circumnavigate this, we propose here to use a slightly stronger version of the s-rule, which will be described below. The strong s-rule in general will not find all colorings, however we conjecture that those satisfying the strong s-rule ({\it strong colorings}), also satisfy the s-rule. The main advantage of this strong s-rule is that it can be efficiently implemented. 
We have tested our algorithm with success on all the theories studied in \cite{Cabrera:2018jxt} and on several families of theories (for low $N$) from \cite{Akhond:2020vhc}, as well as a large class of $SU(N)$ models with $\bm{AS}$ and $\bm{F}$ \cite{BBESII}. However, we repeat:

\paragraph{Disclaimer.} The {\tt{Mathematica}} code will in general only find a subset of the colorings, i.e. of the magnetic quivers, namely those satisfying the strong s-rule. This is indicated in the output by the label {\it strong coloring}. \\

In this appendix, we first explain how to use the notebook and read the results it produces, illustrating with the example of section \ref{sec:ColoringRules}. 
Then we discuss the notion of the \emph{strong s-rule}, which in the code replaces the s-rule of Definition \ref{DefinitionSrule}. We explain the reasons for this substitution, and how this substitution promotes the pruning operation to a more important role.

\subsection{User Guide to the Notebook}

\begin{figure}
\begin{center}
\includegraphics[width=\textwidth]{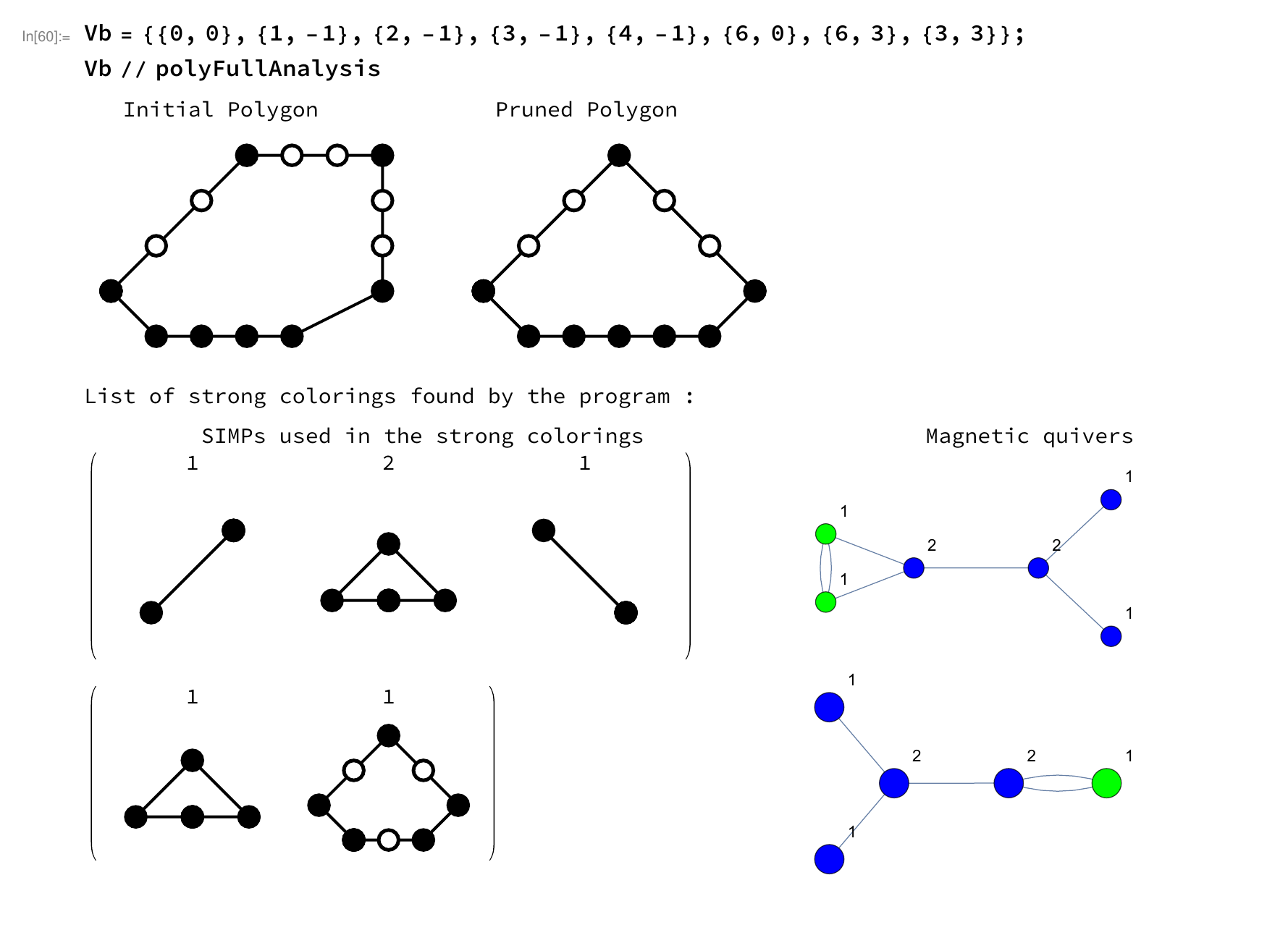}
\end{center}
\caption{Screenshot of the execution of the \texttt{polyFullAnalysis} function on the polygon (\ref{polyMathematica}).  }
\label{figMathematica}
\end{figure}

The companion notebook is divided into two sections. The first section contains all the functions that implement the algorithm. Details concerning some of these functions are given in comments inside the code. The most important function is called \texttt{polyFullAnalysis}. It takes a (not necessarily strictly) convex polygon as its argument. The polygon is encoded as a list of pairs of integers, which are the coordinates of the black vertices, given in counter clockwise order. For instance the polygon (\ref{MainExStart}) is encoded in {\tt{Mathematica}} by the list 
\begin{equation}
\label{polyMathematica}
\texttt{Vb = \{\{0,0\},\{1,-1\},\{2,-1\},\{3,-1\},\{4,-1\},\{6,0\},\{6,3\},\{3,3\}\}} \, . 
\end{equation}
Note that only the black vertices are listed, and that every vertex is listed exactly once. The starting point is irrelevant, as is the choice of coordinates for the first vertex (the result is not affected by a global translation). With the \texttt{draw} function it is possible to obtain a drawing of the polygon. Executing the functions \texttt{polyFullAnalysis} on the polygon \texttt{Vb} prints several graphs, reproduced in Figure \ref{figMathematica}, which we now discuss. 

The first diagram gives a drawing of the polygon given as an input. It distinguishes between black and white dots. The second diagram (which may or may not be identical to the first) is a "pruned" version of the initial polygon (see section \ref{app:pruning}). For the remaining part of the algorithm, it is this pruned version which is considered. 

Below appears a table which lists the strong colorings of the pruned polygon found by the program. We warn the reader that due to the algorithmic complications discussed in the next subsections, there are instances where this list might not be exhaustive (see below). Each strong coloring is represented by a line in the table. In each line: 
\begin{itemize}
\item The first column describes the strong coloring by listing the SIMPs (these are the analogs of the IMPs of Definition \ref{DefinitionSrule}, where the s-rule has been replaced by the strong s-rule of Definitions \ref{def:strongsruletriangles} and \ref{def:strongsrulemultigons}), the number above each tile denoting its multiplicity. It can be checked that the two strong colorings in Figure \ref{figMathematica} correspond to the two colorings identified in equations (\ref{MainExColor1}) and (\ref{MainExColor2}). 
\item The second column of the table gives the magnetic quiver associated to the strong coloring. For convenience, the balanced nodes in the magnetic quivers are painted in blue. The other nodes are painted in green. 
\item The last column gives the adjacency matrix of the magnetic quiver. This is a symmetric matrix where each column corresponds to a gauge node; the rank of the gauge node is given on the diagonal, while the connectivity with the other nodes is given outside the diagonal. In some rare occurrences, the magnetic quiver drawn in the second column is very symmetric and some of the links might be difficult to distinguish on the graph. In those cases, one should check the adjacency matrix, which is unambiguous. 
\end{itemize}

\subsection{The Strong s-Rule}

We now come to the caveat discussed above. While it is easy to check whether a given polygon is a tile in the sense of Definition \ref{def:Tiles}, to our knowledge, there is no efficient algorithm to check whether a given GTP obeys the s-rule as defined by Definition \ref{DefinitionSrule}. In order to overcome this obstacle, we decided to implement a slightly stronger version of the s-rule, which is discussed below. We conjecture that any polygon which obeys the strong s-rule also obeys the s-rule, but the converse is not true (see example below). As a consequence, the algorithm implemented in the notebook might incorrectly discard some colorings, and therefore give an incomplete list of magnetic quivers for a given polygon. How frequent this situation appears to be is the topic of the next subsection. 

The strong s-rule is defined as follows. We call \emph{multigon} a polygon with at least four edges. A GTP is either a triangle or a multigon, and in order to be as close as possible to the s-rule, the definition of the strong s-rule is not the same for triangles and for multigons. 

{\definition[Strong s-rule for triangles]
\label{def:strongsruletriangles}
Let $T$ be a triangle GTP without internal edges, defined by edges $E_\alpha = \lambda_\alpha L_\alpha$ and partitions $\{\mu_{\alpha}\}$. We say that $T$ satisfies the strong s-rule minimally if the partitions are maximal, $\{\mu_{\alpha}\} = \{\lambda_{\alpha}\}$ (i.e. the interior of the edges contains only white dots) and 
\begin{equation}
\lambda_\alpha \lambda_\beta |\mathrm{det} (L_\alpha , L_\beta)| \geq (\lambda_\gamma)^2
\end{equation}
for all $\alpha \neq \beta \neq \gamma \neq \alpha$. Thus, the s-rule and the strong s-rule agree on these triangles.
}

For multigons, the intuition behind the strong s-rule resides in the notion of \emph{depth} defined below. Basically, an edge containing white dots can not belong to a polygon which is not deep enough in the direction transverse to this edge, as in the dual web (and after suitable $\mathrm{SL}(2,\mathbb{Z})$ transformation) this would mean there are not enough NS5 branes to end all the D5 branes coming from the same D7 brane. The difficulty in formalizing this intuition comes from the fact that it is non-local: the white dots on the other edges need to be taken into account to compute the depth of a GTP in a given direction. This gives rise to the notion of \emph{strong depth}. 

{\definition[Strong s-rule for multigons]
\label{def:strongsrulemultigons}
Let $T$ be a GTP without internal edges, defined by edges $E_\alpha = \lambda_\alpha L_\alpha$ and partitions $\{\mu^{\alpha}\}$. We assume that $T$ is a multigon. Along each edge the number of black and white nodes are $b_\alpha$ and $w_\alpha=\lambda_\alpha-b_\alpha-1$ respectively. The depth of an edge $E_\alpha$ with respect to $E_\beta$ is
\be
\label{depth}
d_{\alpha;\beta}=\left|\det\left(L_\alpha,\sum_{\gamma=\alpha}^\beta L_\gamma\right)\right|\,,
\ee
where the sum is extended cyclically. From this we define the depth of $E_\alpha$ as
\be
d_\alpha = \max_\beta d_{\alpha;\beta}\,.
\ee
Furthermore, the edge $E_{\beta^\star}$ with maximal depth from $E_\alpha$ is determined by
\be
\beta^\star_\alpha =  \text{argmax}_\beta\ d_{\alpha;\beta}\,.
\ee
We can check that $\beta^\star_\alpha$ is unique unless $T$ contains an edge with $L_\beta=-L_\alpha$. In this case, any choice of $\beta^\star$ yields the same result.
Next, define the quantities
\be
f^+_\alpha=\sum_{\gamma=\alpha}^{\beta^\star_\alpha} w_\gamma \left|\det\left(L_\alpha,L_\gamma\right)\right|\,, \qquad
f^-_\alpha=\sum_{\gamma=\beta^\star_\alpha}^{\alpha} w_\gamma \left|\det\left(L_\alpha,L_\gamma\right)\right|\,,
\ee
which differ only by the details of the summation. Then we define the strong depths
\be
\label{strongDepth}
s_\alpha = d_\alpha - \max_{\pm} f^{\pm}_\alpha\,, \qquad k_\alpha=\left\lfloor\frac{\lambda_\alpha}{\mu_\alpha}\right\rfloor\,, \quad r_\alpha= \lambda_\alpha \mod \mu_\alpha\,.
\ee
We say that $T$ obeys the strong s-rule minimally if,
\be
\label{strongSrule}
\{\underbrace{s_\alpha\dots,s_\alpha}_{k_\alpha},r_\alpha\} = \{\mu_{\alpha}\}\,,
\ee
for all $E_\alpha$.
}

As before, we say that a GTP $T$ is irreducible if there is not decomposition
\be
T= T_1 \oplus T_2\,,
\ee
such that $T_1$ and $T_2$ obey the \emph{strong} s-rule.

\paragraph{SIMPs} 
Recall that a GTP which is irreducible with respect to the s-rule and satisfies the s-rule minimally was called an IMP (see Definition \ref{DefinitionSrule}). Similarly, a GTP which is irreducible with respect to the strong s-rule and satisfies the strong s-rule minimally will be called a SIMP (for Strong Irreducible and Minimal Polygon). 

\paragraph{The weakness of the strong s-rule}
The strong s-rule suffers from several weaknesses. One obvious weakness is the fact that the formula for the strong depth (\ref{strongDepth}) depends only on the \emph{number} of white dots on the other, and not on the way they are spaced on the edge. A related point, which should not come as a surprise, is that for multigons the s-rule and strong s-rule are not equivalent. This is made manifest by the example below: 
\be \label{ExNonEquivalence}
\begin{tikzpicture}[x=.8cm,y=.8cm]
\draw[step=.8cm,gray,very thin] (0,0) grid (2,3);
\draw[ligne](0,0)--(1,0)--(2,3)--(0,3)--(0,0);
\node[bd] at (0,0) {}; 
\node[bd] at (1,0) {}; 
\node[bd] at (2,3) {}; 
\node[bd] at (0,3) {}; 
\node[bd] at (0,2) {}; 
\node[wd] at (1,3) {}; 
\node[wd] at (0,1) {}; 
\end{tikzpicture}
\ee
One can easily check that this polygon satisfies the s-rule by finding a consistent tiling. However it does not satisfy the strong s-rule, as the vertical edge on the left has depth $d_\alpha = 2$, strong depth $s_\alpha = 1$ and $\{1,1,1\} \neq \{2,1\}$, violating the condition (\ref{strongSrule}). 

Despite these caveats, the strong s-rule presents some obvious advantages, most notably the fact that it is efficiently implemented algorithmically, and this is why we chose to include it on our code. Because of this, the pruning operation acquires a much more central importance, and is the topic of the next subsection. 

\subsection{Pruning}
\label{app:pruning}

The definition of the strong s-rule is not invariant under the pruning operation introduced in section \ref{sec:Pruning}.  
Empirically we find that the accuracy of the strong s-rule improves, and in many cases yields the same result as the s-rule, when the polygon has been pruned, prior to computing the coloring. We now give the precise definition corresponding to the intuition given in section \ref{sec:Pruning}. 

{\definition[Pruning] \label{def:pruning}
Let $P$ be a GTP with edges $E_\alpha = \lambda_\alpha L_\alpha$ and partitions $\mu_{\alpha,i}$. A pruning of $P$ along $\mu_{\alpha,j}$ in direction $s=\pm1$ defines a new GTP
\be
P'=(P,\mu_{\alpha,j},s)\,,
\ee
as follows.
Divide the edges $E$ of $P$ into two parts $E_\pm$ with
\be
\pm \det(L_\alpha,L_\beta) > 0 \qquad \forall L_\beta \in E_\pm\,.
\ee
$E_\alpha$ belongs to neither of the $E_\pm$ and we choose $-L_\alpha\in E_+$.\footnote{Note that the definition does not depend on this choice.}
Then, the edges $E'$ of $P'$ are given by
\begin{enumerate}
\item $E_{s}$
\item $\{\mu_{\alpha,1},\dots,\widehat{\mu_{\alpha,j}},\dots,\mu_{\alpha,{b_\alpha+1}}\} L_\alpha$
\item $\bigcup_{E_\beta \in E_{-s}}\{\mu_{\beta,i}\} \left(L_\beta + \left|\det(L_\alpha,L_\beta)\right| L_\alpha\right)$
\item $\left(-\mu_{\alpha,j} + \sum_{E_\beta \in E_{-s}} \left|\det(L_\alpha,L_\beta)\right| \lambda_\beta\right) \left(-L_\alpha\right)$\,.
\end{enumerate}
If $\left(-\mu_{\alpha,j} + \sum_{E_\beta \in E_{-s}} \left|\det(L_\alpha,L_\beta)\right| \lambda_\beta\right)\leq0$ the pruning is disallowed.
}
\\
\\
In the interest of gaining some intuition for how the slopes $L_\beta$ change in a pruning, consider a simple example, where we move an edge segment $L_\alpha=(0,-1)$ from the vertical edge on the left to the right.\footnote{Note that this setup can always be achieved by global $SL(2,\Z)$-invariance.} The crucial constraint is that the GTP still has to close after this movement. Then, $E_\pm$ represent the lower/upper part of the GTP and we can keep one of these, $E_s$ fixed. However, the slopes for all edges in $E_{-s}$ map to
\be
L_\beta=(x_\beta,y_\beta) \to (x_\beta,y_\beta-|x_\beta|)\,,
\ee
i.e. they are tilted towards $L_\alpha$. Note that the total horizontal extension does not change, whereas the vertical one increases. This ensures that we can add an edge segment along $(-L_\alpha)=(0,1)$ such that the pruned GTP still closes. The length of this new segment is given by the increase of vertical extension from the $L_\beta$ minus the length if the removed edge segment.

In general, performing the pruning along either the positive or negative direction, $s=\pm 1$, results in a set of GTPs that are related by an $SL(2,\Z)$ transformation. In the web this choice reflects the two possible ways of exchanging two 7-branes with 5-branes attached to them, such that the total monodromy is conserved.

{\proposition[Positive and Negative Monodromy Transitions]
Let $P'_\pm=(P,\mu_{\alpha,j},s=\pm)$ be the two different GTPs obtained from $P$ by pruning along $\mu_{\alpha,j}$ with either choice of $s=\pm$. Then,
\be
P'_+ \cong P'_-\,,
\ee
up to $SL(2,\Z)$ transformation. If $L_\alpha=(x_\alpha,y_\alpha)$ then the relevant $SL(2,\Z)$ transformation is
\be
\phi_{L_\alpha} = \begin{pmatrix} 1- x_\alpha y_\alpha & x_\alpha^2 \\ -y_\alpha^2 & 1+ x_\alpha y_\alpha \end{pmatrix}\,,
\ee
that leaves $L_\alpha$ (and $-L_\alpha$) invariant.
}\\

It is also important to note that pruning defines an equivalence class of GTPs, since the process is reversible.  

{\proposition[Reversibility of Pruning]
Let $P'=(P,\mu_{\alpha,j},s)$ be a pruning of $P$. Then
\be
P=(P',\mu_{{-\alpha},j},-s)\,, \qquad \mu_{{-\alpha},j} = \left(-\mu_{\alpha,j} + \sum_{E_\beta \in E_{-s}} \left|\det(L_\alpha,L_\beta)\right| \lambda_\beta\right)\,,
\ee
where $\mu'_{-\alpha,j}$ lies along the edge $-L_\alpha$. Thus, pruning is reversible and all GTPs that can be obtained from each other by pruning are equivalent.
}\\

\paragraph{Implementation in the code.} The {\tt Mathematica} notebook contains a function \texttt{prune} which takes a polygon in argument, and returns another polygon obtained from the original one after one or more operations of pruning. The algorithm tries to minimize the slopes of the edges and the diameter of the polygon by systematically applying all possible pruning operations, and stops when a stable minimum is reached. Details can be found in the notebook.

\providecommand{\href}[2]{#2}\begingroup\raggedright\endgroup

\end{document}